\documentclass[rmp,twocolumn,showpacs,preprintnumbers,floatfix]{revtex4}

\usepackage{graphicx} 
\usepackage{dcolumn} 
\usepackage{bm} 
\usepackage{epsf}
\usepackage{times}
\usepackage{amsmath,amssymb}
\usepackage{float}
\begin{document}

\title{Hadron structure at low $Q^2$.}
\thanks{This review is dedicated to the memory of the late Peter Brix
(MPI Heidelberg), the Nestor of low $Q^2$ physics.}
\author{Dieter Drechsel}
\author{Thomas Walcher}
\affiliation{
  Institut f{\"u}r Kernphysik, Johannes Gutenberg-Universit{\"a}t Mainz\\
  D-55099 Mainz, Germany
}
\date{\today}
\begin{abstract}
This review deals with the structure of hadrons, strongly interacting many-body
systems consisting of quarks and gluons. These systems have a size of about
1~fm, which shows up in scattering experiments at low momentum transfers $Q$ 
in the GeV region. At this scale the running coupling constant of Quantum
Chromodynamics (QCD), the established theory of the strong interactions,
becomes divergent. It is therefore highly intriguing to explore this theory in
the realm of its strong interaction regime. However, the quarks and gluons
can not be resolved at the GeV scale but have to be studied through their 
manifestations in the bound many-body systems, for instance pions, nucleons 
and their resonances. The review starts with a short overview of QCD at low 
momentum transfer and a summary of the theoretical apparatus describing the 
interaction of hadrons with electrons and photons. In the following sections 
we present the experimental results for the most significant observables
studied with the electromagnetic probe: form factors, polarizabilities, 
excitation spectra, and sum rules. These experimental findings are compared 
and interpreted with various theoretical approaches to QCD, such as 
phenomenological models with quarks and pions, dispersion relations as a 
means to connect observables from different experiments, and, directly based 
on the QCD lagrangian, chiral perturbation theory and lattice gauge theory.

\end{abstract}

\maketitle
\tableofcontents
\section{INTRODUCTION}
\label{sec:I}
Hadrons are composite systems with many internal degrees of freedom. The
strongly interacting constituents of these systems, the quarks and gluons are
described by Quantum Chromodynamics (QCD). This theory is asymptotically free,
that is, it can be treated in a perturbative way for very large values of the
four-momentum transfer squared, $Q^2$
~\cite{Gross:1973id,Gross:1973ju,Politzer:1973fx}. However, the binding forces
become increasingly strong if the momentum transfer decreases towards the
region of about 1~GeV, which is the natural habitat of nucleons and pions. In
particular, the ``running'' coupling constant of the strong interaction,
$\alpha_s(Q^2)$, is expected to diverge if $Q^2$ decreases to values near
$\Lambda^2_\text{QCD} \approx(250~\text{MeV})^2$, which defines the ``Landau
pole'' of QCD. This behavior is totally different from Quantum Electrodynamics
(QED), for which the coupling constant $\alpha_\text{em}(Q^2)$ diverges for
huge momentum transfers at the Planck scale, corresponding to $Q^2\approx
10^{38}~\text{GeV}^2$ or $10^{-35}$~m, much below any distance ever to be
resolved by experiment. On the contrary, the Landau pole of QCD corresponds to
a resolution of the nucleon's size, somewhat below 1~fm or $10^{-15}$~m. This
is the realm of non-perturbative QCD, in which quarks and gluons appear as
clusters confined in the form of color-neutral hadrons. As of today it is an
open question whether this confinement can be derived directly from QCD or
whether it is a peculiarity of a strongly interacting many-body system or based
on some deeper grounds. Therefore, the study of QCD in the non-perturbative
domain serves less as a check of QCD per se, but is concerned with the highly
correlated many-body system ``hadron'' and its effective degrees of freedom.\\

Quantum Chromodynamics is a non-abelian gauge theory developed on the basis of
quarks and gluons~\cite{Gross:1973id,Weinberg:1973un,Fritzsch:1973pi}. The
non-abelian nature of this theory gives rise to a direct interaction among the
gluons, and the forces among the quarks are mediated by the exchange of gluons
whose chromodynamic vector potential couples to the vector current of the
quarks. If massless particles interact via their vector current, the helicity
(handedness or chirality) of the particles is conserved. The nucleon is
essentially made of the light $u$ and $d$ quarks plus a small admixture of $s$
quarks, with masses $m_u=1.5$ to 3.0~MeV, $m_d=3$ to 7~MeV, and $m_s=(95 \pm
25)$~MeV~\cite{Yao:2006px}. In the zero mass limit, these light quarks can be
classified according to their chirality by the group SU(3)$_R\otimes$
SU(3)$_L$. Several empirical facts give rise to the assumption that this
symmetry is spontaneously broken down to its vectorial subgroup, and in
addition the finite quark masses cause an explicit symmetry breaking.  The
spontaneously broken symmetry is a most remarkable feature of QCD, because it
can not be derived from the Lagrangian. This is quite different from the
explicit symmetry breaking, which is put in by design through the finite quark
masses in QCD and appears in a similar way in the Higgs sector. As a result one
obtains the conserved vector currents $J_{\mu}^a$ and the only partially
conserved axial vector currents $J_{5\mu}^a$,
\begin{equation}
J_{\mu}^a=\bar{q}\gamma_{\mu}\frac{\lambda^a}{2}q\ , \ \ \
J_{5\mu}^a=\bar{q}\gamma_{\mu}\gamma_5\frac{\lambda^a}{2}q\ , \label{eq:1.1}
\end{equation}
where $q$ are Dirac spinors of point-like (light) quarks and
$\gamma_{\mu},\gamma_5$ the appropriate Dirac matrices. The quantities
$\lambda^a$, $a=1\ ...\ 8$ denote the Gell-Mann matrices of SU(3) describing
the flavor structure of the 3 light quarks, and $\lambda^0$ is the unit matrix.
The photon couples to the quarks by the electromagnetic vector current
$J_{\mu}^\text{em}\sim J_{\mu}^{(3)}+\frac{1}{\sqrt{3}}J_{\mu}^{(8)}$,
corresponding to isovector and isoscalar interactions respectively. The weak
neutral current mediated by the $Z^0$ boson couples to the $3^\text{rd}$,
$8^\text{th}$, and $0^\text{th}$ components of both vector and axial currents.
While the electromagnetic current is always conserved,
$\partial^{\mu}J_{\mu}^\text{em}=0$, the axial current is exactly conserved
only for massless quarks. In this limit there exist conserved charges $Q^a$ and
axial charges $Q_5^a$, which are connected by commutation relations. The
corresponding ``current algebra'' predated QCD and was the basis of various
low-energy theorems (LETs), which govern the low-energy behavior of
(nearly) massless particles.\\

The puzzle we encounter in the physics of hadrons is the following: The
massless quarks appearing in the QCD Lagrangian must conserve the axial
currents. The nucleons should eventually emerge from the same Lagrangian as
massive many-body systems of quarks and gluons. However, the conservation of
the axial current in the Wigner-Weyl mode would require a vanishing axial
coupling constant for these massive nucleons, which is ruled out by the
observed $\beta$ decay. A solution of this puzzle was given by Goldstone's
theorem. At the same time as the ``three-quark system'' nucleon becomes massive
by means of the QCD interaction, the vacuum develops a nontrivial structure due
to finite expectation values of quark-antiquark pairs (condensates
$\langle\bar{q}q\rangle$), and so-called Goldstone bosons are created,
$\bar{q}q$ pairs with the quantum numbers of pseudoscalar mesons. These
Goldstone bosons are massless, and together with the massive nucleons they
warrant the conservation of the axial current. Because the quarks are not
really massless, the chiral symmetry is slightly broken in nature. As a
consequence also the physical Goldstone bosons acquire a finite mass, in
particular the pion mass $m_{\pi}$ follows to lowest order from the
Gell-Mann-Oakes-Renner relation,
\begin{equation}
m_{\pi}^2f_{\pi}^2 = -(m_u+m_d)\langle\bar{q}q\rangle + \cdots \, ,
\label{eq:1.3}
\end{equation}
with the condensate $\langle\bar{q}q\rangle\approx -(225~{\mbox{MeV}})^3$ and
$f_{\pi}\approx93~$MeV the pion decay constant. Since the pions are now
massive, the corresponding axial currents are no longer conserved and the
4-divergence of the axial current becomes finite,
\begin{equation}
\partial^{\mu}J_{5\mu}^a\approx -f_{\pi}m_{\pi}^2\phi_{\pi}^a\, ,
\label{eq:1.4}
\end{equation}
where $\phi_{\pi}^a$ describes the local pion field. In other words the charged
pion decay $\pi^+\rightarrow\mu^++\nu_{\mu}$ and $\pi^-\rightarrow
\mu^-+\bar{\nu}_{\mu}$ proceeds via coupling to the axial current of
Eq.~(\ref{eq:1.4}). While the charged pions decay weakly with a life-time of
$2.6\times10^{-8}$~s, the neutral pion decays much faster, in
$8.4\times10^{-17}$~s, by means of the electromagnetic interaction,
$\pi^0\rightarrow\gamma+\gamma$. This provides an additional source for the
neutral pion or the axial current with index 3,
\begin{equation}
\partial^{\mu}J_{5\mu}^3 = \frac{\alpha_\text{em}}{\pi}\
\vec{E}\cdot\vec{B}\ , \label{eq:1.7}
\end{equation}
where $\vec{E}$ and $\vec{B}$ are the electromagnetic fields. We note that the
scalar product of the two electromagnetic fields is a pseudoscalar. This decay
can be mediated by a triangle of intermediate quark lines, and therefore it is
often called the triangle anomaly. It is ``anomalous'' because such processes
do not appear in classical theories but only in quantum field theories through
the renormalization procedure (Wess-Zumino-Witten term). The analogous anomaly
in QCD is obtained from Eq.~(\ref{eq:1.7}) by replacing the electromagnetic by
the corresponding color fields, $\vec{E}_c$ and $\vec{B}_c$, $\alpha_\text{em}$
by the strong coupling constant $\alpha_s$, and with an additional factor 3 for
$u$, $d$, and $s$ quarks,
\begin{equation}
\partial^{\mu}J_{5\mu}^{0} = 3\ \frac{\alpha_s}{\pi}\
\vec{E}_c\cdot\vec{B}_c \ . \label{eq:1.8}
\end{equation}
As a consequence also $J_{5\mu}^{0}$ is not conserved, not even
for massless quarks (``$U_A(1)$ anomaly'').\\

Unfortunately, no ab-initio calculation can yet describe the intriguing but
complicated world of the confinement region. In principle, lattice gauge theory
should have the potential to describe QCD directly from the underlying
Lagrangian. This theory discretizes QCD on a four-dimensional space-time
lattice and approaches the physical world in the continuum limit of vanishing
lattice constants~\cite{Wilson:1974sk,Kogut:1974ag}. However, these
calculations can as yet only be performed with $u$ and $d$ quark masses much
larger than the ``current quark masses'' mentioned above, and therefore also
the pion mass turns out much too large. As a consequence the Goldstone
mechanism, the abundant production of sea quarks is much suppressed. Lattice
gauge theory has progressed considerably over the past decade and further
progress is foreseen by both improved algorithms and increased computing power.
For recent developments see the following references:
\cite{Alexandrou:2006pr,Alexandrou:2006ru,Boinepalli:2006xd,Gockeler:2006uu,Edwards:2006qx}.
Semi-quantitative agreement has been reached for ratios of masses and magnetic
moments for the hadrons, there also exist predictions for nucleon resonances
and electromagnetic form factors in qualitative agreement with the data.
However, some doubt may be in order whether such procedure can ever fully
describe the pionic degrees of freedom in hadronic physics, particularly in the
context of pion production and similar reactions.\\

A further ab-initio calculation is chiral perturbation theory (ChPT), which has
been established by \textcite{Weinberg:1978kz} in the framework of effective
Lagrangians and put into a systematic perturbation theory by
\textcite{Gasser:1984gg,Gasser:1983yg}. This theory is based on the chiral
symmetry of QCD, which is however expressed by effective degrees of freedom,
notably the Goldstone bosons. Because of the Goldstone mechanism, the threshold
interaction of pions and other Goldstone bosons is weak not only among
themselves but also with the nucleons. Furthermore the pion mass is small and
related to the small quark masses $m_u$ and $m_d$ according to
Eq.~(\ref{eq:1.3}). Based on these grounds, ChPT has been set up as a
perturbation theory in the parameters $p:=(p_1,p_2,...;m_u,m_d)$, where $p_i$
are the external 4-momenta in a particular (Feynman) diagram. Chiral
perturbation theory has been applied to many photoinduced reactions by
\textcite{Bernard:1991rt,Bernard:1995dp,Bernard:1993bg} in the 1990's. As a
result several puzzles have been solved and considerable insight has been
gained. However, ChPT can not be renormalized as QED by adjusting a few
parameters to the observables. Instead, the appearing infinities must be
removed order by order in the perturbation series. This renormalization
procedure gives rise to a growing number of low-energy constants (LECs)
describing the strength of all possible effective Lagrangians consistent with
the symmetries of QCD, at any given order of the perturbation series. These
LECs, however, can not (yet) be derived from QCD but must be fitted to the
data, which leads to a considerable loss of predictive power with increasing
order of perturbation. A further problem arises in the nucleonic sector because
of the large nucleon mass $M$, which is of course not a small expansion
parameter. The latter problem was solved by heavy baryon ChPT, a kind of
Foldy-Wouthuysen expansion in $1/M$. This solution was however achieved at the
expense of approximating the relativistic description by a non-relativistic
one. Over the past decade new schemes have been developed, which provide a
consistent expansion within a manifestly Lorentz invariant
formalism \cite{Becher:1999he,Kubis:2000zd,Schindler:2003xv,Fuchs:2003qc}. For
recent reviews of ChPT see the work of \textcite{Scherer:2002tk} and
\textcite{Bernard:2007zu}.\\

If quarks and gluons are resolved at high momentum transfer, they are
asymptotically free and their momentum distribution can be described by
evolution functions as derived from perturbative expansions (``higher
twists''). This domain has been studied in great detail ever since the
discovery of parton scaling at the end of the 1960's. Such investigations have
given confidence in the validity of the QCD Lagrangian. Systems of heavy quarks
(charm, bottom, top) can be well described by effective field theories based on
QCD. However, these approaches are less effective for systems of light quarks
(up, down, strange), for which the sea quarks and notably pionic degrees of
freedom become very important. In order to incorporate the consequences of
chiral symmetry, a plethora of hybrid models with quarks and pions has been
developed. Quark models have been quite successful in predicting the resonance
spectrum of the nucleon as well as the electromagnetic decay and excitation of
these resonances. However, they have problems to describe the spectrum and the
size of the nucleon at the same time. We do not dwell on these models in the
review but occasionally refer to them at later stages. \\

On the following pages we concentrate on hadronic structure investigations with
the electromagnetic probe, that is electron and photon scattering as well as
electro- and photoproduction of mesons. A broader account can be found in the
book of \textcite{Thomas:2001kw}. In section~\ref{sec:II} we give a brief 
introduction to the formalism relevant for these studies. The following
section~\ref{sec:III} summarizes the information on the form factors of
nucleons and pions. Another bulk property of the hadrons is their
polarizability, which can be determined by Compton scattering as discussed in
section~\ref{sec:IV}.  These global properties are related to the excitation
spectrum of the particles, meson production at threshold, resonances and
continuum backgrounds as detailed in section~\ref{sec:V}. Finally, we examine
the origin and relevance of several sum rules in section~\ref{sec:VI}. In all
of these fields, there has been a rapid evolution over the past years triggered
by high-precision experiments. These have been made possible by a new
generation of electron accelerators with high current, high duty-factor, and
highly polarized beams in combination with improved target and detection
techniques, notably for polarized particles. In many of the presented phenomena
we recover the role of the pion as an effective degree of freedom at low energy
and momentum transfer to the nucleon. It is therefore the leitmotiv of this
review to look at the hadrons as interesting and complicated many-body systems
whose direct description by QCD proper is a major challenge for particle
physics over the years to come.


\section{The electromagnetic interaction with hadrons}
\label{sec:II}
\subsection{Kinematics}\label{sec:II.1}
Let us consider the kinematics of the reaction
\begin{equation}\label{eq:2.1}
e(k_1)+N(p_1)\rightarrow e(k_2)+N(p_2)\ ,
\end{equation}
with $k_i=(\omega_i,\vec{k}_i)$ and $p_i=(E_i,\vec{p}_i)$ denoting the
four-momenta of an electron $e$ with mass $m$ and a nucleon $N$ with mass $M$.
The 4-momenta are constrained by the on-shell conditions $p_1^2=p_2^2=M^2, \,
k_1^2=k_2^2=m^2$, and by the conservation of total energy and momentum,
$k_1+p_1=k_2+p_2$. In order to make Lorentz-invariance manifest, it is useful
to express the amplitudes in terms of the 3 Mandelstam variables
\begin{equation}\label{eq:2.2}
s=(k_1+p_1)^2\ \ ,\ \ t=(k_2-k_1)^2\ \ ,\ \ u=(p_2-k_1)^2\ .
\end{equation}
Due to the mentioned constraints, these variables fulfill the relation
$s+t+u=2(m^2+M^2)$, and therefore we may choose $s$ and $t$ as the two
independent Lorentz scalars. For reasons of symmetry, the center-of-mass (cm)
system is used in the following. The 3-momenta of the particles cancel in this
system, and therefore $s=(\omega_{\text{cm}}+E_{\text{cm}})^2=W^2$, i.e., the
Mandelstam variable $s$ is the square of the total cm energy $W$. Furthermore,
the initial and final energies of each particle are equal, and hence
$t=-(\vec{k}_2-\vec{k}_1)^2_{\text{cm}}$ is related to the 3-momentum transfer
in the cm system. From these definitions it follows that physical processes
occur at $s> (m+M)^2$ and $t<0$. Because of the smallness of the fine-structure
constant $\alpha_\text{em} \approx 1/137$, it is usually sufficient to treat
electron scattering in the approximation that a single photon with momentum
$q=k_1-k_2=(\omega , \vec{q})$ is transferred to the hadronic system. We call
this particle a space-like virtual photon $\gamma^{\ast}$, because $t=q^2<0$,
i.e., the space-like component of the 4-vector $q$ prevails. Since $t$ is
negative in the physical region of electron scattering, it is common use to
describe electron scattering by the positive number $Q^2=-q^2$. This contrasts
the situation in pair annihilation, $e^+e^-\rightarrow\gamma^{\ast}$, which
produces a time-like virtual photon with $q^2=m_{\gamma^{\ast}}^2>0$. The above
considerations can be applied to real Compton scattering (RCS),
\begin{equation}\label{eq:2.3}
\gamma(k_1)+N(p_1)\rightarrow\gamma(k_2)+N(p_2)\ ,
\end{equation}
by replacing $m_{1,2}^2=k_{1,2}^2 \rightarrow 0$ and to virtual Compton
scattering (VCS),
\begin{equation}\label{eq:2.4}
\gamma^{\ast}(k_1)+N(p_1)\rightarrow\gamma(k_2)+N(p_2)\ ,
\end{equation}
by replacing $m_1^2=k_1^2 \rightarrow q^2<0$ and $m_2^2=k_2^2 \rightarrow 0$.\\

Let us now turn to the spin degrees of freedom. The virtual photon with
momentum $\vec{q}$ carries a polarization described by the vector potential
$\vec{A}$, which has both a transverse component, $\vec{A}_T\perp \vec{q}$, as
in the case of a real photon, and a longitudinal component
$\hat{q}\cdot\vec{A}$, which is related to the time-like component $A_0$ by
current conservation, $q\cdot A=\omega A_0-\vec{q}\cdot\vec{A}=0$. Since the
electron is assumed to be highly relativistic, its spin degrees of freedom are
described by the helicity $h=\vec{s}\cdot\hat{k}=\pm \textstyle{\frac{1}{2}}$,
the projection of the spin $\vec{s}$ on the momentum unit vector $\hat{k}$. In
the following we denote the polarization of the incident electron by $P_e = 2 h
= \pm 1$, for example, $P_e=1$ describes a beam of fully polarized right-handed
electrons. The polarization vector $\vec{P}$ of a target or recoil nucleon is
represented in a coordinate system with the z-axis pointing in the direction of
the virtual photon, $\hat{e}_z=\hat{q}$, the y-axis perpendicular to the
electron scattering plane, $\hat{e}_y \sim \hat{k}_1 \times \hat{k}_2$, and the
x-axis ``sideways'', i.e., in the scattering plane and on the side of the
outgoing electron.\\

The scattered electron probes the charge and magnetization distributions of the
hadronic system via the interaction of the electromagnetic currents, which
leads to a transition matrix element $\mathcal{M} = \sum _{\mu} j_{\mu}(e)
J^{\mu}(h)$. If we neglect higher order QED corrections, the electron is a
Dirac point particle with its current given by $j_{\mu} = -e \bar{u}_2
\gamma_{\mu} u_1$, where $\gamma_{\mu}$ are Dirac matrices and $u_i$ Dirac
spinors characterized by the quantum numbers $i= \{ \vec{k}_i, h_i \}$. In the
one-photon exchange approximation, the cross section is then obtained by the
square of the transition matrix element multiplied by phase space factors,
\begin{equation}\label{eq:2.5}
d \sigma \sim \sum_{\text{spins}} |\mathcal{M}|^2 \sim \sum_{\text{spins}}
\eta_{\mu\nu}(e)W^{\mu\nu}(h) \, ,
\end{equation}
where $\eta_{\mu\nu}=j_{\mu}j_{\nu}^{\ast}$ can be calculated
straightforwardly. By varying the incident electron energy and the scattering
angle as well as the polarizations of the respective particles, it is then
possible to enhance or suppress specific components of the hadronic tensor
$W_{\mu\nu}=J_{\mu}J_{\nu}^{\ast}$, and thus to study different aspects of the
hadronic structure in a model-independent way. For further details and a
general introduction to the structure of hadrons and nuclei, we refer to the
book of \textcite{Boffi:1996si} (see also \cite{Boffi:1993gs}).
\subsection{Elastic electron scattering}
\label{sec:II.2}
The hadronic current for elastic electron scattering off the nucleon is given
by the most general form for the vector current with the same
spin-$\frac{1}{2}$ particle before and after the collision,
\begin{equation}\label{eq:2.6}
J_{\mu} = \bar{u}_{p_2} \left ( \gamma_{\mu}\,
F_1(Q^2)+i\frac{\sigma_{\mu\nu}q^{\nu}}{2M} \, F_2(Q^2) \right ) u_{p_1}\ ,
\end{equation}
where $u_{p_1}$ and $u_{p_2}$ are the 4-spinors of the nucleon in the initial
and final states, respectively. The first structure on the rhs of
Eq.~(\ref{eq:2.6}) is the Dirac current, which describes the finite size of the
nucleon by the Dirac form factor $F_1(Q^2)$. The second term reflects the fact
that the internal degrees of freedom also produce an anomalous magnetic moment
$\kappa$ whose spatial distribution is described by the Pauli form factor
$F_2(Q^2)$. These form factors are normalized to $F_1^p(0)=1$, $F_2^p
(0)=\kappa_p=1.79$ and $F_1^n (0)=0$, $F_2^n (0)=\kappa_n=-1.91$ for proton and
neutron,
respectively.\\

From the analogy with non-relativistic physics, it is seducing to associate the
form factors with the Fourier transforms of the charge and magnetization
densities. The problem is that the charge distribution $\rho(\vec{r})$ has to
be calculated by a 3-dimensional Fourier transform of the form factor as
function of $\vec{q}$, whereas the form factors are generally functions of
$Q^2=\vec{q}\ ^2-\omega^2$. However, there exists a special Lorentz frame, the
Breit or brick-wall frame, in which the energy of the (space-like) virtual
photon vanishes. This can be realized by choosing, for example,
$\vec{p}_1=-\frac {1}{2} \vec{q}$ and $\vec{p}_2=+\frac {1}{2} \vec{q}$ leading
to $E_1=E_2=(M^2+\frac {1}{4}\vec{q}\ ^2)^{1/2}$, $\omega=0$, and $Q^2=\vec{q}\
^2$. Equation~(\ref{eq:2.6}) takes the following form in this
frame~\cite{Sachs:1962dd}:
\begin{equation}\label{eq:2.7}
J_{\mu}  = \left ( G_E(Q^2)\ ,\ i\frac{\vec{\sigma}\times\vec{q}}{2M} G_M(Q^2)
\right ) \ ,
\end{equation}
where $G_E(Q^2)$ stands for the time-like component of $J_{\mu}$ and hence is
identified with the Fourier transform of the electric charge distribution,
while $G_M(Q^2)$ appears with a structure typical for a static magnetic moment
and hence is interpreted as Fourier transform of the magnetization density. The
Sachs form factors $G_E$ and $G_M$ are related to the Dirac form factors by
\begin{eqnarray}
G_E (Q^2) & = & F_1 (Q^2) - \tau F_2 (Q^2)\ ,\nonumber \\
G_M (Q^2) & = & F_1 (Q^2) + F_2 (Q^2)\ , \label{eq:2.8}
\end{eqnarray}
where $\tau=Q^2/4M^2$ is a measure of relativistic (recoil) effects. Although
Eq.~(\ref{eq:2.8}) is a covariant definition, the Sachs form factors can only
be Fourier transformed in a special frame, namely the Breit frame, with the
result
\begin{eqnarray}\label{eq:2.9}
G_E (\vec{q}\ ^2)& = &\int\rho(\vec{r}) e^{i\vec{q}\cdot\vec{r}}
d^3\vec{r} \\
& = & \int\rho(\vec{r})d^3\vec{r}-\frac{\vec{q}\ ^2}{6}
\int\rho(\vec{r})\vec{r}\ ^2 d^3\vec{r} +\ ...\ ,\nonumber
\end{eqnarray}
where the first integral yields the total charge in units of $e$, i.e., 1 for
the proton and 0 for the neutron, and the second integral defines the square of
the electric root-mean-square (rms) radius, $\langle r^2\rangle_E$. We note
that each value of $Q^2$ requires a particular Breit frame, i.e., the
information on the charge distribution is taken from an infinity of different
frames, which is then used as input for the Fourier integral in terms of
$G_E(\vec{q}\ ^2)$. Therefore, the density $\rho(\vec{r})$ is not an observable
that we can ``see'' in any particular Lorentz frame but only a mathematical
construct in analogy to a ``classical'' charge distribution. The problem is
that an ``elementary'' particle has a small mass such that recoil effects,
measured by $\tau$, and size effects, measured by $\langle r^2\rangle$, become
comparable and can not be separated in a unique way. The situation is
numerically quite different for a heavy nucleus, in which case the
size effects dominate the recoil effects by orders of magnitude.\\

Because the hadronic current is completely defined by Eq.~(\ref{eq:2.6}), any
observable for elastic electron scattering can be uniquely expressed in terms
of the two form factors. In particular the unpolarized differential cross
section is given by \cite{Rosenbluth:1950dd}
\begin{equation}\label{eq:2.10}
\frac{d\sigma}{d\Omega} =\left (\frac{d\sigma}{d\Omega} \right )_0 \left (
\frac{G_E^2+\tau G_M^2}{1+\tau}+2\tau\tan^2\frac{\theta}{2}G_M^2 \right )\, ,
\end{equation}
with $(d\sigma/d\Omega)_0$ the cross section for electron scattering off a
point particle and $\theta$ the scattering angle of the electron in the
laboratory (lab) system. Equation~(\ref{eq:2.10}) gives us the possibility to
separate the form factors by variation of $\tan^2\frac{\theta}{2}$ while
keeping $Q^2$ constant. In fact the data should lie on a straight line
(``Rosenbluth plot'') with a slope that determines the magnetic form factor
$G_M$. However, there are limits to this method, in particular if one of the
form factors is very much smaller than the other. In such cases a
double-polarization experiment can help to get independent and more precise
information. Such an experiment requires a polarized electron beam and a
polarized target, or equivalently the measurement of the nucleon's polarization
in the final state. The measured asymmetry takes the form \cite{Arnold:1980zj}
\begin{equation}\label{eq:2.11}
  \mathcal {A} = - P_e \frac{\sqrt{2\tau\varepsilon(1-\varepsilon)}\
G_EG_MP_x+\tau\sqrt{1-\varepsilon^2}\ G_M^2\ P_z} {\varepsilon\ G_E^2+\tau\
G_M^2}\ ,
\end{equation}
where $\varepsilon=1/[ 1+2(1+\tau) \tan^2 \frac{\theta}{2}]$ is the transverse
polarization of the virtual photon. In particular we find that the
longitudinal-transverse interference term, appearing if the nucleon is
polarized perpendicularly (sideways) to $\vec{q}$, is proportional to $G_EG_M$,
while the transverse-transverse interference term, appearing for polarization
in the $\vec{q}$ direction, is proportional to $G_M^2$. The ratio of both
measurements then determines $G_E/G_M$ with high precision, because most
normalization and efficiency factors cancel.
\subsection{Parity violating electron scattering}
\label{sec:II.3}
In the previous section we have tacitly assumed that the interaction between
electron and hadron is mediated by the virtual photon and therefore parity
conserving. With this assumption the polarization of only one particle does not
yield any observable effect. However, it is also possible to exchange a Z$^0$
gauge boson, although this is much suppressed in the low-energy region because
of the large mass $M_{Z^0}= 91$\,GeV. This boson couples to electrons and
nucleons with a mixture of vector and axial vector currents typical for the
weak interaction. If the Z$^0$ is emitted from one of the particles by the
vector coupling and absorbed by the other one by the axial vector coupling, it
produces a parity-violating asymmetry that can be observed if one of the
particles (typically the incident electron) is polarized. The coupling of the
Z$^0$  to the electron involves the current
\begin{equation}\label{eq:2.12}
\tilde {j}_{\mu} = g' \bar{u}_2 [\gamma_{\mu}(1-4
\sin^2\Theta_W)-\gamma_{\mu}\gamma_5] u_1\, ,
\end{equation}
where $\Theta_W $ is the Weinberg angle and $g'$ a weak coupling constant.
Because $\sin^2\Theta_W \approx 0.23$, the vector current in
Eq.~(\ref{eq:2.12}) is largely suppressed compared to the axial vector part
containing the $\gamma_5$ factor. The corresponding weak hadronic current can
be parameterized as follows:
\begin{eqnarray}
\tilde {J}_{\mu} & = & \bar{u}_{p_2} [ \gamma_{\mu}\, \tilde
{F}_1(Q^2)+i\frac{\sigma_{\mu\nu}q^{\nu}}{2M}\, \tilde {F}_2(Q^2) \nonumber \\
&&  + \gamma_{\mu} \gamma_5 \, \tilde {G}_A(Q^2)] u_{p_1}\, ,\label{eq:2.13}
\end{eqnarray}
where the tilde signifies the coupling to the Z$^0$. The weak Sachs form
factors are defined as in Eq.~(\ref{eq:2.8}), and the cross sections and
asymmetries are calculated as in the previous section. However, the
contribution of the weak current to the differential cross section is well
below the experimental error bars, and information can only be obtained from
the interference between the electromagnetic and the weak transition
amplitudes. The parity-violating asymmetry $\tilde{ \mathcal {A}} =
(d\sigma^{+} - d\sigma^{-}) / (d\sigma^{+} +d\sigma^{-})$, where $d\sigma^+$
and $d\sigma^-$ are the differential cross sections for incident electrons with
positive and negative helicities, takes the form
\begin{eqnarray}
&&\tilde{ \mathcal {A}} = - \frac{G_F}{4 \pi \alpha \sqrt {2}} \nonumber \\
&&\times \bigg (\frac{\varepsilon G_E\tilde{G}_E+\tau G_M\tilde{G}_M-
\varepsilon '(1-4\sin^2\Theta_W)G_M\tilde{G}_A }
{\varepsilon G_E^2+\tau G_M^2}\bigg )\nonumber \\
&&= {\cal A}^E (\tilde{G}_E) + {\cal A}^M (\tilde{G}_M) + {\cal A}^A
(\tilde{G}_A) \, , \label{eq:2.14}
\end{eqnarray}
with $\varepsilon '=\sqrt {(1-\varepsilon^2) \tau (1+\tau)}$.
\subsection{Pseudoscalar meson electroproduction}
\label{sec:II.4}
\begin{figure}[]
\begin{center}
\includegraphics[width=0.6\columnwidth,angle=0]{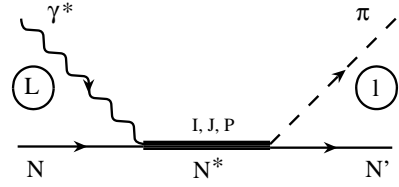}
\end{center}
\caption {Multipole notation for pion photoproduction. See text for further
explanation. \label{fig:II.1}}
\end{figure}

The reaction
\begin{equation}
\gamma^{\ast}(q) + N (p_1) \rightarrow \pi(k) + N(p_2) \label{eq:2.15}
\end{equation}
is described by the transition matrix element $\varepsilon^{\mu}J_{\mu}$, with
$\varepsilon^{\mu}$ the polarization of the (virtual) photon and $J_{\mu}$ the
transition current leading from the nucleon's ground state to a meson-nucleon
continuum. This current can be expressed by 6 different Lorentz structures
constructed from the independent momenta and appropriate Dirac matrices. Since
the photon couples to an electromagnetic vector current and the pion is of
pseudoscalar nature, the transition current appears as an axial 4-vector in the
nucleon sector. The space-like ($\vec{J}$) and time-like ($\rho$) components of
the transition operator take the following form in the hadronic cm frame:
\begin{eqnarray}
\vec{J} & = & \tilde{\sigma} F_1+i(\hat{q}\times\vec{\sigma})
(\vec{\sigma}\cdot\hat{k}) F_2+\tilde{k}(\vec{\sigma}\cdot\hat{q}) F_3
\nonumber \\ && +\tilde{k} (\vec{\sigma}\cdot\hat{k}) F_4+\hat{q}
(\vec{\sigma}\cdot\hat{q}) F_5+\hat{q}
(\vec{\sigma}\cdot\hat{k}) F_6\ , \label{eq:2.16} \\
\rho & = & (\vec{\sigma}\cdot\hat{k}) F_7+(\vec{\sigma}\cdot\hat{q}) F_8 \, ,
\label{eq:2.17}
\end{eqnarray}
with $\hat{q}$ and $\hat{k}$ the 3-momentum unit vectors of virtual photon and
pion, respectively, and $F_1$ to $F_8$ the CGLN amplitudes \cite{Chew:1957tf}.
The structures in front of the $F_i$ are all the independent axial vectors and
pseudoscalars that can be constructed from the Pauli spin matrix $\vec{\sigma}$
and the independent cm momenta $\vec{k}$ and $\vec{q}$. We further note that
$\tilde{\sigma}$ and $\tilde{k}$ are the transverse components of
$\vec{\sigma}$ and $\hat{k}$ with regard to $\hat{q}$.  With these definitions
$F_1$ to $F_4$ describe the transverse, $F_5$ and $F_6$ the longitudinal, and
$F_7$ and $F_8$ the time-like or Coulomb components of the current. The latter
ones are related by current conservation, $\vec{q}\cdot\vec{J}-\omega\rho=0$,
leading to $|\vec{q}|F_5=\omega F_8$ and $|\vec{q}|F_6=\omega F_7$. The CGLN
amplitudes depend on the virtuality of the photon, $Q^2$, as well as the total
hadronic energy $W$ and the pion-nucleon scattering angle $\theta_{\pi}^{\ast}$
in the hadronic cm system. These amplitudes are complex functions, because the
transition leads to a continuum state with a complex phase factor. They can be
decomposed in a series of multipoles (see \textcite{Drechsel:1992pn} for
further details),
\begin{equation}
\mathcal {M}_{l\pm} = \{ E_{l\pm}, \,M_{l\pm}, \,L_{l\pm}, \,S_{l\pm}\}\, ,
\label{eq:2.18}
\end{equation}
where $E_{l\pm}$,  $M_{l\pm}$,  $L_{l\pm}$, and $S_{l\pm}$ denote the
transverse electric, transverse magnetic, longitudinal, and scalar (time-like
or Coulomb) multipoles, in order. The latter two are related by gauge
invariance, $|\vec{q}| L_{l\pm}=\omega S_{l\pm}$, and therefore we may drop the
longitudinal multipoles in the following without loss of generality. The CGLN
multipoles are complex functions of 2 variables, ${\cal M}_{\ell\pm}={\cal
M}_{\ell\pm}(Q^2,W)$. The notation of the multipoles is clarified by
Fig.~\ref{fig:II.1}. The incoming photon carries the multipolarity $L$, which
is obtained by adding the spin 1 and the orbital angular momentum of the
photon. The parity of the multipole is ${\cal P}=(-1)^L$ for $E$, $L$, and $S$,
and ${\cal P}=(-1)^{L+1}$ for $M$. The photon couples to the nucleon with spin
$\frac {1}{2}$ and ${\cal P}=+1$, which leads to hadronic states of spin $J=\,
\mid L\pm\frac{1}{2}\mid$ and with the parity of the incoming photon. The
outgoing pion has negative intrinsic parity and orbital angular momentum $l$,
from which we can reconstruct the spin $J=\, \mid l\pm\frac{1}{2}\mid$ and the
parity ${\cal P}=(-1)^{l+1}$ of the excited hadronic state. This explains the
notation of the multipoles, Eq.~(\ref{eq:2.18}), by the symbols $E$, $M$, and
$S$ referring to the type of the photon, and by the index $l\pm$ with $l$
standing for the pion angular momentum and the $\pm$ sign for the two
possibilities to construct the total spin $J=\, \mid l\pm\frac{1}{2}\mid$ in
the intermediate states.\\

We complete the formalism of pion photoproduction by discussing the isospin.
Since the incoming photon has both isoscalar and isovector components and the
produced pion is isovector, the matrix elements take the form
\begin{equation}
{\cal M}_{l\pm}^{\alpha} = \frac{1}{2}[\tau_{\alpha},\tau_0]{\cal
M}_{l\pm}^{(-)} + \frac{1}{2}\{\tau_{\alpha},\tau_0\}{\cal
M}_{l\pm}^{(+)}+\tau_{\alpha} {\cal M}_{l\pm}^{(0)}\, , \label{eq:2.19}
\end{equation}
where $\tau_{\alpha}$ are the isospin Pauli matrices in a spherical basis,
i.e., $\alpha = \{ +, 0, - \}$. It follows that the intermediate state in
Fig.~\ref{fig:II.1} can only have isospin $I=\frac {1}{2}$ or $I=\frac {3}{2}$.
The 4 physical amplitudes with final states $ \{ p \pi^0, n \pi^0, n \pi^{+}, p
\pi^{-} \}$ are given by linear combinations of the 3 isospin amplitudes. We
should however keep in mind that the isospin symmetry is not exact but broken
by the mass differences between the nucleons $(n, p)$ and pions $(\pi^\pm,
\pi^0)$ as well as explicit Coulomb effects, in particular
near threshold.\\

The calculation of the observables is straightforward but somewhat tedious, and
therefore we choose pion photoproduction at threshold as an illustrative
example. Near threshold the partial wave series may be truncated to s and p
waves, i.e., the transverse multipoles $E_{0^+}$, $M_{1^+}$, $E_{1^+}$, and
$M_{1^-}$. With $P_1=3E_{1^+}+M_{1^+}-M_{1^-}$, $P_2=3E_{1^+}-M_{1^+}+M_{1^-}$,
and $P_3=2M_{1^+}+M_{1^-}$ the differential cross section takes the following
form in the cm frame:
\begin{equation}\label{eq:2.20}
\frac{d\sigma}{d\Omega}(\theta_{\pi}^{\ast}) =\frac {k} {q} \left ( A+B\cos
\theta_{\pi}^{\ast} +C \cos^2 \theta_{\pi}^{\ast}  \right )\, ,
\end{equation}
with $A=|E_{0^+}|^2 + \frac {1}{2}|P_2|^2+\frac {1}{2}|P_3|^2$, $B=2
Re(E_{0^+}^{\ast} P_1)$, and $C=|P_1|^2 -\frac {1}{2}|P_2|^2-\frac
{1}{2}|P_3|^2$. As is to be expected, the s-wave multipole yields a constant
angular distribution, whereas the forward-backward asymmetry is given by the
interference between the s wave and the  p-wave combination $P_1$. The terms in
$\cos^2 \theta_{\pi}^{\ast}$ determine a further p-wave combination, $ \mid
P_2\mid^2+ \mid P_3\mid^2$. A complete experiment requires to measure one
further observable, the photon asymmetry
\begin{eqnarray}
\Sigma(\theta_{\pi}^{\ast}) &=&  \frac{d\sigma^{\bot} -
d\sigma^{\parallel}}{d\sigma^{\bot}
+d\sigma^{\parallel}}\nonumber\\
&=&\frac {k} {2 q}\left( \, |P_2|^2-|P_3|^2 \right) \sin^2 \theta_{\pi}^{\ast}
\bigg / \frac{d\sigma}{d\Omega}(\theta_{\pi}^{\ast})\, , \label{eq:2.21}
\end{eqnarray}
where ${\bot}$ and ${\parallel}$ stand for photon polarizations perpendicular
and parallel to the reaction plane.\\

The theory of meson electroproduction is more involved and we refer the reader
to the literature, see \textcite{Drechsel:1992pn} and references quoted
therein. The scattered electron serves as a source of virtual photons whose
flux $\Gamma_V$ and transverse polarization $\varepsilon$ can be controlled by
varying the electron kinematics. Moreover, we assume that the electron beam is
polarized. The five-fold differential cross section for meson electroproduction
is written as the product of a virtual photon flux factor $\Gamma_V$ and a
virtual photon cross section,
\begin{equation}
\frac{d\sigma}{d\Omega_2 d\epsilon_2 d\Omega_{\pi}^{\ast}}=\Gamma_V \,
\frac{d\sigma}{d\Omega_{\pi}^{\ast}}\, . \label{eq:2.22}
\end{equation}
The electron kinematics is commonly given in the lab system, whereas the
hadrons are described in the hadronic cm system as indicated by an asterisk.
The reaction plane and the electron scattering plane have the same $z$-axis,
but the former is tilted against the latter by the azimuthal angle
$\Phi_{\pi}$. With these definitions, the virtual photon cross section takes
the following form for polarized electrons but unpolarized hadrons:
\begin{eqnarray}\label{eq:2.23}
\frac{d\sigma}{d\Omega_{\pi}^{\ast}} & = &
\frac{d\sigma_T}{d\Omega_{\pi}^{\ast}}  + \varepsilon
\,\frac{d\sigma_L}{d\Omega_{\pi}^{\ast}} + \sqrt{2\varepsilon
(1+\varepsilon)}\,\,\frac{d\sigma_{LT}}{d\Omega_{\pi}^{\ast}}\,
\cos{\Phi_{\pi}} \\
\nonumber &+& \varepsilon\,\frac{d\sigma_{TT}}{d\Omega_{\pi}^{\ast}}\,
\cos{2\Phi_{\pi}} +  P_e \sqrt{2\varepsilon
(1-\varepsilon)}\,\,\frac{d\sigma_{LT'}}
{d\Omega_{\pi}^{\ast}}\,\sin{\Phi_{\pi}}\,.
\end{eqnarray}
Denoting the initial and final electron lab energies by $\epsilon_1$ and
$\epsilon_2$, respectively, the photon lab energy is
$\omega_L=\epsilon_1-\epsilon_2$, and in the same notation the photon lab
three-momentum is given by ${\vec q}_L$. With these definitions the transverse
electron polarization and the virtual photon flux take the form
\begin{equation}\label{eq:2.24}
\varepsilon=\frac{1}{1 + 2\frac{{\vec q}\,^2_L}{Q^2}\tan^2 \frac{\theta}{2}}
\,, \quad \Gamma_V = \frac{\alpha_\text{em}}{2\pi^2}\
\frac{\epsilon_2}{\epsilon_1}\,\frac{K}{Q^2}\ \frac{1}{1-\varepsilon}\,,
\end{equation}
with $K=(W^2-M^2)/2M$ the photon ``equivalent energy'' in the lab frame. The
partial cross sections in Eq.~(\ref{eq:2.23}) are functions of the virtuality
$Q^2$, the pion scattering angle $\theta_{\pi}^{\ast}$, and the total hadronic
cm energy $W$. The first two terms on the rhs of this equation contain the
transverse ($\sigma_T$) and longitudinal ($\sigma_L$) cross sections. The third
and fifth terms yield the longitudinal-transverse interferences $\sigma_{LT}$
and $\sigma_{LT'}$. These terms contain the explicit factors $\cos \Phi_{\pi}$
and $\sin \Phi_{\pi}$, respectively, and an implicit factor
$\sin\theta_{\pi}^{\ast}$ in the partial cross sections, and therefore they
vanish in the direction of the virtual photon. The latter is also true for the
fourth term, which contains the transverse-transverse interference
($\sigma_{TT}$), which is proportional to $\sin^2\theta_{\pi}^{\ast}$ and
appears with the explicit factor $\cos 2\Phi_{\pi}$. These 5 partial cross
sections can be expressed in terms of the 6 independent CGLN amplitudes $F_1$
to $F_6$, or in terms of 6 helicity amplitudes $H_1$ to $H_6$ given by linear
combinations of the CGLN amplitudes. The particular form of the $\Phi_{\pi}$
dependence in Eq.~(\ref{eq:2.23}) is of course related to the fact that the
virtual photon transfers one unit of spin. A close inspection shows that the 5
responses provided by the polarization of the electron can be separated in a
``super Rosenbluth plot''. This requires measuring the polarization
$\varepsilon$ of the virtual photon, the beam polarization $P_e$, and the
angular distribution of the pion with at least one non-coplanar angle
$\Phi_{\pi}$. A double-polarization experiment measuring also the target or
recoil polarizations of the nucleon yields 18 different response functions
altogether. The relevant expressions can be found in the work of
\textcite{Drechsel:1992pn,Knochlein:1995qz}.
\subsection{Resonance excitation}
\label{sec:II.5}
As shown in the previous section, each partial wave is characterized by 3
quantum numbers, orbital angular momentum $\ell$, total angular momentum $J$,
and isospin $I$. Most of these pion-nucleon partial waves show distinct
resonance structures at one or more values of the hadronic cm energy.
Furthermore, there are generally 3 (independent) electromagnetic transitions
between the nucleon and a particular partial wave, an electric, a magnetic, and
a Coulomb transition. Let us consider as an example the most important
resonance of the nucleon, the $\Delta$(1232) with the spectroscopic notation
$P_{33}$, which decays with a life-time of about $0.5 \times 10^{-23}$\,s into
a pion-nucleon state, except for a small electromagnetic decay branch of about
0.5~\%. This intermediate state contains a pion in a p wave, i.e., $\ell=1$ and
${\cal P}=+1$. The indices 33 refer to twice the isospin and spin quantum
numbers, $I=J=\frac{3}{2}$. The electroexcitation of this resonance takes place
by magnetic dipole ($M1$), electric quadrupole ($E2$), and Coulomb quadrupole
($C2$) radiation, which is denoted by the 3 complex functions $M_{1+}^{3/2}$,
$E_{1+}^{3/2}$, and $S_{1+}^{3/2}$. If one neglects the small photon decay
branch, unitarity requires that all 3 electroproduction multipoles carry the
phase $\delta_{1+}^{3/2}(W)$ of the pion-nucleon final state
\cite{Watson:1954dd}. For a stable particle with the quantum numbers of a
$P_{33}$ resonance, \textcite{Rarita:1941dd} have developed a consistent
relativistic theory involving 3 (real) form factors $G_i^{\ast}(Q^2)$. However,
the physical pion-nucleon state has a complex phase factor, the resonance
phenomenon spreads over more than 100\,MeV in excitation energy, and there is
no model-independent way to extract the ``bare'' resonance parameters from the
observables. It is therefore common practice to relate the form factors to the
transition multipoles taken at the resonance position, $W=M_\Delta=1232$\,MeV.
Corresponding to the independent transition multipoles, the following 3 form
factors for the N$\Delta$ transition have been defined:
\begin{eqnarray}
M_{1+}^{3/2} (M_\Delta , Q^2) &=&i \, N \; \frac{q_\Delta (Q^2)}{M} \,
G_M^* (Q^2) \, , \nonumber\\
E_{1+}^{3/2} (M_\Delta , Q^2) &=& -i \,  N \; \frac{q_\Delta (Q^2)}{M} \,
G_E^* (Q^2) \, ,
\label{eq:2.25} \\
S_{1+}^{3/2} (M_\Delta , Q^2) &=& -i \,  N  \; \frac{q_\Delta (Q^2)^2}{2 \, M
M_\Delta} \, G_C^* (Q^2) \, , \nonumber
\end{eqnarray}
with $q_{\Delta}$ and $k_{\Delta}$ the 3-momenta of photon and pion at the
resonance, and $N=\sqrt{3 \alpha_\text{em} /(8 k _\Delta \Gamma_\Delta)}$ a
kinematic factor relating pion photoproduction to total photoabsorption at the
resonance. We note that these definitions divide out the $q$ dependence of the
multipoles at pseudothreshold ($q \rightarrow 0)$ such that the form factors
are finite at this point. Equation~(\ref{eq:2.25}) corresponds to the
definition of \textcite{Ash:1967dd}, the form factors of
\textcite{Jones:1972ky} are obtained by multiplication with an additional
factor, $G^\text{JS}=\sqrt{1+Q^2/(M
+M_\Delta)^2} \; G^{\text{Ash}}$.\\

Because the background becomes more and more important as the energy increases,
the concept of transition form factors is usually abandoned for the higher
resonances. Instead it is common to introduce the helicity amplitudes, which
are uniquely defined for each resonance by matrix elements of the transition
current between hadronic states of total spin $J$ and projection $M$. With the
photon momentum $\vec{q}$ as axis of quantization, the virtual photon can only
transfer intrinsic spin 1 to the hadronic system, with projections $\pm 1$ for
right- and left-handed transverse photons (current components $J_{\pm 1}$) and
0 for the Coulomb interaction (time-like component $\rho$). Using parity and
angular momentum conservation, we find 3 independent helicity amplitudes,
\begin{eqnarray}
A_{1/2} & = & \frac{1}{\sqrt{2K}} \left\langle N^{\ast} (J, \textstyle{\frac{1}{2}}) \mid
J_+ \mid  N (\textstyle{\frac{1}{2}},  - \textstyle{\frac{1}{2}}) \right\rangle \,, \nonumber \\
A_{3/2} & = & \frac{1}{\sqrt{2K}}\left\langle N^{\ast} (J, \textstyle{\frac{3}{2}}) \mid
J_+ \mid  N (\textstyle{\frac{1}{2}}, \textstyle{\frac{1}{2}}) \right\rangle \,, \label{eq:2.26} \\
S_{1/2} & = & \frac{1}{\sqrt{2K}} \left\langle N^{\ast} (J, \textstyle{\frac{1}{2}}) \mid
\rho \mid N (\textstyle{\frac{1}{2}},  \textstyle{\frac{1}{2}}) \right\rangle \, .\nonumber
\end{eqnarray}

In particular we note that the amplitude $A_{3/2}$ exists only for resonances
with $J \ge \frac{3}{2}$, and neither does this amplitude exist for a free
quark. Hence asymptotic QCD predicts that $A_{3/2}$ should vanish in
the limit of large momentum transfer, $Q^2 \rightarrow \infty$. The
electromagnetic multipoles can be expressed by combinations of the helicity
amplitudes. For the $\Delta$~(1232) these relation take the following form:
\begin{eqnarray}
M_{1^+}^{3/2} & \sim & -\frac{1}{2\sqrt{3}}(\sqrt{3}\ A_{1/2} + 3\ A_{3/2})\, , \nonumber\\
E_{1^+}^{3/2} & \sim & -\frac{1}{2\sqrt{3}} (\sqrt{3}\ A_{1/2}-A_{3/2})\, ,\label{eq:2.27}\\
S_{1^+}^{3/2} & \sim & -\frac{1}{\sqrt{2}} S_{1/2}\, . \nonumber
\end{eqnarray}
It is interesting to observe that asymptotic QCD predicts the following
multipole ratios in the limit $Q^2 \rightarrow \infty$:
\begin{eqnarray}
R_{EM} &=& \frac {{\rm {Im}}\,E^{3/2}_{1+}}{{\rm {Im}}\,M^{3/2}_{1+}}
\bigg|_{W=M_\Delta} \rightarrow 1 \,,\nonumber\\
R_{SM} &=& \frac {{\rm {Im}}\,S^{3/2}_{1+}}{{\rm {Im}}\,M^{3/2}_{1+}}
\bigg|_{W=M_\Delta} \rightarrow {\rm {const}}. \label{eq:2.28}
\end{eqnarray}
In these relations, the multipoles are evaluated at resonance, defined by
the energy for which the real part passes through zero (K-matrix pole). This
should happen at the same energy for all 3 multipoles as long as the
Fermi-Watson theorem is valid.
\subsection{Dispersion relations}
\label{sec:II.6}
Dispersion relations (DRs) play an important role in the following sections.
They are based on unitarity and analyticity and, by proper definitions of the
respective amplitudes, fulfil gauge and Lorentz invariance as well as other
symmetries. The analytic continuation in the kinematic variables allows one to
connect the information from different physical processes and thus to check the
consistency of different sets of experiments. As demonstrated in
section~\ref{sec:IV}, DRs are prerequisite to determine the polarizabilities of
the hadrons from Compton scattering, and several sum rules are derived in
section~\ref{sec:VI} by combining DRs with low-energy theorems. Most of these
techniques are very involved and we have to refer to the literature. Therefore
we only give an overview of the dispersive approach for the nucleon form
factors, which are discussed in more detail in the following
section~\ref{sec:III}.\\

\begin{figure}[]
\begin{center}
\includegraphics[width=0.9\columnwidth,angle=0]{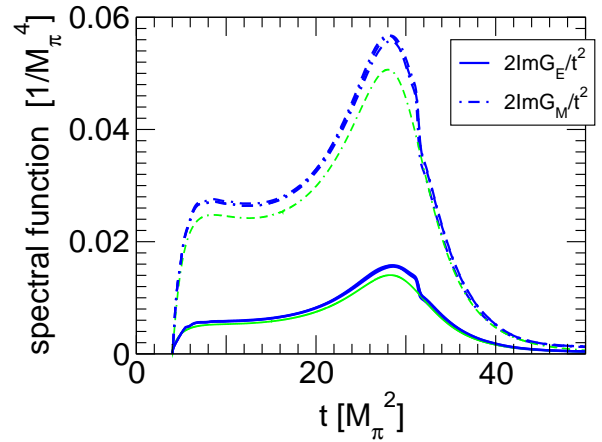}
\end{center}
\caption{The isovector spectral functions in units of $m_{\pi}^{-4}$. Solid
lines: $2\text{Im}G_E^v(t)/t^2$, dashed lines: $2\text{Im}G_M^v(t)/t^2$. The
thin lines are from the work of \textcite{Hohler:1974eq}, the thick lines
include modern data for the pion form factor. The figure is from 
\cite{Belushkin:2005ds}.}
\label{fig:disp_spectrum}
\end{figure}

Let G(t) be a generic (electromagnetic) form factor describing the ground state
of the nucleon. The real and imaginary parts of G(t) are then related by DRs.
Assuming further an appropriate high-energy behavior, these amplitudes fulfill
an unsubtracted DR in the Mandelstam variable $t$,
\begin{equation}\label{eq:2.29}
\text{Re} ~G(t) = \frac {1}{\pi} \, \int_{t_{0}}^{\infty} dt' \; \frac
{\text{Im}~G(t')} {t'-t-i\epsilon}\,,
\end{equation}
where $t_0$ is the lowest threshold for the electroproduction of pions by
$e^+\,e^-$ pair annihilation. These form factors can be measured by electron
scattering for space-like momentum transfer ($t=-Q^2<0$) and by collider
experiments for time-like momentum transfer ($t>4M^2$). The imaginary part or
spectral function, ${\rm {Im}}~G(t)$, vanishes in the space-like region, and
therefore the $i\epsilon$ in Eq.~(\ref{eq:2.29}) can be dropped for elastic
electron scattering. However we note that Eq.~(\ref{eq:2.29}) defines the real
part of the form factor in both the space-like and time-like regions, provided
that the spectral function is sufficiently well known. The dispersive formalism
also yields information on proton and neutron at the same time as is evident
from the following reasoning. The spectral function can be obtained from the
two-step process $\gamma^{\ast} \rightarrow X \rightarrow N~{\bar {N}}$, with
$X$ a hadronic state with the quantum numbers of the photon. In the usual
notation of these quantum numbers with isospin $I$, $G$-parity, spin $J$,
parity $P$, and $C$-parity,  the isoscalar photon has $I^G(J^{PC})=0^-(1^{--})$
and the isovector photon $I^G(J^{PC})=1^+(1^{--})$. The lightest hadronic
system $X$ in the intermediate state is a pion pair, which has even G-parity
and therefore contributes only to the isovector current. This part of the
spectral function is therefore composed of (I) the vertex $\gamma^{\ast}
\rightarrow \pi~{\bar {\pi}}$ given by the pion form factor $F_{\pi}(t)$, here
in the time-like region and therefore a complex function, and (II) the process
$\pi~{\bar {\pi}}\rightarrow N~{\bar {N}}$. The latter process is needed in the
unphysical region, which can however be reached by analytic continuation of the
p-wave amplitudes for pion-nucleon scattering \cite{Hoehler:1983dd}. As a
result, the two-pion contribution to the spectral function can be constructed
from $t_0=4m_{\pi}^2$ up to about 1~GeV$^2$ as
\begin{eqnarray}
{\rm {Im}}~G_E^v(t) &=& \frac {q_t^3}{M \sqrt{t}} \, F_{\pi}(t)^{\ast}f_+^1(t)
\,,\nonumber\\
{\rm {Im}}~G_M^v(t) &=& \frac {q_t^3}{\sqrt{2t}} \, F_{\pi}(t)^{\ast}f_-^1(t)
\,,\label{eq:2.30}
\end{eqnarray}
with $q_t=\sqrt{t/4-m_{\pi}^2}$ the pion momentum in the intermediate state and
$f_{\pm}^1(t)$ the p-wave $\pi \pi \rightarrow N~{\bar {N}}$ amplitude. The
spectral functions for the Sachs form factors are plotted in
Fig.~\ref{fig:disp_spectrum}. The figure shows a rapid rise of the spectral
functions at the two-pion threshold ($t=4~m_{\pi}^2$), because the projection
of the nucleon Born graphs to the p wave yields a singularity on the second
Riemann sheet at $t=3.98~m_{\pi}^2$, just below threshold. Quite similar
results for the two-pion continuum have also been obtained by a two-loop
calculation in ChPT \cite{Kaiser:2003qp}. Furthermore, we observe the strong
peak near $t\approx28~m_{\pi}^2$, which is due to the $\rho$ meson with mass
770\,MeV and a large width.
The spatial distribution of charge and current can be obtained by the
respective form factors in the Breit frame. The starting point is
Eq.~(\ref{eq:2.29}) for space-like $t=-Q^2 \rightarrow -{\vec {q}\,^2}$, which
is Fourier transformed to ${\vec {r}}$-space with the result
\begin{equation}\label{eq:2.31}
\rho (r) = \frac {1}{4\pi^2} \, \int_{t_{0}}^{\infty} dt \; {\rm {Im}}~G(t)
\frac {e^{-\sqrt{t}\,r}}{r}\,.
\end{equation}
The mean squared radius for a particular region of the spectral function at
$t=\mu^2$, where $\mu$ is the mass of the intermediate state, is given by
$<r^2>=6/\mu^2$. For instance, the onset of the spectral function corresponds
to an rms radius of about 1.7~fm, the $\rho$ meson to about 0.6~fm, and so on.
We conclude that the density at large distances is dominated by the lightest
intermediate states. As a consequence, the tail of the density distribution at
large radii should take a Yukawa form, $e^{-\mu r}/r$, with $\mu_v=2 m_{\pi}$
and $\mu_s=3 m_{\pi}$ for the isovector and isoscalar form factors,
respectively. It is therefore natural to identify the ``pion cloud'' with the
two-pion contribution to the spectral function, which remains after subtraction
of the $\rho$ peak from the spectral function of Fig.~\ref{fig:disp_spectrum}.
Whereas the isovector spectral function can be constructed from the available
experimental information up to $t\approx 1~\text{GeV}^2$, the higher part of
the spectrum has be modeled from information about resonances and continua.
Because the isoscalar spectral function contains at least 3 pions in the
intermediate state, it can not be obtained directly from experimental data. In
the region below $t\approx 1~\text{GeV}^2$ it is dominated by the $\omega$ and
$\Phi$ mesons, the three-pion continuum has been shown to couple only weakly,
see \textcite{Belushkin:2006qa} and references to earlier work.

\section{Form factors}
\label{sec:III}
\begin{figure*}[]
\parbox{\columnwidth}
{\center
\includegraphics[width=0.6\columnwidth,angle=-90]{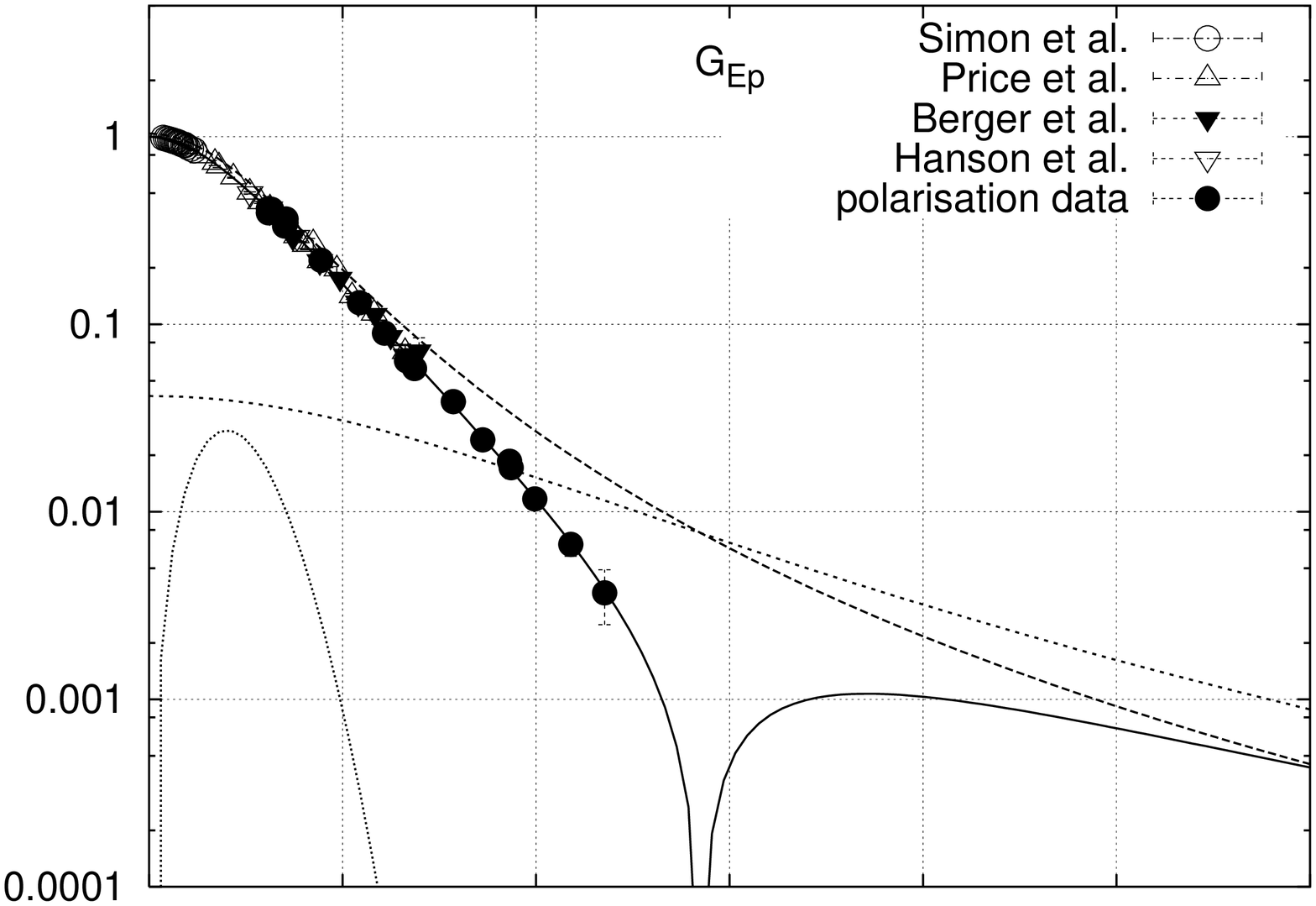}\\[5mm]
\includegraphics[width=0.6\columnwidth,angle=-90]{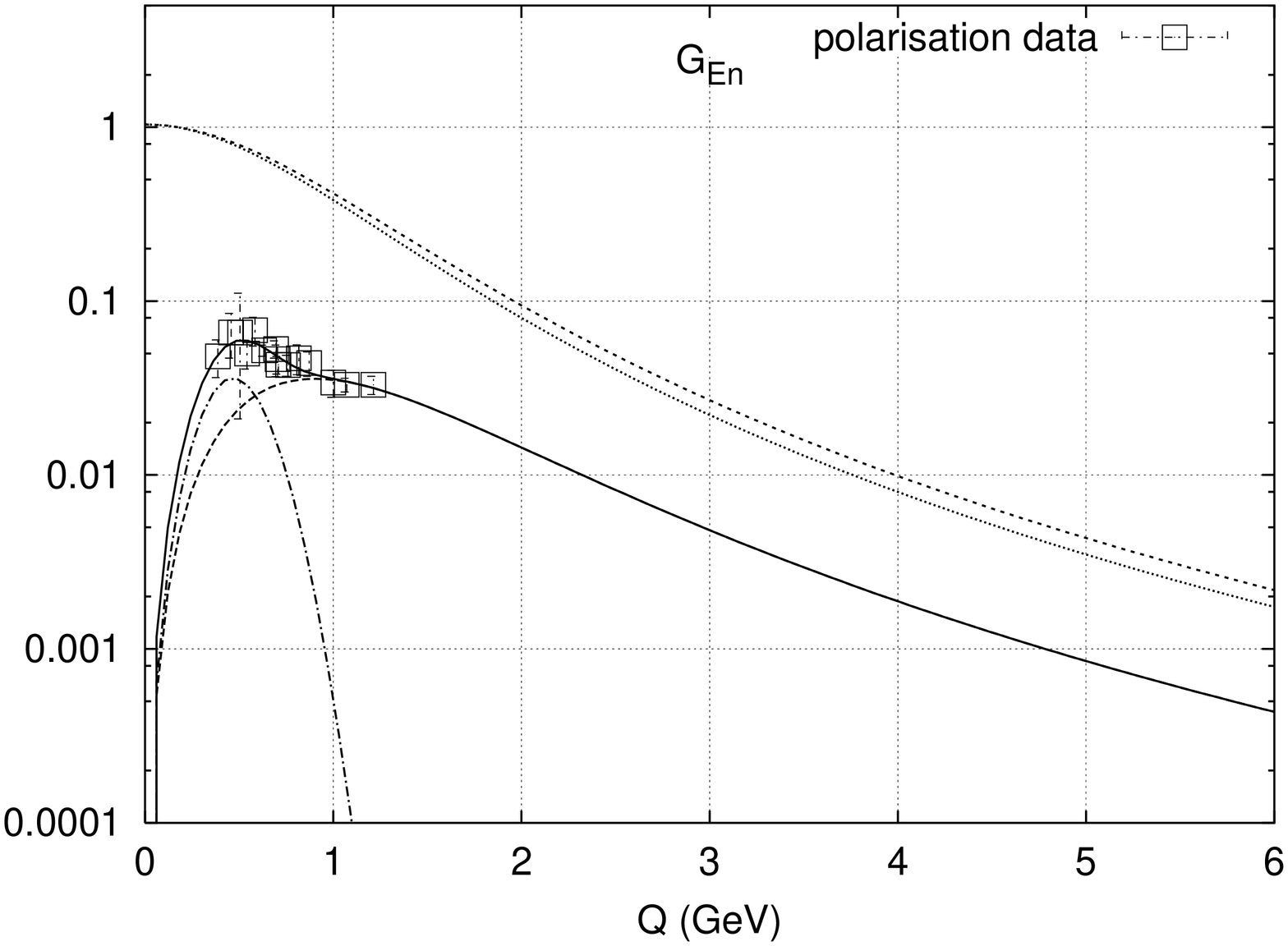}\\[5mm]
} \hfill
\parbox{\columnwidth}
{\center
\includegraphics[width=0.6\columnwidth,angle=-90]{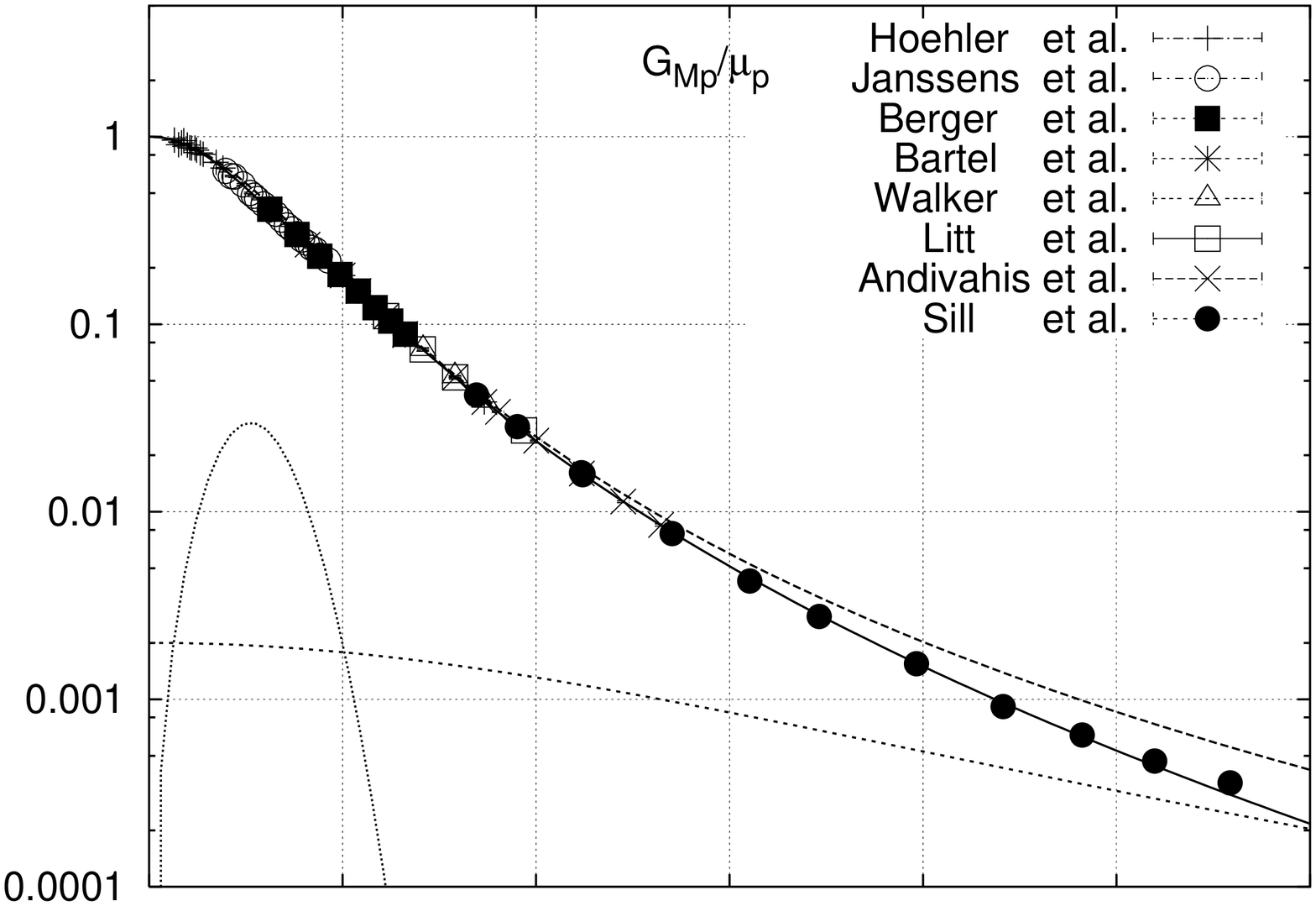}\\[5mm]
\includegraphics[width=0.6\columnwidth,angle=-90]{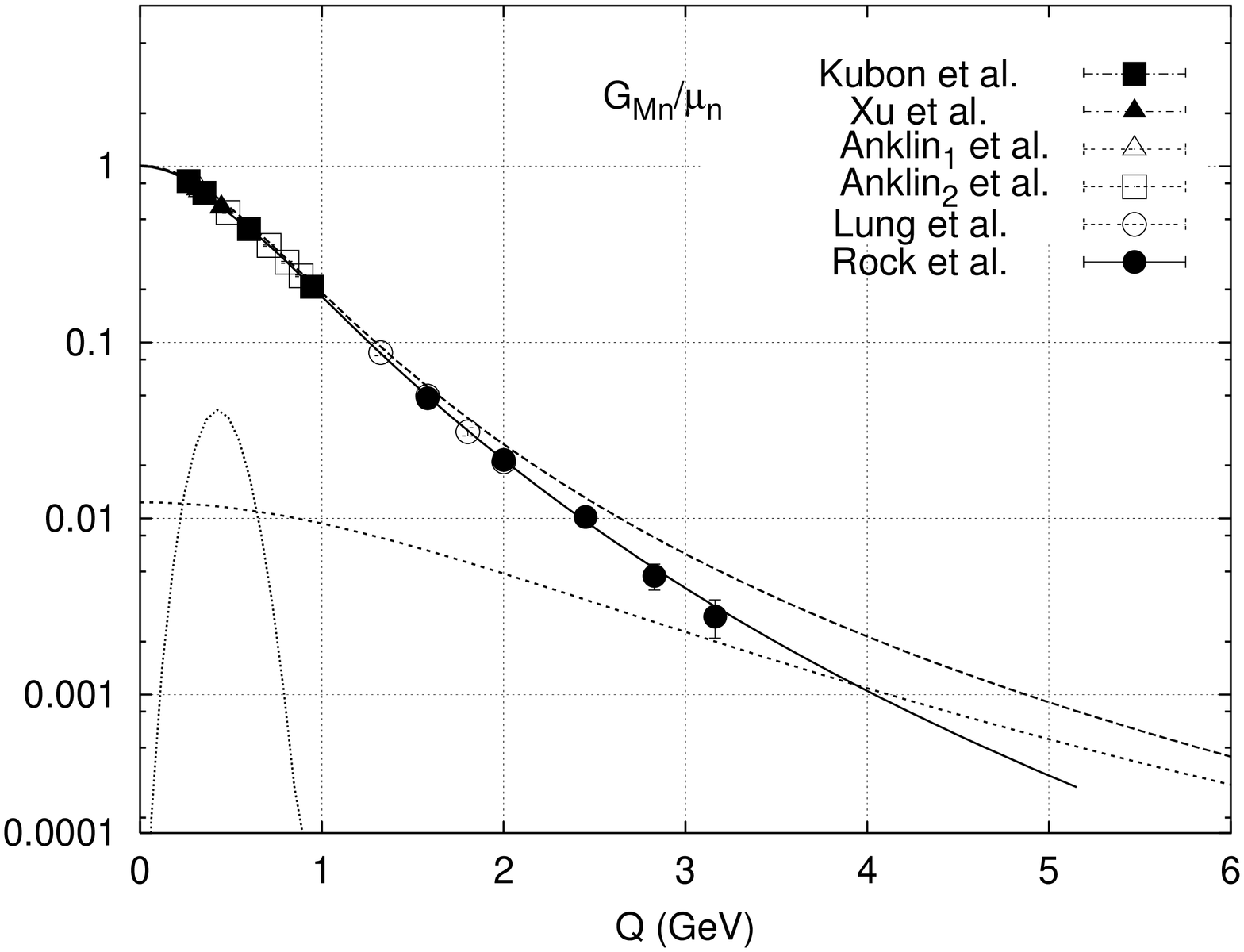}\\[5mm]
} \caption{The world data for the 4 nucleon form factors according to
\textcite{Friedrich:2003iz}. The solid line is the phenomenological fit given
by Eqs.~(\ref{eq:3.2}) and (\ref{eq:3.3}). The other lines show the two dipole
contributions and the bump/dip term of the fit separately. Note that the
absolute value is plotted for negative quantities. See
\textcite{Friedrich:2003iz} for a detailed listing of the data.}
\label{fig:ff_of_q}
\end{figure*}
Ever since \textcite{Hofstadter:1956a,Hofstadter:1957wk} first determined the
size of the nucleon, it has been taken for granted that the nucleon's
electromagnetic form factors follow the shape of a dipole form, with some minor
deviations and, of course, modified for the vanishing charge of the neutron.
This form was conveniently parameterized as
\begin{equation}
G(Q^2) = \frac{1}{(1+Q^2/\Lambda_{D}^2)^2}\,, \label{eq:3.1}
\end{equation}
with $\Lambda_{D}\approx~0.84~$GeV a universal parameter. Because this
parameter is close to the mass of the $\rho$ meson, it was assumed that the
nucleon structure is dominated by a vector meson cloud which was described by
the ``vector dominance model''. This idea was of course in conflict with the
quark model after its establishment in the 1970's and many attempts were made
to reconcile these conflicting models.\\

In order to set the scene, let us recall the following properties of proton,
neutron, and heavier baryons:
\begin{itemize}
\item
The complexity of these strongly interacting many-body system reflects itself
in the finite size in space, the anomalous magnetic moment, and the
continuum of excited states with strong resonance structures. These three
aspects are of course closely related, and can be connected quantitatively 
in some cases by sum rules as detailed in section~\ref{sec:VI}.
\item
Because of the approximate SU(3) symmetry of $u$, $d$, and $s$ quarks, the
nucleon forms a doublet in isospin with strangeness $S=0$ in an octet of states.
The other partners are: the $\Lambda^0$ and the triplet {$\Sigma^+, \Sigma^0,
\Sigma^-$} with $S=-1$, and the {$\Xi^0, \Xi^-$} with $S=-2$. The strange
baryons decay weakly with a typical mean life of about $10^{-10}$~s. Because
the {$\Sigma^0$} can also decay into the $\Lambda^0$ by photoemission, its mean
life is only about $10^{-19}$~s. The most important resonance of the nucleon,
the $\Delta(1232)$ appears as an isospin quadruplet with strangeness $S=0$ in a
decuplet of states. The partners of the $\Delta(1232)$ are: an isospin triplet
{$\Sigma^+(1385), \Sigma^0(1385), \Sigma^-(1385)$} with $S=-1$, a doublet
{$\Xi^0(1530), \Xi^-(1530)$} with $S=-2$, and the $\Omega^-(1672)$ with
$S=-3$. The latter lives about $10^{-10}$~s, because it can only decay weakly.
All the other particles in the decuplet decay by the strong interaction with a
mean life of order $10^{-23}$~s.
\item
The size effect reflects itself in the form factors of elastic electron
scattering. Because of the life time, only the proton and, with some caveat,
the neutron can be studied as a target. With the dipole form of
Eq.~(\ref{eq:3.1}) and the definition of an rms radius according to
Eq.~(\ref{eq:2.9}), the result of \textcite{Hofstadter:1956a} was
$r_E^p=0.81$~fm for the charge distribution of the proton. In the mean time
several new experiments have led to the larger radius of $r_E^p \approx
0.88$~fm. The form factors of the strange baryons can in principle be measured
by scattering an intense beam of these particles off the atomic electrons of
some nuclear target. Such an experiment was performed by the SELEX
Collaboration at the Fermi Lab Tevatron with a high-energetic $\Sigma^-$
beam \cite{Eschrich:2001ji}. The result is a first datum on a hyperon radius,
$r_E^{\Sigma^-}=(0.78 \pm 0.08 \pm 0.05)$~fm, distinctly smaller than the
accepted value for the proton.
\item
The ``normal'' magnetic moment of a particle i expected for a pointlike
Fermion is given by $eQ_i/(2 M_i)$, with $M_i$ the mass and $e\,Q_i$ the
charge of the particle. If the magnetic moments $\mu_i$ are given in units 
of the nuclear magneton $[\mu_N]=e/(2 M_p)$, the values for proton and neutron,
$\mu_p=2.79~[\mu_N]$ and $\mu_n=-1.91~[\mu_N]$, signal a large isovector
anomalous magnetic moment of the nucleon. From electron scattering we also know
that the rms radius of the magnetization distribution is quite similar to the
radius of the charge distribution. Because of their long mean life, the
magnetic moments of 5 other octet baryons and, in addition, the one of the
$\Omega^-(1672)$ in the decuplet are known from spin precession experiments.
Without going in the details, also these particles have large anomalous
moments. As an example, even the $\Omega^-$, a configuration of 3 $s$ quarks
with the large mass of 1.672~GeV, has $\mu_\Omega=-2.02~[\mu_N]$ compared to a
``normal'' magnetic moment of $-0.56~[\mu_N]$. In order to get more information
about the decuplet, several experiments were performed to measure the magnetic
moment of the $\Delta^{++}$ as a subprocess of radiative pion-nucleon
scattering \cite{Nefkens:1977eb,Bosshard:1991zp} and of the $\Delta^{+}$ as a
subprocess of radiative pion photoproduction on the proton
\cite{Kotulla:2003pm}. The results for the magnetic moments are in
qualitative agreement with quark model predictions but still with large model
errors \cite{Pascalutsa:2007wb}.
\end{itemize}

At the turn of the century new surprising results put the nucleon form factors
into focus once more. These new results became possible through the new
generation of cw electron accelerators with sources of high-intensity polarized
beams combined with progress in target and recoil polarimetry. As summarized in
section~\ref{sec:II.2} the measurement of asymmetries allows one to determine
both form factors even if they are of very different size. This situation
occurs in two cases: (I) Because of its vanishing total charge but large
anomalous magnetic moment, the neutron's electric form factor is very much
smaller than the magnetic one, at least for small and moderate momentum
transfer. (II) As shown by Eq.~(\ref{eq:2.10}), the magnetic form factor
$G_M(Q^2)$ appears with a factor $\tau=Q^2/4M^2$, whereas $G_E(Q^2)$ is
suppressed by a factor $1/(1+\tau)$. As a consequence, $G_E(Q^2)$ is less well
determined by the Rosenbluth plot if $Q^2$ becomes large. Even though this was
known, it was to the great surprise of everybody when asymmetry measurements
showed a dramatic deviation from previous results based on the Rosenbluth
separation and, at the same time, from the dipole shape of the proton form
factors \cite{Jones:1999rz,Gayou:2001qd}. Another open question concerns the
behavior of the form factors at low 4-momentum transfer, such as
oscillations at very small $Q^2$ and conflicting results for the rms radius of
the proton. All these experimental findings have caused an intense theoretical
investigation that has been summarized by several recent review papers
\cite{Gao:2003ag,Hyde-Wright:2004gh,Perdrisat:2006hj,Arrington:2006zm}. In the
present work we concentrate on the low momentum transfers and, therefore, the
phenomena for $Q^2 \geq 1~\text{GeV}^2$ will only be discussed briefly.
\subsection{Space-like electromagnetic form factors of the nucleon}
\label{sec:III.1}
The new results in the field of space-like form factors have been obtained at
basically 3 facilities, the cw electron accelerators CEBAF at the Jefferson Lab
 \cite{Cardman:2006xt} and Mainz Microtron \cite{Jankowiak:2006yc}, and the
electron stretcher ring at MIT/Bates \cite{Milner:2006xs}. All these facilities
provide an intense beam of polarized electrons. The second essential for
measuring the asymmetries was the development of polarized targets and
polarimeters to determine the polarization of the recoiling particles. For
details of the new accelerators, targets, and particle detectors as well as
pertinent references, we refer the reader to  a recent review of
\textcite{Hyde-Wright:2004gh}. In this context it is however important to
realize that there exist no free neutron targets, but only targets with
neutrons bound in a nucleus. Therefore, any analysis of the data requires
theoretical models to correct for the binding effects, which include
initial-state correlations, meson exchange and other two-body currents with
intermediate resonance excitation of the nucleons, and final-state interactions
while the struck nucleon leaves the target. In this situation, we infer that
the deuteron provides the most trustworthy neutron target, because its
theoretical description is far more advanced than in the case of heavier
nuclei. Of course, measurements with heavier nuclei, in particular polarized
$^3He$ targets, provide complementary information and are interesting
for their own sake.\\

In Fig.~\ref{fig:ff_of_q} we display the nucleon form factors as functions of
$Q$. We choose this somewhat unusual presentation in order to emphasize the
small $Q^2$ region. The data base shown is from \textcite{Friedrich:2003iz},
which meanwhile has been complemented by results from the following references:
\cite{Crawford:2006rz,Bermuth:2003qh,Glazier:2004ny,Plaster:2005cx,
Ziskin:2005ab,Anderson:2006jp}. The phenomenological fit shown by the solid
line in Fig.~\ref{fig:ff_of_q} is composed of two dipoles and a bump/dip
structure. The dipole form is given by
\begin{equation}
G_s(Q^2) = \frac{a_{10}}{(1+Q^2/a_{11})^2} + \frac{a_{20}}{(1+Q^2/a_{21})^2}
\,, \label{eq:3.2}
\end{equation}
and the bump/dip structure, seen at $Q^2=Q_b^2 \approx 0.2~\text{GeV}^2$ on top
of the smooth dipoles, is parameterized as
\begin{equation}
G_b(Q^2) =a_b\,Q^2\,\left( e^{\textstyle -\frac{1}{2}(\frac{Q -
Q_b}{\sigma_b})^2 } + e^{\textstyle -\frac{1}{2}(\frac{Q +
Q_b}{\sigma_b})^2}\right ) \, . \label{eq:3.3}
\end{equation}
We note that this ansatz provides an even function of $Q$ as required by
general arguments. A similar form has been introduced by \textcite{Sick74} in
r-space in order to obtain a non-singular function for his model-independent
analysis of nuclear charge distributions.\\

\begin{figure}[]
\begin{center}
\includegraphics[width=0.6\columnwidth,angle=-90]{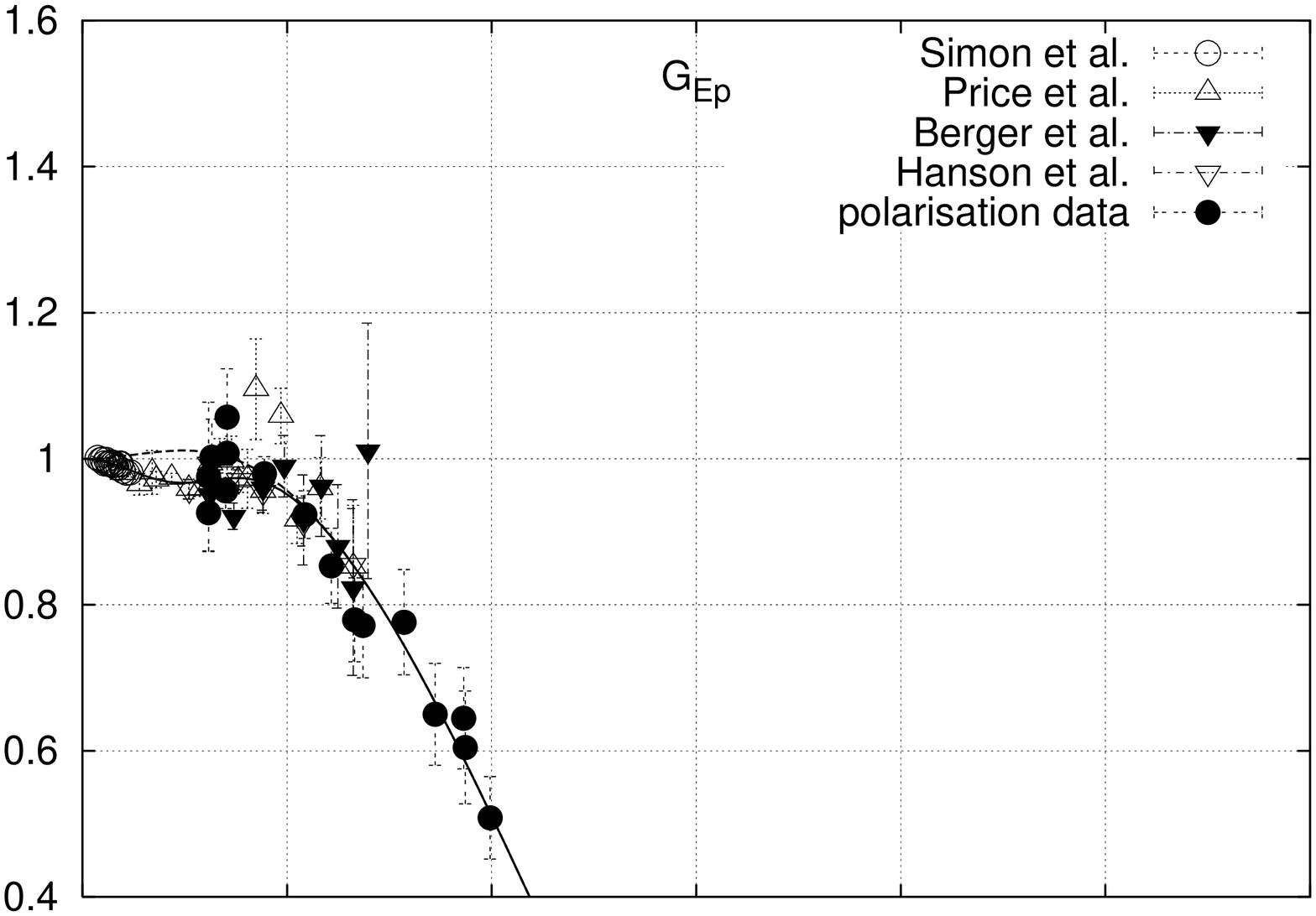} \\
\vspace*{-0.5cm}
\includegraphics[width=0.6\columnwidth,angle=-90]{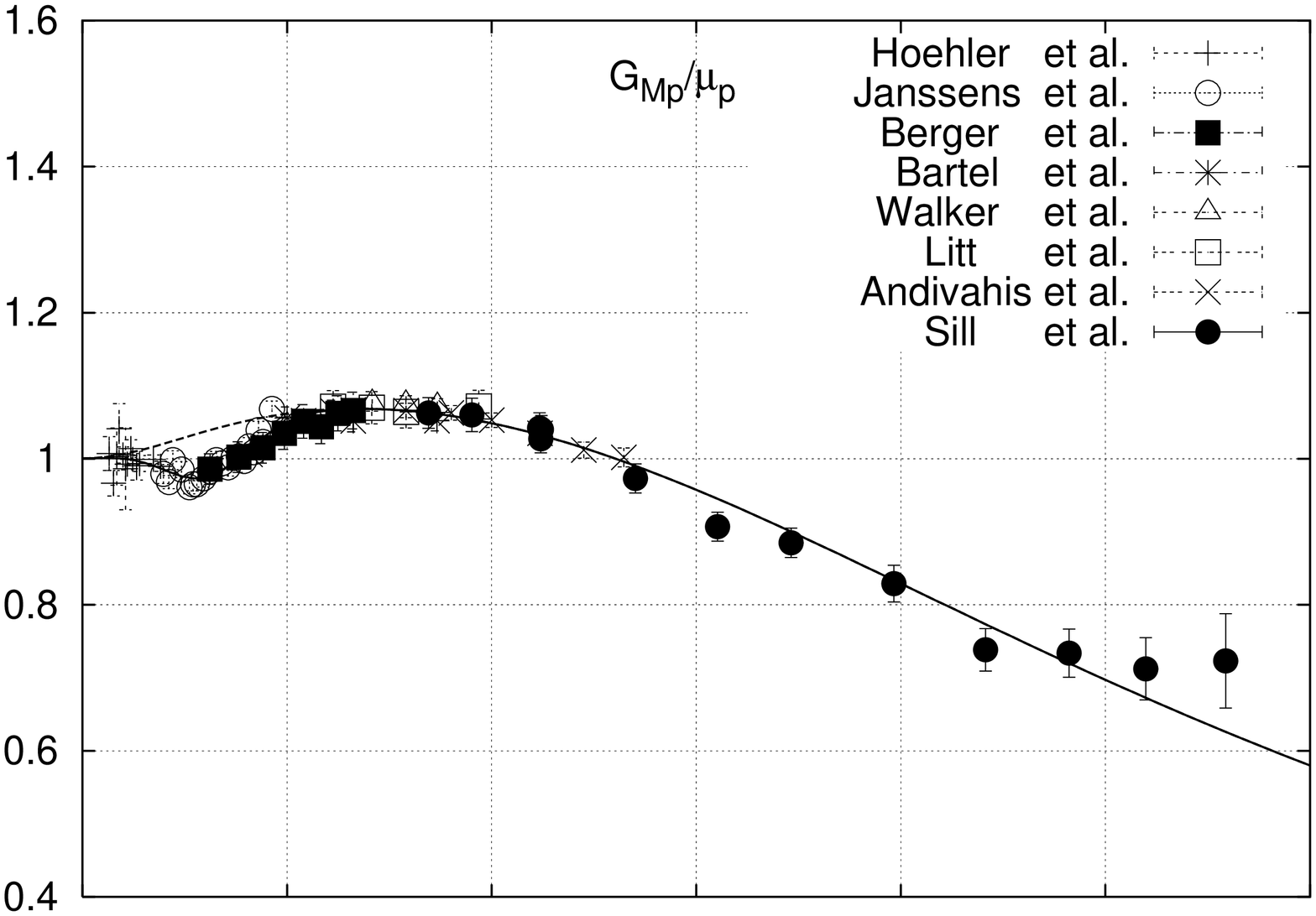} \\
\vspace*{-0.5cm}
\includegraphics[width=0.6\columnwidth,angle=-90]{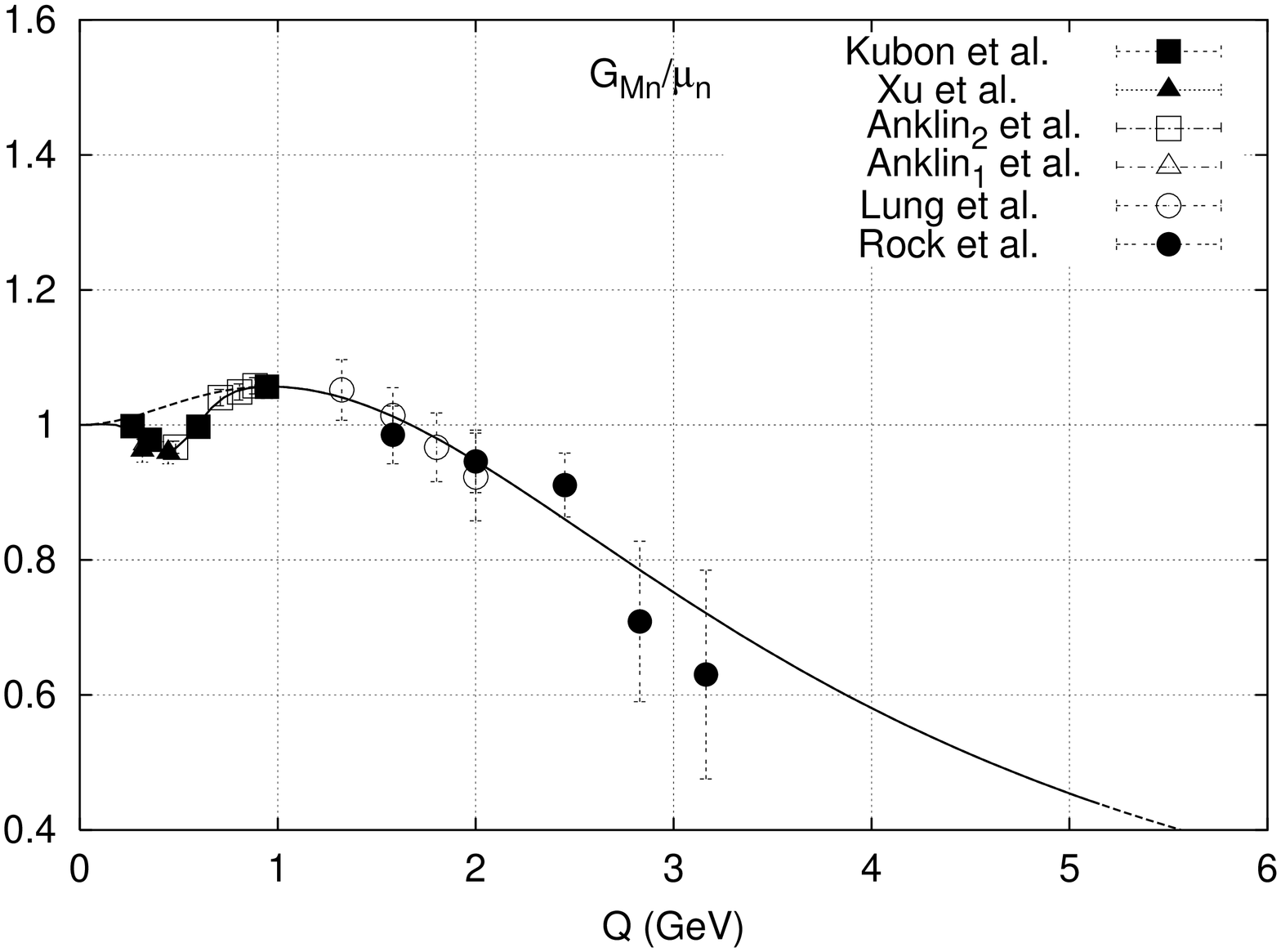}
\end{center}
\caption{The nucleon form factors $G_E^p$, $G_M^p/\mu_p$, and $G_M^n/\mu_n$ as
function of $Q$, divided by the standard dipole form factor of
Eq.~(\ref{eq:3.1}). The full line represents the fit of
\textcite{Friedrich:2003iz}, the broken line shows the smooth contribution of
the two dipoles.} \label{fig:ff_over_dipole}
\end{figure}
\begin{figure}[htp]
\begin{center}
\includegraphics[width=1.0\columnwidth,angle=0]{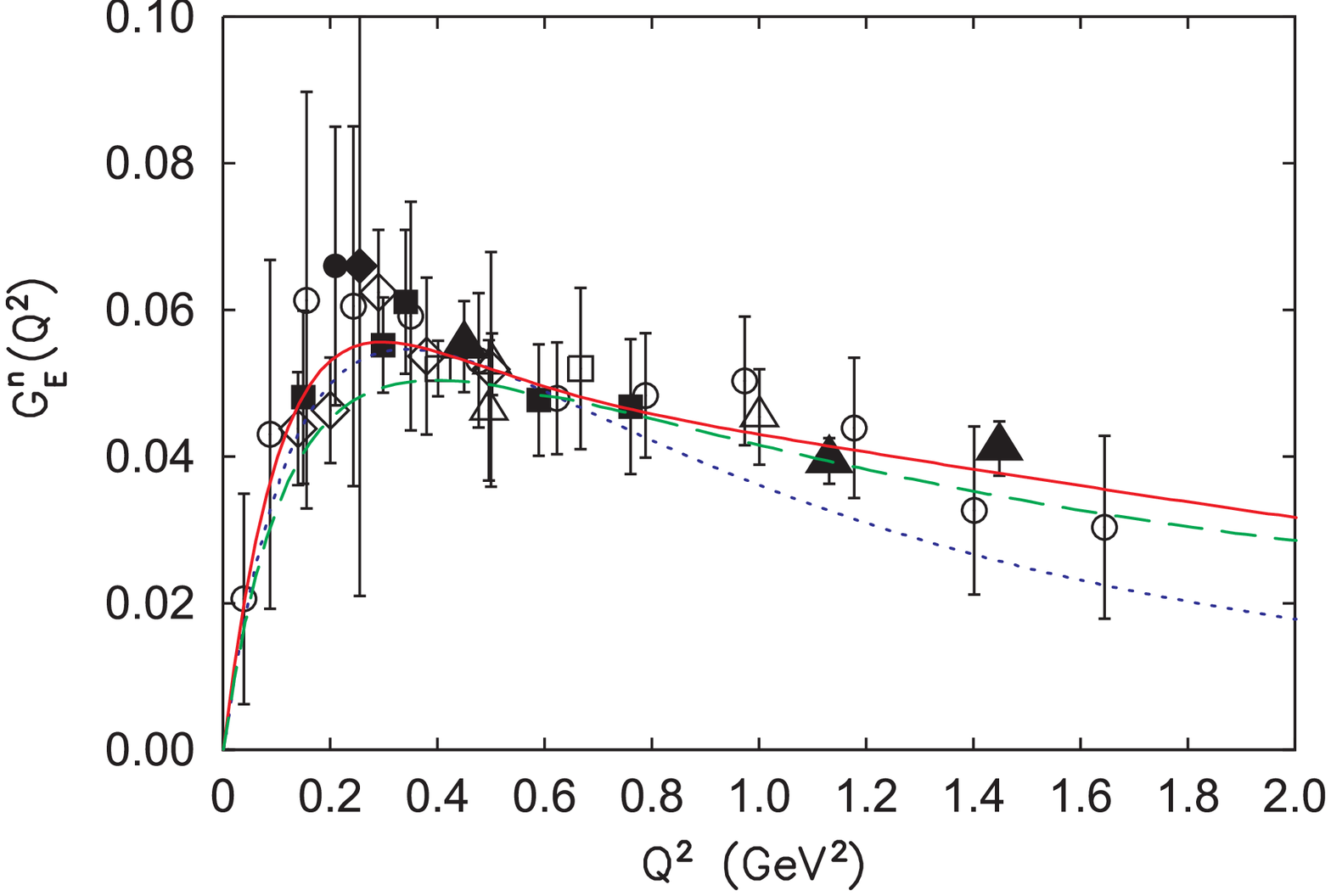}
\end{center}
\caption{The world data for the electric form factor of the neutron, $G_E^n$,
as function of $Q^2$. Solid squares: $d(\vec{e},e'\vec{n})p$ at MAMI,
\textcite{Ostrick:2006jd,Glazier:2004ny}, solid triangles:
$d(\vec{e},e'\vec{n})p$ at Jlab, \textcite{Madey:2003av}, solid diamond:
$d(\vec{e},e'\vec{n})p$ at BATES, \textcite{Eden:1994ji}, solid circle:
$\vec{d}(\vec{e},e'n)p$ at NIKHEF, \textcite{Passchier:1999cj}, open triangles:
$\vec{d}(\vec{e},e'n)p$ at Jlab, \textcite{Warren:2003ma}, open diamonds:
$\vec{d}(\vec{e},e'n)p$ at BATES, \textcite{Ziskin:2005ab}, open squares:
$^3\vec{He}(\vec{e},e'n)pp$ at MAMI, \textcite{Rohe:2005qz}. The open circles
are derived from the deuteron quadrupole form factor \cite{Schiavilla:2001qe}.
Solid line: new fit of the phenomenological model of 
\textcite{Friedrich:2003iz}, dashed line: the result of dispersion
relation \cite{Belushkin:2006qa,Meissner:2007tp}, dotted line: the original
Galster fit of Eq.~(\ref{eq:3.4}). The new fit includes all 
the shown data except for the preliminary of \textcite{Ziskin:2005ab} and 
the values derived by \textcite{Schiavilla:2001qe}. The data in the figure 
were updated by J.~Friedrich and the theory curves complemented by L.~Tiator.} 
\label{fig:GEn_all+Ingo}
\end{figure}

Figure~\ref{fig:ff_of_q} shows that the bump/dip structure is of the order 3\%
and only visible for $Q^2<1~\text{GeV}^2$. Therefore we have magnified the data
and their structure by the linear plot of Fig.~\ref{fig:ff_over_dipole}, which
shows the form factors divided by the standard dipole form factor of
Eq.~(\ref{eq:3.1}). The electric form factor of the neutron is quite special
because of its vanishing charge, which results in an overall small value of
$G_E^n$. Therefore we have plotted this form factor in a different way in
Fig.~\ref{fig:GEn_all+Ingo}, which displays the published world data as
measured with polarized electrons. We note, however, that the results of
\textcite{Schiavilla:2001qe} have been deduced from an analysis of the deuteron
quadrupole form factor $F_{C2}$, which requires a careful investigation of the
model dependence due to the nucleon-nucleon potential. The error bars of these
data are therefore not statistical but indicate the (systematic) model error.
The combined data shown in Fig.~\ref{fig:GEn_all+Ingo} clearly support the
existence of the bump structure at $Q^2 \approx 0.2~\text{GeV}^2$ as in the
previous cases. The solid line in this figure is the result of a new fit with 
the phenomenological model given by Eqs.~(\ref{eq:3.2}) and (\ref{eq:3.3}). 
The dashed line in this figure is the parameterization first given by 
\textcite{Galster:1971kv},
\begin{equation}
G_E^n(Q^2) = \frac{a_G~\tau}{(1+ b_G~\tau)} \cdot \frac{1}{(1+
Q^2/\Lambda_D^2)^2} \, , \label{eq:3.4}
\end{equation}
with $a_G=-\mu_n$ and $b_G=5.6$. The result from dispersion theory is displayed
by the dotted line. Neither dispersion theory nor the Galster fit reproduce
the data at low $Q^2$.\\

\begin{figure}[]
\begin{center}
\includegraphics[width=1.0\columnwidth,angle=0]{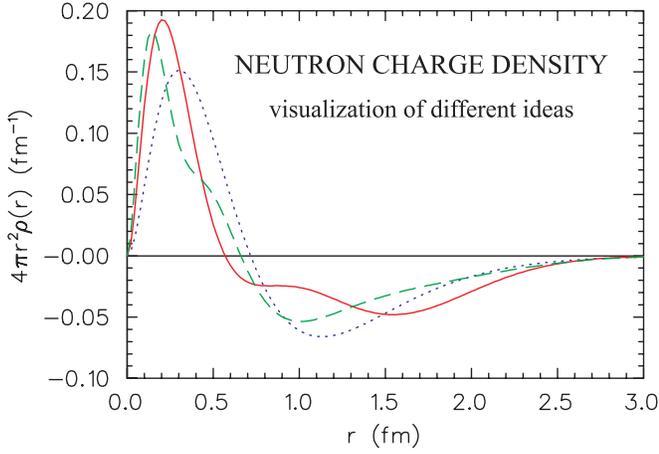}
\end{center}
\caption{The charge density distribution for the neutron, $4\pi r^2 \rho_E^n$,
as function of $r$. The notation for the lines corresponds to that of the
form factors shown in Fig.~\ref{fig:GEn_all+Ingo}. Figure by courtesy of 
J.~Friedrich and L.~Tiator.} 
\label{fig:rho_n}
\end{figure}

The Fourier transform of $G_E^n$ is of particular interest, because the overall
charge of the neutron must vanish. A finite charge distribution is therefore a
definite sign of correlations among the charged constituents, for example,
between the $u$ quark and the two $d$ quarks of the constituent quark model.
The charge distribution $\rho_E^n$ is displayed in Fig.~\ref{fig:rho_n} for the
3 fits to the neutron factor shown in the previous figure.
We recall the arguments of subsection~\ref{sec:II.2} that the Fourier transform
can only be obtained in the Breit or ``brick wall'' frame, which however is a
different Lorentz frame for different values of $Q^2$. The visualization of the
Fourier transforms as charge and magnetization distributions in r-space is
therefore only approximately correct if the momentum transfer is small compared
to $|Q^2|=4M^2$, which defines the threshold of nucleon pair production at
time-like momentum transfer. With this caveat in mind, we may interpret the
charge distribution of Fig.~\ref{fig:rho_n} by the dissociation of the neutron
in a proton and a pion, i.e., as a negative ``pion cloud'' around a positive
core. As we see from the figure, the pion cloud found
by \textcite{Friedrich:2003iz} extends to large radii. It is important to
realize that this result does not depend on a model assumption but is borne out
by a statistically satisfactory reproduction of the data. We also note that the
signal of the pion cloud is empirically present in all 4 form factors.\\

In a recent paper, \textcite{Miller:2007uy} has found a negative density in the
center of the neutron from an analysis of generalized parton distributions as
function of the impact parameter $ b$. The charge distribution $\rho_1^n(b)$
is then defined by the 2-dimensional Fourier transform of the Dirac form factor
$F_1^n(Q^2)$. We note that this does not contradict the results shown in
Fig.~\ref{fig:rho_n}. In fact our results for the 3-dimensional Fourier
transform of $F_1^n(Q^2)$ agree very much with the findings of
\textcite{Miller:2007uy}: a negative density in the center, positive values for
$0.4$~fm$<r<1.2$~fm, and a negative tail for the larger distances.
\subsection{Time-like electromagnetic form factors of the nucleon}
\label{sec:III.2}
\begin{figure}[]
\begin{center}
\includegraphics[width=0.9\columnwidth,angle=0]{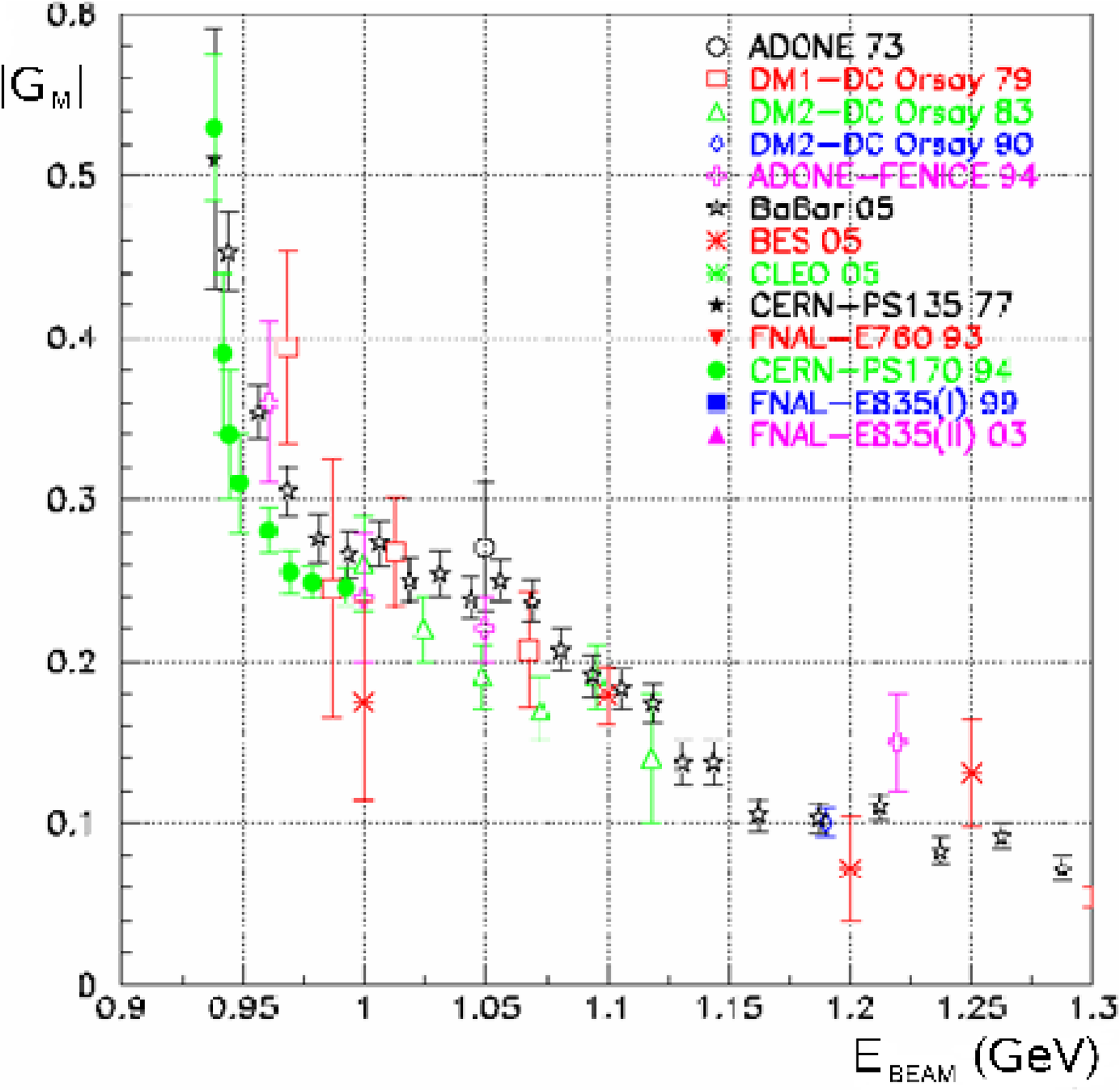}
\end{center}
\caption{The world data for the time-like form factor of the proton as function
of the beam energy $E_{\text{beam}}$, extracted with the assumption $|G_E| =
|G_M|$. Open symbols: data from $e^+e^- \rightarrow p\bar{p}$
\cite{Antonelli:1996xn,Delcourt:1979ed,Bisello:1983at,Bisello:1990rf,
Castellano:1973wh,Antonelli:1998fv,Ablikim:2005nn,Pedlar:2005sj,
Aubert:2005cb}, solid symbols: data from $\bar{p}p \rightarrow e^-e^+$
\cite{Bassompierre:1977ks,Bardin:1994am,Armstrong:1992wq,Ambrogiani:1999bh,
Andreotti:2003bt}. The figure is from \textcite{Rossi:2006}.}
\label{fig:ff_proton_tl}
\end{figure}
\begin{figure}[]
\begin{center}
\includegraphics[width=0.9\columnwidth,angle=0]{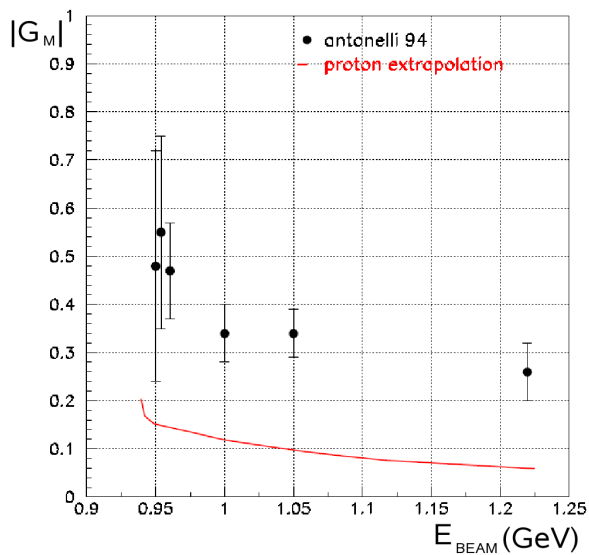}
\end{center}
\caption{The world data for the time-like form factor of the neutron as
function of the beam energy $E_{\text{beam}}$, extracted with the assumption
$|G_E| = |G_M|$. The data are obtained by the FENICE collaboration from the
reaction $e^+e^- \rightarrow n\bar{n}$. The solid line is a perturbative QCD
extrapolation using the proton data \cite{Antonelli:1998fv,Antonelli:1994kq}.
The figure is from \textcite{Rossi:2006}.} \label{fig:ff_neutron_tl}
\end{figure}
Let us next discuss the form factors for time-like momentum transfer, i.e.,
positive values of the Mandelstam variable $t=q^2=-Q^2>0$, see
subsection~\ref{sec:II.1} for definitions. By inspection we find that the
previously defined dipole form factors have poles $t=\Lambda_D^2 \approx
M_V^2$, that is, the phenomenological fits ``predict'' the existence of vector
mesons in the time-like region. The form factors in the space-like and
time-like regions are connected by analyticity and unitarity, and therefore
also the knowledge of the time-like form factors is mandatory for a complete
understanding of the nucleon
\cite{Baldini:1998qn,Geshkenbein:1974gm,Hammer:2006mw,
Mergell:1995bf,Belushkin:2006qa}. The time-like photons are obtained in
collider experiments by the reaction $e^+e^-\rightarrow N\,{\bar {N}}$ for
$t=(2E_\text{beam})^2>4M^2$. Whereas the space-like form factors are real, the
time-like form factors are complex functions because of the strong interaction
between the produced hadrons. However, the unpolarized cross section in the
time-like region only depends on the absolute values of the two form factors,
$|G_E(t)|$ and $|G_M(t)|$. In order to get information on the relative phase
between the form factors, polarization experiments are required. Unfortunately,
the present data basis does not even allow for a Rosenbluth separation.
Therefore the data are analyzed with the assumption $|G_E(t)| = |G_M(t)|$,
which follows from Eq.~(\ref{eq:2.8}) at threshold, $t=4M^2$, but is of course
not expected to hold for higher beam energies.\\

Figure~\ref{fig:ff_proton_tl} displays a compilation of the proton data known
so far. These data cover the range of $4M^2<t<6.8~\text{GeV}^2$. The figure
shows an overall falloff with the beam energy for $G_M^p$, somewhat faster than
$1/t^2$, and some structure near $E_{\text{beam}}= 1.05$\,GeV or 
$t=4.4~\text{GeV}^2$, which may indicate a resonance in that region. We note
that the decrease of the form factor is in qualitative agreement with
perturbative QCD, which requires a falloff like $|t|^{-2}$ for both space-like 
and time-like photons. However, a comparison shows that the space-like form 
factor at $t=-Q^2=-16~\text{GeV}^2$ is already about a factor 3 smaller than
the time-like form factor at $t=+16~\text{GeV}^2$, that is, asymptotia is
still far away. A look at Fig.~\ref{fig:ff_neutron_tl} tells us that our
knowledge about the neutron's time-like form factors is still far from
satisfactory. We hope that the currently planned experiments will improve on
the precision of the time-like form factors and, in particular, also determine
their relative phases, which is absolutely necessary in order to get the full
information on the structure of the nucleon \cite{Rossi:2006}.
\subsection{Theoretical considerations}
\label{sec:III.3}
The electromagnetic form factors encode information on the wave functions of
the charged constituents in a bound system. However, in the case of the hadrons
we face severe obstacles to get a real grip on the elementary quarks. As has
been mentioned in section~\ref{sec:I}, only two ab-initio approaches exist to
describe QCD in the confinement phase, chiral perturbation theory and lattice
gauge theory. Chiral perturbation theory is restricted to small values of the
momenta. Moreover, if extended to higher order in the perturbation series, ChPT
loses predictive power, because the number of unknown low energy constants
increases. Lattice gauge theory, on the other hand, is still hampered by the
use of large quark masses. This has the consequence that the pionic effects
appearing at low momentum transfer are underestimated. Beyond these two
approaches, which are in principle exact realizations of QCD, a plethora of
``QCD inspired'' models with quarks and pions has been developed. The problems
are twofold:
\begin{itemize}
\item
Starting directly from QCD, one would have to use the small $u$ and $d$ quark
masses of order 10~MeV. The many-body system is therefore highly relativistic
from the very beginning. However, a typical constituent quark model (CQM) has
quarks with masses of several hundred MeV. It is therefore obvious that these
entities are many-body systems of quarks and gluons by themselves. In any case,
the constituent quarks wave functions have to ``boosted'' if hit by the virtual
photon. However, there exists no unique scheme to boost a strongly interacting
relativistic many-body system.
\item
In view of the small current mass of the quarks, the interaction as mediated by
gluon exchange inevitably produces a considerable amount of quark-antiquark
admixture. These effects have to be modeled by properties of the constituent
quarks, such as mass and form factor, see \textcite{DeSanctis:2005kt}, or by
explicitly introducing a meson cloud of the ``bare'' constituent
quarks \cite{Faessler:2005gd}.
\end{itemize}
\begin{figure}[]
\begin{center}
\includegraphics[width=0.7\columnwidth,angle=90]{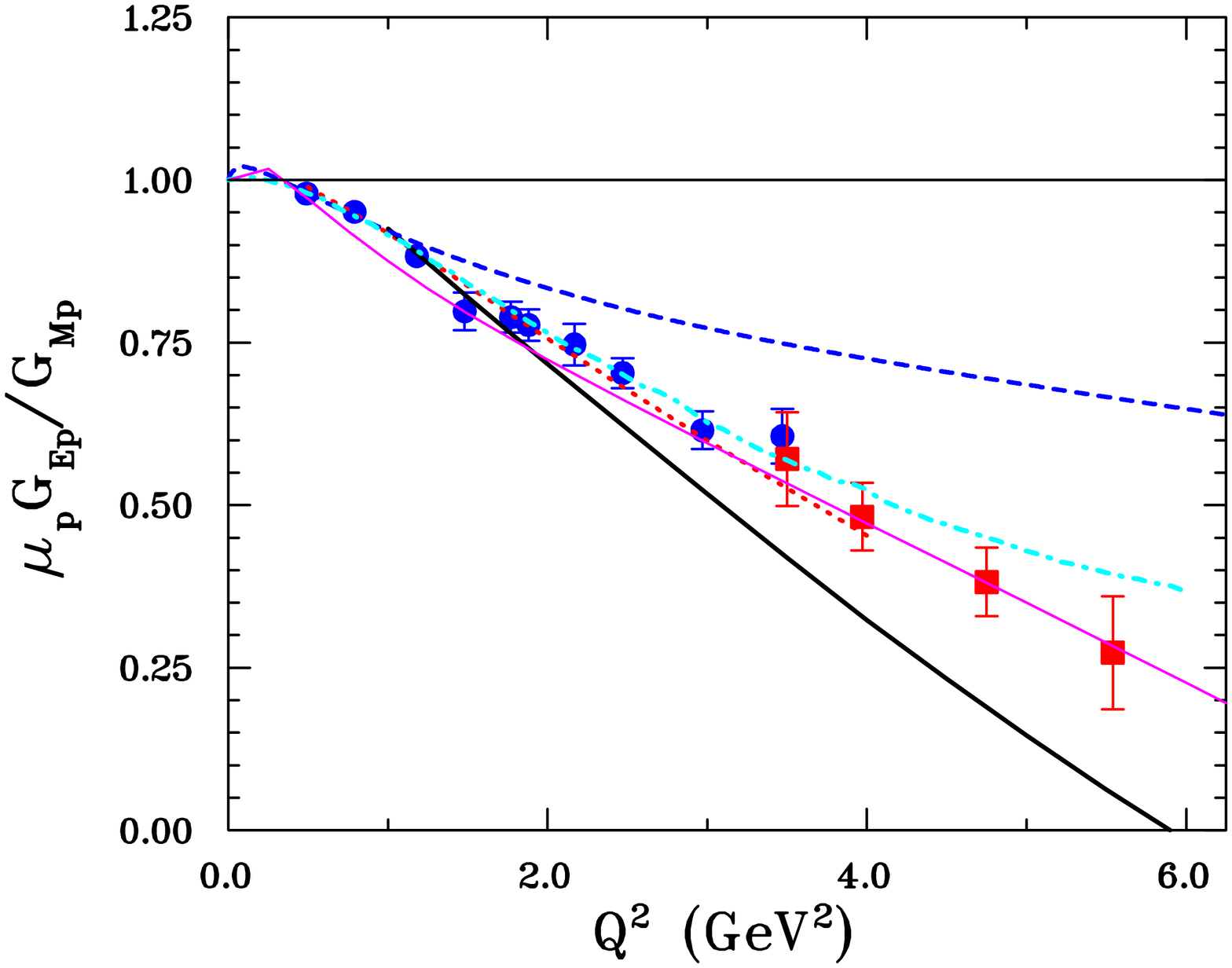}
\end{center}
\caption{The ratio of the electric and magnetic form factors of the proton,
$\mu_p\,G_E^p/G_M^p$, compared to relativistic calculations in the framework of
the CQM. Dotted line: light-cone front CQM \cite{Chung:1991st}, thick solid
line: light-cone front CQM \cite{Frank:1995pv}, dashed-dotted line: light-cone
front CQM with point-like constituent quarks~
\cite{Cardarelli:1995dc,Cardarelli:2000tk}, dashed line: Goldstone boson
exchange between point-like constituent quarks \cite{Boffi:2001zb}, thin solid
line: covariant spectator CQM \cite{Gross:2004nt}. The data have been taken at
Jefferson Lab by \textcite{Punjabi:2005wq} (circles) and
\textcite{Gayou:2001qd} (squares). The figure is from
\textcite{Perdrisat:2006hj}.} \label{fig:gepgmp_jlab_vsCQM}
\end{figure}
Since this article is dedicated to the low-$Q^2$ domain, it suffices to
consider some models useful in this region. A more detailed discussion of the
wide range of models is given by \textcite{Perdrisat:2006hj}. The traditional
model of the nucleon is the CQM with quark masses $m_q \approx M/3$. Except for
the smallest momentum transfers, the quark wave functions have to be
relativized, which is usually done by relativistic boosts of the single-quark
wave functions. Figure~\ref{fig:gepgmp_jlab_vsCQM} shows the result of several
calculations for the ratio $G_E^p/G_M^p$ compared to the recent data from the
Jefferson Lab obtained with a double-polarization experiment. The rapid falloff
of this ratio was a real surprise, because previous experiments without
polarization did not find big deviations from the dipole fit for both form
factors. The solution of problem was explained by two-photon effects, which are
usually small but turn out to become large in special cases, see
\textcite{Carlson:2007sp} and \textcite{Arrington:2007ux} for recent reviews.
Even though the figure shows only a small selection of models, quite different
results are obtained by similar models, depending on the properties of the
constituent quarks, their interaction, and the boosting mechanism. We are also
not aware of many predictions for the rapid drop of the ratio, before the JLab
double-polarization data were obtained. In any case, the shown models describe
the data qualitatively well, and they support the expected zero crossing of the
electric form factor at $Q^2 \approx 7~\text{GeV}^2$. The zero crossing is also
predicted from a Poincar\'{e} covariant Faddeev calculation describing the
nucleon as a correlated quark-diquark system, generated with an interaction
fitted to the structure of mesons \cite{Alkofer:2004yf,Holl:2005zi}. On the
other hand, the models without explicit Goldstone bosons, in particular pions,
can not describe the region of $Q^2 \lessapprox 0.5~\text{GeV}^2$, in which the
pionic degrees of freedom
play a decisive role.\\

The inclusion of a pion cloud into quark models of the nucleon started with the
``little bag model'' \cite{Brown:1979ui} and was first applied to the nucleon
form factors in the form of the ``cloudy bag
model'' \cite{Thomas:1982kv,Lu:1997sd,Miller:2002qb,Miller:2002ig}. More
recently, \textcite{Pasquini:2007iz} studied a system of valence quarks 
surrounded by a meson cloud with light-cone wave functions. They found
distinct features of such a cloud below $Q^2=0.5~{\text {GeV}}^2$, however 
without a pronounced bump-dip structure. Another model incorporating quarks
and Goldstone bosons is the ``chiral quark soliton model'' 
\cite{Diakonov:1985eg,Diakonov:1987ty}. The nucleon form factors were
calculated within this framework by \textcite{Christov:1995hr}. The linear
decrease of the ratio $G_E^p/G_M^p$ with $Q^2$ was shown to follow from this
model quite naturally 
\cite{Holzwarth:1996xq,Holzwarth:2002ga,Holzwarth:2005re}. However,
in all of these models the bump/dip structure of the form factors is not
in focus. On the other hand, \textcite{Faessler:2005gd} showed that this
structure can be reproduced within a chiral quark model by a cloud of
pseudoscalar mesons. One should also keep in mind that usually the authors do
not give a complete description of the data. For example, a cross section ratio
may be obtained in agreement with the data, whereas the model fails to describe
the individual cross sections. On the contrary, the parameterization of
\textcite{Friedrich:2003iz} covers the full $Q^2$ domain up to 7~GeV$^2$, and
therefore the charge distribution derived from this work is directly based on
the experimental data. As is obvious from Fig.~\ref{fig:rho_n}, the neutron
charge distribution $\rho_E^n(r)$, as defined by the Fourier transform of the
electric Sachs form factor $G_E^n(Q^2)$, is positive in the interior region and
negative for radii larger than about 0.7~fm. However, it certainly takes a
model to quantify the separation into components, say a core and a ``pion
cloud''. As an example, there is no unique way to break the spectral function
of Fig.~\ref{fig:disp_spectrum} into parts belonging to the two-pion continuum
and heavier intermediate states like the $\rho$ meson. On the other side, it is
also evident that the tail of the density at large radii is determined
by the lightest hadron, the pion.\\

Since the size of the bump/dip signal found by \textcite{Friedrich:2003iz} (FW)
is in conflict with calculations using dispersion
relations \cite{Belushkin:2006qa,Meissner:2007tp} (BHM), it is worthwhile to
discuss the differences more closely.
\begin{itemize}
\item
The fit of FW describes the data in the space-like region with a very
good $\chi^2/\text{dof} \approx 0.9$ for about 160 degrees of freedom (dof),
because the fitting function is designed for the space-like data. The
dispersion relations try to reproduce both the space-like and the time-like
form factors by use of all the available spectral data for the involved
hadrons, and therefore the fits are much more constrained. As a result, the
fits of BHM have the much larger $\chi^2/\text{dof} \approx 1.8$ with
$\text{dof} \approx 200$. Such large values of $\chi^2$ have to be taken with
great caution since they are somewhat outside the range of the validity of
statistics. In particular, the statistical probability $P(\chi^2/\text{dof} > 
1.8,\,\text{dof} \approx 200)$ is smaller than $10^{-10}$. This leaves the
usual suspects: the problem is in the
data \cite{Belushkin:2006qa,Meissner:2007tp}, the dispersion relation have
still an incomplete input, or both data and theory have problems. In any case,
the 1-$\sigma$ bands of \textcite{Belushkin:2006qa} and
\textcite{Meissner:2007tp} derived by increasing the absolute $\chi^2$ by 1 are
not meaningful if the $\chi^2/\text{dof}$ is as much off as $1.8$.
\item
In order to obtain the bump/dip structure at $Q^2 \approx 0.2~\text{GeV}^2$,
BHM would have to include two more ``effective'' poles: an additional isoscalar
pole near the (isoscalar) $\omega$ meson, but with the opposite sign and twice
the strength of the $\omega$, and a weaker isovector structure close to the
mass of 3 pions, which is the threshold of the isoscalar channel. With these
modifications, also BHM could obtain a $\chi^2/\text{dof} \approx 0.9$. There
is however no evidence for such structures in ${e^+}-{e^-}$ collisions nor are
such objects known to interact with the nucleon, and therefore BHM discard
these fits.
\item
The electric rms radius of the proton, $r_E^p$, is another piece of evidence
showing some peculiarity around $0.2~\text{GeV}^2$. From a fit to all the
available low $Q^2$ data, \textcite{Sick:2003gm} finds the radius $r_E^p =
(0.95 \pm 0.018)$~fm . On the other hand, FW obtain $r_E^p = 0.794$~fm without
and $r_E^p = 0.858$~fm with the bump/dip structure. According to
\textcite{Rosenfelder:1999cd}, Coulomb and recoil corrections have to be added
to these results, which leads to $r_E^p = 0.876 \pm 0.015$~fm in accord with
results from Lamb shift measurements \cite{Udem:1997xx}. (For an overview of
the results from atomic physics and their interpretation, we refer to the work
of \textcite{Karshenboim:1998ky} and \textcite{Carlson:2007sp}.) However, as in
all previous work based on dispersion relations, the electric rms radius of the
proton also turns out to be small in the work of BHM,  $r_E^p = 0.844$~fm or
even smaller.
\item The dip structure reported by FW for the proton corresponds to a bump
structure obtained for the neutron at a similar value of $Q^2$. This change of
sign makes sense, because the pion cloud couples to the isovector photon. As
mentioned before, \textcite{Kopecky:1997rw} obtained a mean square radius
$\langle r^2 \rangle_E^n= -(0.115 \pm 0.004)~\text{fm}^2$ from low-energy
neutron scattering off $^{208}$Pb, however this extraction is certainly model
dependent. BHM get $\langle r^2 \rangle_E^n= -0.118~\text{fm}^2$, in
agreement with \textcite{Kopecky:1997rw}. FW take $\langle r^2 \rangle_E^n =
-0.115$\,fm as a fixed parameter or obtain $\langle r^2 \rangle_E^n =
-0.147$\,fm in the new fit of the analytical form of the phenomenological
model.
\item
It follows from dispersion relations that the tail of the charge distributions
at large radii has a Yukawa shape with the mass of the lightest intermediate
state, that is the 2 pion masses for the isovector and 3 pion masses for the
isoscalar densities. Hence it would take a considerable cancellation of positive
and negative structures in the lower part of the spectral function if one wants
to shift the pion cloud to rms radii above 1.7~fm. This is in conflict with the
bump/dip structure of Eq.~(\ref{eq:3.3}), which results in a considerable
amount of charge in the ``pion cloud''  above 1.7~fm, as seen in
Fig.~\ref{fig:rho_n}.
\item
The fit of FW is restricted to the space-like form factors. This approach can
not be extended to the time-like region, and another purely empirical fit
would make little sense in view of the restricted data basis in the this
region. The dispersion relations, on the other hand, are built just to make
the connection between the two regions, and the results of BHM give a good 
overall description in both domains. However, they miss a structure at 
$E_{\text{beam}} \approx 1.05\,$GeV or $t \approx 4.4~\text{GeV}^2$.
\end{itemize}
In concluding these arguments, we mention that dispersion theory and FW agree
on the dip seen for the magnetic form factors of both proton and neutron at
$Q^2 \approx 0.2~\text{GeV}^2$. There is also qualitative agreement that the
charge and the magnetization in the surface region of the nucleon, $r
\gtrapprox 1$\,fm, are dominated by the pion cloud, which reaches much beyond
the rms radius of the proton. It remains a challenge for both experiment and 
theory to answer the raised questions concerning the distributions of charge 
and magnetization inside an nucleon, which we consider a key aspect of the 
nucleon structure.
\subsection{Weak form factors of the nucleon}
\label{sec:III.4}
\subsubsection{Axial form factor of the nucleon}
\label{sec:III.4.1}
\begin{figure}[]
\begin{center}
\includegraphics[width=0.55\columnwidth,angle=90]{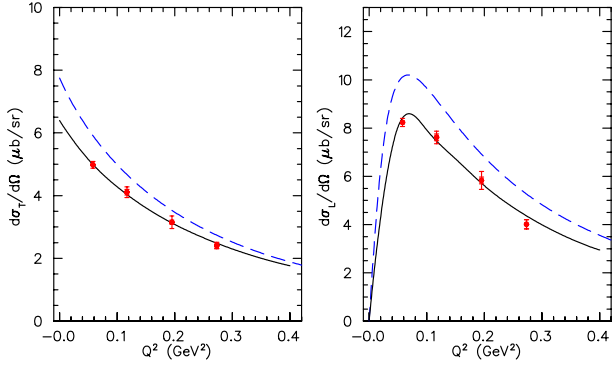}
\end{center}
\caption{The transverse and longitudinal cross sections for charged pion
production as function of $Q^2$. The data are from \textcite{Baumann:2005da}.
Dashed lines: cross sections predicted by MAID2007 \cite{Drechsel:2007if}, 
solid lines: fit to the data with MAID2007 (See text for details.) 
Figure by courtesy of L.~Tiator.}
\label{fig:sig_pi_T}
\end{figure}

The axial current of the nucleon can be studied by anti-neutrino and neutrino
scattering, pion electroproduction, and radiative muon capture, see
\textcite{Bernard:2001rs,Gorringe:2002xx} for recent reviews. The (isovector)
axial current between nucleon states takes the form
\begin{equation}\label{eq:3.8}
A_{\mu} = \bar{u}_{p_2} \left ( \gamma_{\mu}\,
G_A(Q^2)+i\frac{(p_2-p_1)_{\mu}}{2M} \, G_P(Q^2) \right )\gamma_5\, u_{p_1}\,.
\end{equation}
As for the vector current, Eq.~(\ref{eq:2.6}), there appear two form factors,
the axial form factor $G_A$ and the induced pseudoscalar form factor $G_P$. A
linear combination of the form factors $G_A$ and $G_P$ is related to the
pion-nucleon form factor $G_{\pi\,N}$ by the PCAC relation. The experimental
information about the induced pseudoscalar form factor is limited. The data are
mostly obtained from muon capture by the proton,
$\mu^- + p \rightarrow n + \nu_{\mu}$. This determines the value of $G_P$ at
$Q^2=0.88\,m_{\mu}^2\approx0.01$~GeV$^2$, which is usually described by the
induced pseudoscalar coupling constant, $g_P=\frac
{m_{\mu}}{2M}G_P(Q^2=0.88\,m_{\mu}^2)$. A recent experiment at PSI yielded the
value $g_P=7.3\pm1.1$ \cite{Andreev:2007wg}, in agreement with the result from
heavy baryon ChPT \cite{Bernard:2001rs}, $g_P=8.26\pm0.16$, and manifestly
Lorentz-invariant ChPT \cite{Schindler:2006it}, $g_P=8.29\pm 0.7$, with an
estimated error stemming mostly from the truncation of the chiral expansion.
The axial form factor $G_A$ is usually parameterized in the dipole form of
Eq.~(\ref{eq:3.1}), with a parameter $\Lambda_A$ called the ``axial mass'',
\begin{equation}
G_A(Q^2) = \frac{g_A}{(1+Q^2/\Lambda_A^2)^2}\, \label{eq:3.9}
\end{equation}
with $g_A=1.2695(29)$ \cite{Yao:2006px}. A recent (corrected) global average of
the axial mass as determined by neutrino scattering has been given by
\textcite{Budd:2003wb},
\begin{equation}\label{eq:3.10}
\Lambda_A^{\nu} = (1.001 \pm 0.020)~\text{GeV}\,.
\end{equation}
However, quite a different value, $\Lambda_A^{\nu} = (1.20 \pm
0.12)~\text{GeV}$ has been derived by the K2K Collaboration from quasi-elastic
$\nu_{\mu}n \rightarrow \mu^-p $ in oxygen nuclei \cite{Gran:2006jn}. The axial
form factor has also been studied by pion electroproduction 
\cite{Baumann:2005da}. The Rosenbluth separation of these
data is shown in Fig.~\ref{fig:sig_pi_T}. The results are in agreement with
an earlier experiment by \textcite{Liesenfeld:1999mv}, but are complemented by
a data point at the very low momentum transfer of $Q^2=0.058~\text{GeV}^2$. An
exact Rosenbluth separation is prerequisite, because the transverse cross
section $\sigma_T$ is sensitive to $G_A$ and the longitudinal cross section to
the pion form factor $F_{\pi}$, which is discussed in
subsection \ref{sec:III.5}. These electroproduction data have been analyzed
with MAID2007 as follows: (I) The cross sections $\sigma_T$ and $\sigma_L$ of
MAID were normalized to the data by factors 0.825 and 0.809, respectively, and
(II) the axial dipole mass $\Lambda_A$ and the corresponding ``mass'' for the
monopole form of the pion form factor were fitted. The result is
$\Lambda^{\pi}_A= (1.028\pm 0.025)$~GeV. However, this value has to be
corrected for the ``axial mass discrepancy'', $\Lambda^{\pi}_A -
\Lambda^{\nu}_A \approx 0.055$~MeV, which is due to loop corrections
\cite{Bernard:1992ys}. With this correction, the electroproduction data
of \textcite{Baumann:2005da} yield
\begin{equation}\label{eq:3.11}
\Lambda_A^{\text{corr}} = (0.973 \pm 0.025)~\text{GeV}\,,
\end{equation}
which agrees with the corrected value from neutrino scattering given by
Eq.~(\ref{eq:3.10}), but disagrees with both the previous result of
\textcite{Liesenfeld:1999mv} and the measurement of \textcite{Gran:2006jn}. In
view of the relatively large normalization factor applied to the
electro-production data, it would be helpful to check the normalization at
$Q^2=0$ by pion photoproduction.
\subsubsection{Strangeness content of the nucleon}
\label{sec:III.4.2}
As outlined in subsection~\ref{sec:II.3}, the parity violating component of
electron scattering provides access to the weak form factors $\tilde{G}_E$ and
$\tilde{G}_M$. These form factors are related to the strangeness content of the
nucleon by the universality of the electroweak interaction with the quarks. For
a detailed derivation of the strange form factors and their experimental
determination see, e.g., the review of \textcite{Beck:2001yx}. Because the
strangeness in the nucleon appears only through the presence of the heavy $s
\bar{s}$ pairs, these observables are of great importance for our understanding
of the nucleon in terms of large vs. small scales. The strangeness content is
related to the $\sigma$ term, which has been derived from pion-nucleon
scattering at the (unphysical) Cheng-Dashen point, $s=u=M^2,\,t=2m_{\pi}^2$
\cite{Thomas:2001kw,Sainio:2001bq}. This term is a direct measure of the chiral
symmetry breaking in QCD, the chiral properties of the strong interactions, and
the impact of sea quarks on the nucleon's properties. Its relation to the
strangeness contribution is given by
\begin{equation}\label{eq:3.12}
\sigma = \frac{\langle N|\bar{m}(\bar{u}u+\bar{d}d-2\bar{s}s)|N\rangle}
{1-y}\,,
\end{equation}
where $\bar{m}=(m_u+m_d)/2$ is the average of the $u$ and $d$ quark masses,
and $y$ is a measure for the scalar strange quark content of the nucleon,
\begin{equation}\label{eq:3.13}
y = \frac{2 \langle N|\bar{s}s|N\rangle}{\langle
N|\bar{u}u+\bar{d}d|N\rangle}\,.
\end{equation}
From a recent detailed analysis, \textcite{Pavan:2001wz} found the value $y
\approx 0.46$, indeed a surprisingly large strangeness content in the nucleon,
whereas a much smaller value was obtained in earlier work \cite{Sainio:2001bq}.
These inconsistencies were a very strong motive to study the strangeness
content with the electromagnetic probe. At large momentum transfer,
$1~\text{GeV}^2 \leqslant Q^2 \leqslant 100~\text{GeV}^2$, the strangeness
contribution has been derived from unpolarized deep-inelastic lepton scattering
at the Fermi Lab Tevatron \cite{Bazarko:1994tt}. The momentum fraction of the
sea quarks carried by the strange quarks extracted is
\begin{equation}\label{eq:3.14}
\kappa = \frac{\langle x(s(x) + \bar{s}(x))\rangle}{\langle x(\bar{u}(x) +
\bar{d}(x))\rangle} \approx 0.5 \,,
\end{equation}
or about 3\% of the total nucleon momentum. If this contribution is
extrapolated to large spatial scales by the quark evolution, a rather small
value is obtained.\\

\begin{figure}[]
\begin{center}
\includegraphics[width=0.90\columnwidth,angle=0]{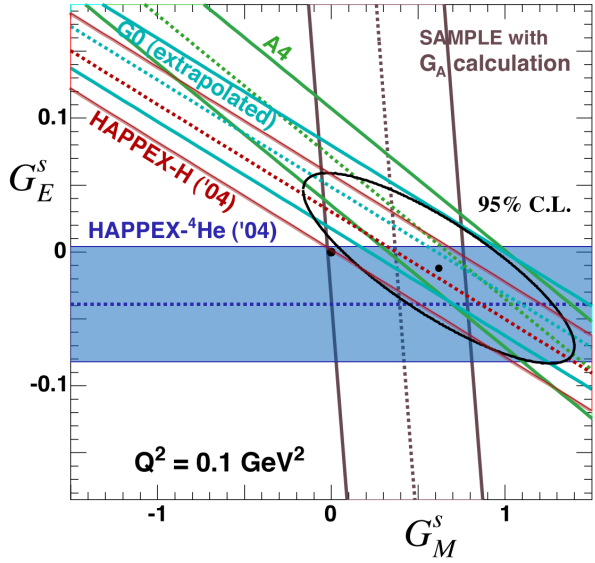}
\end{center}
\caption{The strangeness form factors $G^s_E$ and $G^s_M$ at $Q^2 =
0.1~$GeV$^2$ as obtained by the SAMPLE, A4, HAPPEX, and G0 experiments. The
bands represent the one-sigma (statistical plus systematical) error of the
individual experiments, the ellipse is the combined two-sigma area for all
measurements. The figure is from  \textcite{Kowalski:2006xx}.}
\label{fig:GEs_vers_GMs_ellipse}
\end{figure}

\begin{figure}[]
\begin{center}
\includegraphics[width=0.95\columnwidth,angle=0]{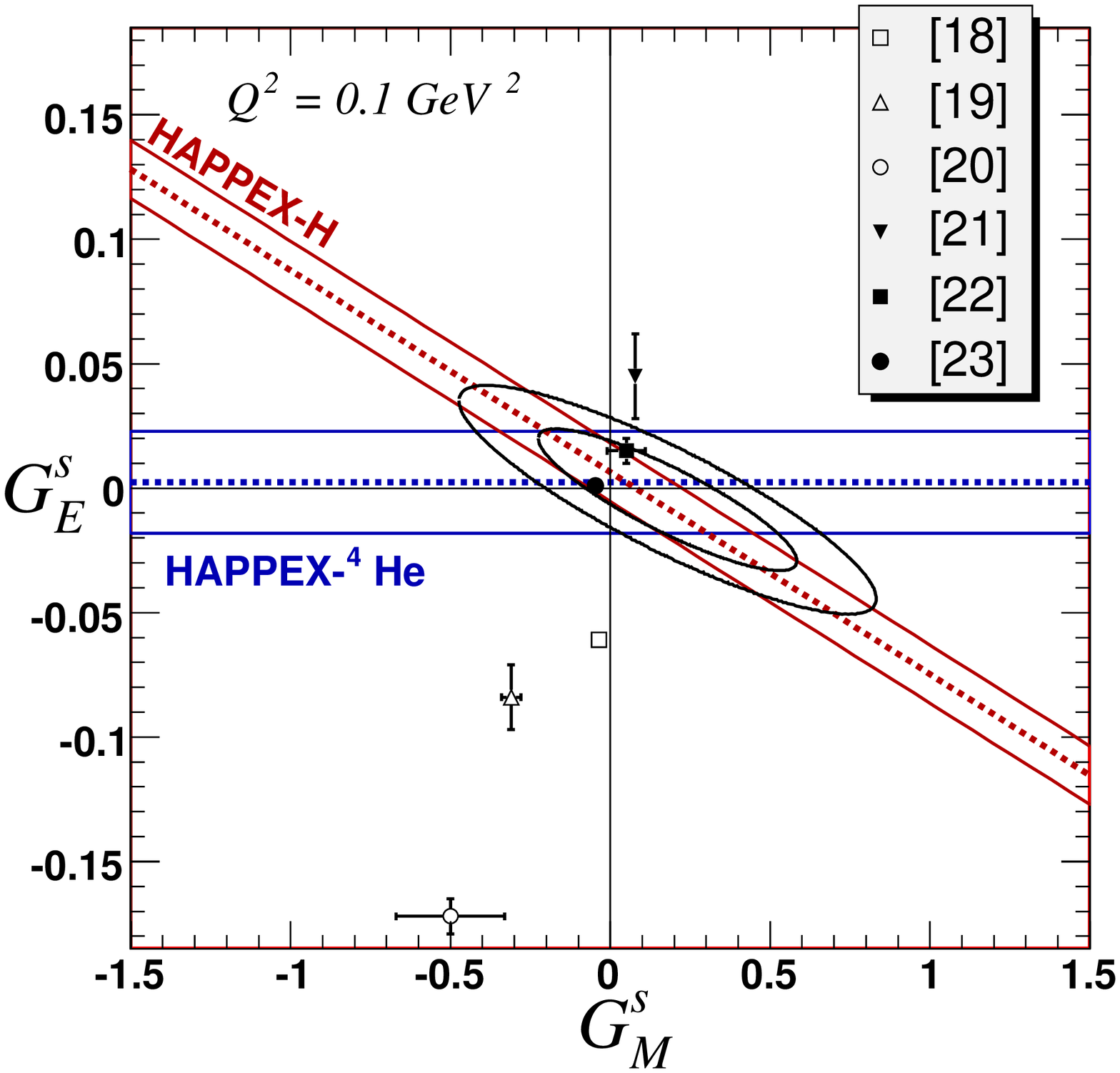}
\end{center}
\caption{The one-sigma band of HAPPEX-II together with the one- and two-sigma
ellipses derived by \textcite{Acha:2005my}. The theoretical predictions with
their estimate errors are identified by the numbers given in the figure: [18]
\textcite{Park:1991fb}, [19] \textcite{Hammer:1995de}, [20]
\textcite{Hammer:1999uf}, [21] \textcite{Silva:2001st}, [22]
\textcite{Lewis:2002ix}, [23] \textcite{Leinweber:2004tc,Leinweber:2006ug}.}
\label{fig:2005_happex}
\end{figure}

\begin{figure}[]
\begin{center}
\includegraphics[width=0.95\columnwidth,angle=0]{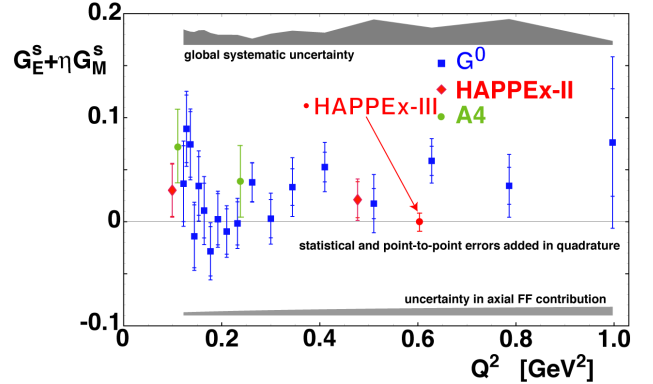}
\end{center}
\caption{The form factor combination $G^s_E(Q^2) + \eta G^s_M(Q^2)$ as obtained
by kinematical extrapolation from A4, HAPPEX~II, and G0 experiments. The point
``HAPPEX~III'' indicates the error bar for a planned measurement. The figure is
from \textcite{Kowalski:2006xx}.} \label{fig:GEs+eta_GMs_fct_Q2}
\end{figure}

On the theoretical side a plethora of nucleon models have usually predicted
strangeness form factors of considerable size, see for example the review of
\textcite{Beck:2001dz}. These considerations have initiated an intense
experimental program at several laboratories. These activities started at the
Bates/MIT laboratory with the SAMPLE experiment, which first proved that it is
feasible to measure the small asymmetries of order $10^{-6}$ in
parity-violating electron scattering \cite{Spayde:2003nr,Kowalski:2006xx}. This
experiment was based on a particular technique using Cherenkov detectors
developed previously for a parity-violation experiment at the Mainz linac
\cite{Heil:1989dz} and an improvement of the SLAC polarized electron source
\cite{Souder:1990ia}. At the Mainz Microtron MAMI, the A4 collaboration built
a Cherenkov detector consisting of 1022 PbF$_2$ crystals, which in conjunction
with electronics allowing for on-line identification of electromagnetic
clusters, made it possible to count single events \cite{Maas:2006xy}.
Furthermore, two experiments were performed at the Jefferson Lab. The first of
these experiments (HAPPEX) used the two-spectrometer set-up of Hall~A taking
advantage of a pair of septum magnets for measurements at very small scattering
angles and low momentum transfers. This project was passing through different
phases of improvement. HAPPEX-I measured on a hydrogen target
\cite{Aniol:2004hp} at $Q^2 = 0.48~\text{GeV}^2$ only. In this geometry the
combination $G^s_E + 0.392~G^s_M=0.014 \pm 0.020 \pm 0.010$ was determined. In
the next step HAPPEX-II measured on both hydrogen \cite{Aniol:2005zg} and
helium targets \cite{Aniol:2005zf}. The nucleus $^4$He is quite special as a
target, because only the electric form factor can contribute due to its zero
total spin. The results of HAPPEX are compared to those of other collaborations
in Fig.~\ref{fig:GEs_vers_GMs_ellipse}. Each measurement gives an error band in
the plot of $G^s_E$ versus $G^s_M$. The common error ellipse indicates values
for $G_E(0.1~$GeV$^2)$ and $G_M(0.1~$GeV$^2)$ that are consistent with zero but
at variance with most theoretical predictions, however, not incompatible with
the experiments obtained with the other methods mentioned above.
Recently the second phase of HAPPEX-II was completed, with the result of a much
improved precision \cite{Acha:2005my}. These results are compared with several
theoretical predictions in Fig.~\ref{fig:2005_happex}. As is obvious from the
figure, the strangeness form factors are centered about zero, whereas most of
the models predict large values. The only theoretical results compatible with
these experiments are from lattice gauge calculations with chiral extrapolation
to the physical pion mass
\cite{Lewis:2002ix,Leinweber:2004tc,Leinweber:2006ug}. The second JLab
experiment was performed by the G0 collaboration. This collaboration has built
an eight-sector superconducting toroidal magnetic
spectrometer \cite{Armstrong:2005hs}. Figure~\ref{fig:GEs+eta_GMs_fct_Q2}
displays the $Q^2$ dependence of the world data including the G0 results. From
this figure we get the impression of a small but finite value for that
particular combination of the two strangeness form factors.\\
\subsection{Form factor of mesons}
\label{sec:III.5}
Because mesons are unstable particles, their form factors can not be measured
directly by lepton scattering but have to be obtained by more indirect methods.
In the following we concentrate on the form factor of the charged pion, however
similar methods can also be used to measure the form factors of heavier mesons 
or rare decays \cite{Guidal:1997hy,Vanderhaeghen:1997ts}. At present, only the
data for the pion are precise enough to allow for a reliable extraction of the
form factors over a large region of momentum transfer. There exist two
experimental methods to overcome the missing target problem. The first method
is the scattering of relativistic mesons with dilated lifetime on atomic
electrons, which are then identified by measuring the recoil of the struck
electron. This method is limited to relatively small momentum transfer, $Q^2
<0.5~\text{GeV}^2$. As a consequence, this method is essentially sensitive
to the rms radius of the free pion, $r_{\pi} = \sqrt{\langle r^2
\rangle_{\pi}}$, which is related to the mass parameter $\Lambda_{\pi}^2$ in
the usual monopole form as follows:
\begin{equation}
F_{\pi}(Q^2) = \frac{1}{1+Q^2/\Lambda_{\pi}^2} = 1 - \frac{1}{6}\, Q^2\,\langle
r^2 \rangle_{\pi} + {\mathcal {O}}~(Q^4)\,. \label{eq:3.15}
\end{equation}
From an experiment at the CERN SPS,
\textcite{Amendolia:1983di,Amendolia:1986wj} derived the rms charge radius of
the pion, $r_{\pi} = (0.663 \pm 0.006)~$fm. At the same time also the kaon form
factor was measured \cite{Amendolia:1986ui}. The rms charge radius of the
charged kaon was found to be $r_K = (0.58 \pm 0.05)~$fm, somewhat smaller than
the pion radius, which is to be expected because of the heavier strange quark 
in the kaon. Because of the small momentum transfer involved, these results
depend only little on the monopole form of the ansatz.\\

\begin{figure}[]
\begin{center}
\includegraphics[width=\columnwidth,angle=0.]{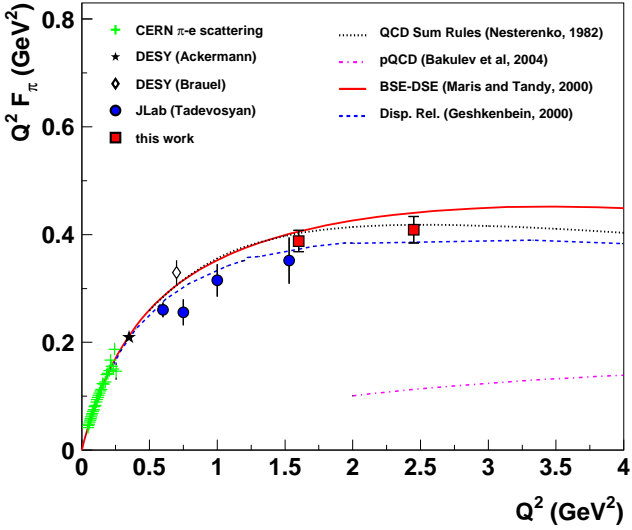}
\end{center}
\caption{The pion form factor $F_{\pi}$ multiplied with $Q^2$. Crosses:
\textcite{Amendolia:1983di,Amendolia:1986wj} ($\pi-e$ scattering at CERN),
asterisks: \textcite{Ackermann:1977rp} (DESY), diamonds:
\textcite{Brauel:1977ra,Brauel:1979zk} (DESY), circles:
\textcite{Tadevosyan:2007yd} (Jefferson Lab), and squares:
\textcite{Horn:2006tm} (Jefferson Lab). The experimental results are compared
to four model calculations as indicated by the lines. Figure from
\textcite{Horn:2006tm}.} \label{fig:ff_pion}
\end{figure}

The second possibility to study the pion form factor is given by electron
scattering on the pion cloud of the nucleon, which is part of the reaction
$p(e,e'\pi^+)n$. The obvious problem is that the initially bound pion is off
its mass-shell and that many other diagrams contribute as well. The idea is
therefore to study this process in kinematic regions for which the t-channel
pion exchange is dominant. In principle one should extrapolate the cross
section to the pion pole, which however lies at the unphysical 4-momentum
transfer $t =m_{\pi}^2$. At very high momentum transfer, $Q^2 \gg 1~$GeV$^2$,
this extrapolation can be performed within the Regge model
\cite{Vanderhaeghen:1997ts}. Following this approach, \textcite{Horn:2006tm}
have recently determined $F_{\pi}$ by a Rosenbluth separation of the
longitudinal and transverse cross sections at the Jefferson Lab. In this way
they extracted two precise values of the pion form factor at $Q^2 =
1.60~\text{GeV}^2$ and $2.45~\text{GeV}^2$, which are shown together with
previous data and several model calculations in Fig.~\ref{fig:ff_pion}. Beside
the model dependence on the Regge analysis, another problem arises because the
data at the higher momentum transfer do not follow the monopole form, i.e., the
value of $\Lambda_{\pi}^2$ differs by as much as 7\% from $\Lambda_{\pi}^2$ =
0.53~GeV$^2$ as obtained by \textcite{Amendolia:1986wj}. This leads to an
inconsistency of about 1 standard deviation. At small values of $Q^2$ one can
also try to derive the rms radius from pion electroproduction as pointed out in
subsection~\ref{sec:III.4.1}. A previous result yielded $\Lambda^2_{\pi} =
0.425~\text{GeV}^2$, equivalent to an rms radius
$r_{\pi}=(0.74\pm0.03)~$fm \cite{Liesenfeld:1999mv}. This result is clearly at
variance with the value of \textcite{Amendolia:1986wj}. A reason for this
discrepancy was given by \textcite{Bernard:2000qz,Bernard:2001rs} in terms of
the loop corrections for the longitudinal s-wave multipole $L_{0+}^{(-)}$,
which dominates the cross section $\sigma_L$ at small $Q^2$. In fact, these
loop corrections increase the downward slope of $L_{0+}^{(-)}$ substantially,
such that
\begin{equation}
\langle r^2 \rangle_{\pi}[p(e,e'\,\pi^+)n] =\langle r^2\rangle_{\pi}
+0.266~{\text{fm}}^2 \,. \label{eq:3.16}
\end{equation}
With the pion radius according to \textcite{Amendolia:1983di,Amendolia:1986wj},
$\langle r^2\rangle_{\pi}=0.440$~fm$^2$, the electroproduction experiment
should therefore measure $\langle r^2 \rangle_{\pi}[p(e,e'\,\pi^+)n]
=0.706~\text{fm}^2$ or an ``effective'' rms radius of 0.84~fm, which is
surprisingly close to the nucleon radius. The fit with MAID07 to the data of
\textcite{Baumann:2005da} yields $\Lambda^2_{\pi} = (0.386\pm
0.042)~\text{GeV}^2$, which is smaller than the result of
\textcite{Liesenfeld:1999mv}. If we include the loop correction, we obtain
$r_{\pi}=(0.78\pm0.04)$~fm, which is quite close to the prediction of ChPT. Of
course, there are weak points in our reasoning. In the first place, there is
the already discussed overall reduction of the MAID model to fit the data.
Second, the s-wave yields only about half of the measured cross section, and
much smaller loop corrections are expected for the higher partial waves.
Nevertheless we may conclude that the virtual constituent pion looks quite
different from the free pion and appears, in this particular experiment, nearly
as large as the nucleon.\\

Because the neutral pion is its own antiparticle, its form factor vanishes
identically. However, the reaction $\gamma*(Q^2) + \gamma \rightarrow \pi^0$
can be studied as function of $Q^2$. This provides information on the
transition form factor $F_{\gamma*\gamma\pi^0}$ of the Wess-Zumino-Witten
anomaly defined by Eq.~(\ref{eq:1.7}) of section~\ref{sec:I}. As shown in
section~\ref{sec:IV} this anomaly is quite important in Compton scattering and
in particular for the spin polarizability of the nucleon. The transition form
factor  $F_{\gamma*\gamma\pi^0}$ was measured at the $e^+\,e^-$ collider at
Cornell by the CLEO collaboration \cite{Gronberg:1997fj} and at the PETRA
storage ring by the CELLO collaboration \cite{Behrend:1990sr}. Analyzed with a
monopole form factor as given by Eq.~(\ref{eq:3.15}), these experiments yielded
the parameter $\Lambda_{\gamma*\gamma\pi^0}=(776\pm 10\pm 12\pm 16)$~MeV, and
similarly for the corresponding transition form factor of the $\eta$ meson
$\Lambda_{\gamma*\gamma\eta}=(774\pm 11\pm 16\pm 22)$~MeV. Both results confirm
the prediction of the vector dominance model assuming that the virtual photon
coexists with the neutral $\rho$ meson, which decays in a pion or an $\eta$ and
a real photon. The $Q^2$ dependence of the transition is then simply given by
the propagator of the $\rho$, that is, $\Lambda \approx m_{\rho}= 775.5$~MeV
for both reactions. With the same arguments as for the ordinary form factors,
we can turn this value in a transition radius $r_{\rm WZW}=0.62$~fm.

\section{Polarizabilities}
\label{sec:IV}
The polarizability measures the response of a system to a quasi-static
electromagnetic field. In particular the energy of a homogeneous and isotropic
system is described by the  electric ($\alpha_{E1}$) and magnetic
($\beta_{M1}$) dipole polarizabilities. In the case of a macroscopic system
these polarizabilities are related to the dielectric constant $\varepsilon$ and
the magnetic permeability $\mu$. The classical theory of Lorentz describes the
dispersion in a medium in terms of electrons bound by a harmonic force. In the
presence of a static and uniform electric field $\vec{E}_0$, the Hamiltonian of
the harmonic oscillator takes the form
\begin{equation} \label{eq:4.0.1}
H=\frac{\vec{p}\,^2}{2m}+\frac{m\omega_0^2}{2}\,\vec{r}\,^2+{\tilde e}\,
\vec{r}\cdot \vec{E}_0 \, . \end{equation}
We note that Gaussian units are used in this subsection in order to concur with
the standard notation of classical electrodynamics, i.e., the fine structure
constant is related to the charge by $\alpha_{\rm em}={\tilde e}^2\approx
1/137$. In all other parts of this review we use Heaviside-Lorentz units, i.e.,
$\alpha_{\rm em}=e^2/(4\,\pi) \approx 1/137$. Substituting $\vec {r}=\vec{r}\,'
+\Delta\vec{r}$, where $\Delta \vec{r}= -{\tilde e}\vec{E}_0/(m \omega_0^2)$ is
the displacement due to the electric field, we may rewrite Eq.~(\ref{eq:4.0.1})
as
\begin{equation} \label{eq:4.0.2}
H=\frac{\vec{p}^2}{2m}+\frac{m\omega_0^2}{2}\,\vec{r}\,'^2+\Delta E\, ,
\end{equation}
that is, the applied electric field has induced a dipole moment $\vec {d}$ and
an energy shift $\Delta E$,
\begin{equation} \label{eq:4.0.3}
\vec {d} = -{\tilde e} \Delta \vec{r} = \frac{\alpha_{\rm em}}
{m\omega_0^2}\,\vec{E}_0\ , \ \ \ \Delta E=-\frac{\alpha_{\rm
em}}{2m\omega_0^2}\,\vec{E}_0^2 \, .
\end{equation}
The electric dipole polarizability $\alpha_{E1}$ is obtained by varying the
induced dipole moment or the energy shift with regard to the electric field,

\begin{equation}
\label{eq:4.0.4} \alpha_{E1}=\frac{\delta \vec{d}}{\delta \vec{E}_0} =
-\frac{\delta^2 \Delta E}{(\delta \vec{E}_0)^2}=\frac {\alpha_{\rm
em}}{m\omega_0^2} \, .
\end{equation}
This result is quite general and also valid for quantum mechanical systems. The
energy shift to first order (linear Stark effect) vanishes for a system with
good parity, and if the ground state is spherically symmetric, the second order
(quadratic Stark effect) yields
\begin{equation} \label{eq:4.0.5}
\Delta E = -\alpha_{\rm em} \sum_{n>0}\,\frac{|\langle n|z|0 \rangle|^2}
{\epsilon_n-\epsilon_0}\, \vec{E}_0^2 \, ,
\end{equation}
with the electric field pointing along the $z$-axis, and $\epsilon_n$ the
energies of the eigenstates  $|n>$. Combining Eqs.~(\ref{eq:4.0.4}) and
(\ref{eq:4.0.5}), we obtain the (quasi-static) electric dipole polarizability,
\begin{equation} \label{eq:4.0.6}
\alpha_{E1} = 2\,\alpha_{\rm em}\,\sum_{n>0}\,\frac{|\langle n|z|0 \rangle|^2} 
                    {\epsilon_n-\epsilon_0}\ .\\
\end{equation}
As an example for a classical extended object we quote the polarizabilities of
a small dielectric and permeable sphere with radius $a$~\cite{Jackson:1975dd},
\begin{equation} \label{eq:4.0.7}
\alpha_{E1}=\frac{\epsilon-1}{\epsilon+2}\,a^3\ ,\ \ \
\beta_{M1}=\frac{\mu-1}{\mu+2}\,a^3\ .
\end{equation}
The polarizabilities for a perfectly conducting sphere are obtained from
Eq.~(\ref{eq:4.0.7}) in the limits $\epsilon\rightarrow\infty$ and
$\mu\rightarrow0$, $\alpha_{E1}=a^3$ and $\beta_{M1} = -\frac{1}{2}\,a^3$. Up
to a factor $4\pi/3$, the electric polarizability of a conducting sphere is the
volume of the sphere. Because of the different boundary conditions for the
magnetic field, the magnetic polarizability turns out negative. The induced
currents in the conductor lead to a magnetization opposite to the applied field
according to Lenz's law, i.e., diamagnetism. A permeable sphere can be
diamagnetic $(\mu<1)$ or paramagnetic $(\mu>1)$, in the latter case the
magnetic moments are already preformed and become aligned in the presence of
the external field. Whereas the magnetic polarizabilities of atoms and
molecules are usually very small, $|\mu-1|\lesssim10^{-2}$, the electric
polarizabilities may be quite large compared to the volume. For example, with a
static dielectric constant of $\varepsilon=81$, water is a nearly perfect
conductor, although in the visible range this constant is down to
$\varepsilon=1.8$, corresponding to a refraction index $n=1.34$. A further,
quantum mechanical example is the hydrogen atom. Its ground state has good
parity and spherical symmetry and therefore Eq.~(\ref{eq:4.0.6}) applies. It is
even possible to sum over the excited states and to obtain the closed form
$\alpha_{E1}\,(^1H) = \frac{9}{2}\,a_B^3$, where $a_B$ is the Bohr radius
\cite{Merzbacher:1970dd}. With an rms radius given by $\langle r^2 \rangle 
=3a^2_B$, the equivalent hard sphere has the radius $R=\sqrt{5}\,a_B$, and 
as a result the hydrogen atom is a pretty good conductor, 
$\alpha_{E1}$/volume $\approx1/10$.\\

In the following we report on the polarizabilities of the nucleon (subsections
\ref{sec:IV.1} and \ref{sec:IV.2}) and pion (subsection \ref{sec:IV.3}). Both
particles are very rigid objects. They are held together by strong
interactions, and the applied electromagnetic field can not easily deform the
charge distribution. If compared to macroscopic matter, the nucleon is a
dielectric medium with $\varepsilon\approx1.002$, that is a very good
insulator. Furthermore, magnetic effects are a priori of the same order as the
electric ones, because the charged constituents, the quarks and mesons, move
close to the speed of light. However, the diamagnetic effects of the pion cloud
and the paramagnetic effects of the quark core of the nucleon tend to cancel,
with the result of a relatively small net value of $\beta_{M1}$. The
polarizability of the nucleon can be measured by Compton scattering: The
incoming photon deforms the nucleon, and by measuring the energy and angular
distributions of the outgoing photon one can determine the induced current and
magnetization densities. Particularly interesting is the case of ``virtual
Compton scattering'' (VCS), which yields information on the spatial distribution
of the polarization densities. Furthermore, the nucleon has a spin and
therefore polarized nucleons appear as anisotropic objects. This leads to the
spin or vector polarizabilities  whose closest parallel in classical physics is
the Faraday effect.
\subsection{Real Compton scattering}
\label{sec:IV.1}
The reaction $\gamma (q, \varepsilon)+N(p, \lambda)\rightarrow \gamma (q\, ',
\varepsilon\, ')+N(p\, ', \lambda\, ')$ involves the absorption of an incident
real photon with 4-momentum $q$ and polarization $\varepsilon$ on a nucleon
with 4-momentum $p$ and polarization $\lambda$, leading to a spectrum of
intermediate hadronic states, which finally decay by the emission of a real
photon leaving the nucleon back in its ground state. Typical intermediate
states are shown diagrammatically in Fig.~\ref{fig:RCS_graph}, and the
following Fig.~\ref{fig:RCS_lex} shows the contributions of these diagrams to
the differential cross section. For a point Dirac particle only the diagrams
(a) and (b), with a nucleon in the intermediate state would contribute. These
two ``nucleon Born terms'' yield singularities for the (unphysical) kinematics
$s=M^2$ and $u=M^2$, respectively. The differential cross section for such a
point nucleon was first calculated by \textcite{Klein:1929dd}. The predicted
cross section increases surprisingly much by adding the Pauli current, i.e.,
the anomalous magnetic moment of the nucleon. The result is the Powell cross
section \cite{Powell:1949dd}. If we further add the ``pion pole term'' of
Fig.~\ref{fig:RCS_graph}~(f), the cross section falls back towards the
Klein-Nishina values. The pion pole term has a singularity at $t=m_{\pi}^2$, it
results from the decay $\pi^0\rightarrow\gamma+\gamma$ as a consequence of the
axial anomaly derived on general grounds by \textcite{Wess:1971yu} and
\textcite{Witten:1983tw}. This term is often referred to as triangle anomaly,
because the vertex $\pi\gamma\gamma$ can be microscopically described by a
triangular quark loop, a diagram not allowed in any classical theory and only
appearing due to the renormalization process of quantum field theory. As we see
from Fig.~\ref{fig:RCS_lex}, the pion pole term yields a large contribution for
backward angles. All further contributions in Fig.~\ref{fig:RCS_graph} do not
have pole structures, but correspond to excited states in $s$-, $u$- or
$t$-channel processes. As such they yield dispersive contributions that
determine the polarizabilities of the nucleon. If we include only the electric
and magnetic dipole polarizabilities, we obtain the low energy expansion (LEX)
in Fig.~\ref{fig:RCS_lex}. This expansion describes the data only up to a
photon lab energy of about 80~MeV, over a region in which the polarization
effects are small and the data scatter. Therefore, the analysis of the modern
data has been based on dispersion relations whose results are shown by the
solid line in the figure. Clearly the higher order terms become more and more
important with increasing photon energy, particularly after crossing the pion
threshold (seen as a kink at about 150~MeV), from thereon the energy increases
sharply towards the $\Delta (1232)$ resonance.
\begin{figure}[]
\begin{center}
\includegraphics[width=1.0\columnwidth,angle=0]{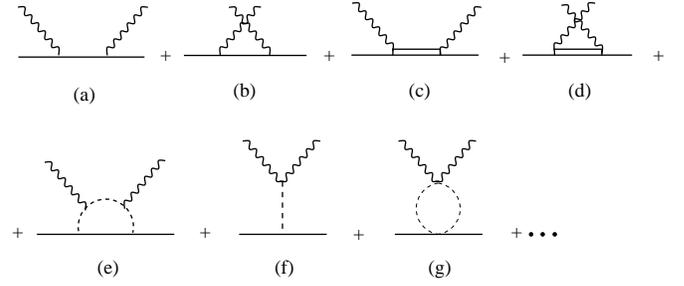}
\end{center}
\caption{Typical intermediate states contributing to Compton scattering off the
nucleon. Upper row: The direct (a) and crossed (b) Born diagrams with
intermediate nucleons, a typical resonance excitation in the $s$-channel (c)
and its crossed version (d). Lower row: mesonic contributions with photon
scattering off an intermediate pion (e), the pion pole diagram (f), and a
correlated two-pion exchange such as the ``$\sigma$ meson'' (g).}
\label{fig:RCS_graph}
\end{figure}
\begin{figure}[]
\begin{center}\hspace{-3.2cm}
\includegraphics[width=0.6\columnwidth,angle=0]{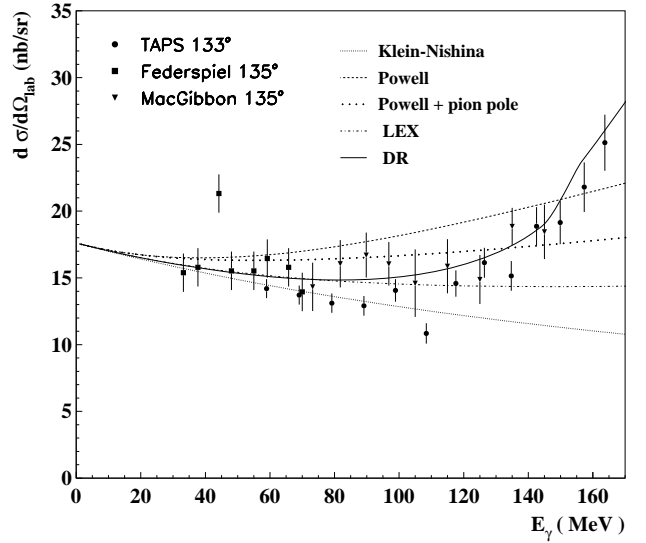}
\vspace{0.4cm}
\end{center}
\caption{Differential cross section for Compton scattering off the proton as
function of the photon lab energy $E_\gamma$ and at scattering angle
$\theta_{\text{lab}}\approx135^{\circ}$. The lines show the results of
fixed-$t$ subtracted dispersion relations (solid), Klein-Nishina (small dots),
Powell (dashed), Powell plus $\pi^0$ pole (large dots), and LEX including the
leading order contributions of $\alpha_{E1}^p$ and $\beta_{M1}^p$
(dashed-dotted). The experimental data are from \textcite{Federspiel:1991yd},
\textcite{MacGibbon:1995in}, and \textcite{OlmosdeLeon:2001zn}. 
Figure from \textcite{Drechsel:2002ar}.}
\label{fig:RCS_lex}
\end{figure}
\subsubsection{Compton amplitudes and polarizabilities}
\label{sec:IV.1.1}
Assuming invariance under parity, charge conjugation and time reversal
symmetry, the general amplitude for real Compton scattering (RCS) can be
expressed by 6 independent structure functions $A_i(\nu,t)$ depending on the
two Lorentz invariant variables $\nu=(s-u)/(4M)$ and $t$, see
Eq.~(\ref{eq:2.2}) for the definitions. These variables are related to the
initial ($E_{\gamma}$) and final ($E_{\gamma}'$) photon lab energies, and to
the lab scattering angle $\theta_{\text{lab}}$ as follows:
\begin{eqnarray} \label{eq:4.1.1}
t &  = & - 4E_{\gamma}E_{\gamma}'\sin^2 \frac{\theta_{\text{lab}}}{2} =
- 2M (E_{\gamma}-E_{\gamma}')\ , \nonumber \\
\nu & = & E_{\gamma} + \frac{t}{4M} = \frac{1}{2} (E_{\gamma}+E_{\gamma}')\ .
\end{eqnarray}
Note that for $\theta_{\text{lab}}=0$  (forward scattering) the Mandelstam
variable $t$ vanishes and the crossing-odd variable $\nu$ is simply the
incident photon lab energy $E_{\gamma}$. \\

The general RCS scattering matrix takes the form
\begin{equation}\label{eq:4.1.2}
T_{fi} = \varepsilon_{\mu} \, \varepsilon_{\nu}'^{\ast} \, \bar{u}_f \,
(p',\lambda'_N) H^{\mu\nu} \, u_i(p,\lambda_N) \, ,
\end{equation}
where $u$ and $\bar u$ are the nucleon spinors. The Compton tensor $H^{\mu\nu}$
contains the hadronic transition currents $J_{\mu}$ and $J_{\nu}$ as well as
the propagation of the intermediate hadronic state. It can be decomposed into a
complete basis of 6 tensor structures constructed from the independent momentum
4-vectors and appropriate Dirac matrices \cite{Prange:1958dd,L'vov:1980wp},
\begin{equation} \label{eq:4.1.3}
H^{\mu\nu}  = \sum_{i=1,6}\mathcal {M}_i^{\mu\nu}A_i (\nu, t)\, .
\end{equation}
For further details we refer to \textcite{Drechsel:2002ar}.\\

Let us now consider the forward scattering of a real photon by a nucleon. The
incident photon is characterized by the Lorentz vectors of momentum,
$q=(|\vec{q}\,|\,,\vec {q}\,)$, and transverse polarization,
$\varepsilon_{\lambda} =(0,\,\vec{\varepsilon}_{\lambda})$, with $q \cdot q=0$
for real photons and $\varepsilon_{\lambda}\cdot q=0$. The corresponding
quantum numbers of the outgoing photon are denoted by primed quantities. If the
incident photon moves in the direction of the z-axis,
$\vec{q}=|\vec{q}\,|\,\hat {e}_z$, the polarization vectors
\begin{equation}\label{eq:4.1.4a}
\vec {\varepsilon}_{\pm} = \mp \frac{1}{\sqrt{2}}\,(\hat {e}_x \pm i\hat{e}_y)
\end{equation}
correspond to circularly polarized light with helicities $\lambda=+1$
(right-handed) and $\lambda=-1$ (left-handed). The forward Compton amplitude
takes the form
\begin{equation}\label{eq:4.1.4b}
T(\nu,\,\theta=0) = \vec{\varepsilon}\,'\,^{\ast}\cdot\vec {\varepsilon} \,
f(\nu)+
i\,\vec{\sigma}\cdot(\vec{\varepsilon}\,'\,^{\ast}\times\vec{\varepsilon}\,)\,g(\nu)\,.
\end{equation}
Because $T$ is invariant under the crossing transformation,
$\varepsilon\,'\,^{\ast}\leftrightarrow\varepsilon$ and $\nu\rightarrow-\nu$,
$f$ must be even and $g$ odd as function of $\nu$. These forward scattering
amplitudes have the low-energy expansion \cite{Low:1954kd,Gell-Mann:1954kc}
\begin{eqnarray}\label{eq:4.1.5}
f(\nu) & = & - \frac{e^2e_N^2}{4\pi M}+ (\alpha_{E1}+\beta_{M1})\,\nu^2 +
\mathcal{O}(\nu^4)\ , \\
g(\nu) & = & - \frac{e^2\kappa_N^2}{8\pi M^2}\,\nu + \gamma_0\nu^3 +
\mathcal{O}(\nu^5) \ , \label{eq:4.1.6}
\end{eqnarray}
with $e_N$ the charge of the nucleon in units of $e$ and $\kappa_N$ the
anomalous magnetic moment in units of nuclear magnetons. The leading
contribution to $f(\nu)$ is the Thomson term familiar from non-relativistic
theory. The term linear in $\nu$ vanishes due to the crossing symmetry, and the
term ${\cal {O}}(\nu^2)$ contains the sum of the scalar polarizabilities giving
information on the internal structure. Being odd under crossing, the spin-flip
amplitude $g(\nu)$ starts with the term $\cal {O}(\nu)$ proportional to the
square of the anomalous magnetic moment, and its next order term is described
by the forward spin polarizability $\gamma_0$. The leading terms for both
amplitudes are obtained from the pole terms typical for a point-like particle,
whereas the polarizabilities are contained in the sub-leading terms. As is
evident from the above equations, the scalar and spin polarizabilities have
different units. In the following all scalar polarizabilities are given in
units of $10^{-4}$~fm$^3$, while the vector or spin polarizabilites have units
of $10^{-4}$~fm$^4$. As will be detailed in section~\ref{sec:VI.1}, the forward
scalar ($\alpha_{E1}+\beta_{M1}$) and forward spin ($\gamma_0$)
polarizabilities of Eqs.~(\ref{eq:4.1.5}) and (\ref{eq:4.1.6}) can be
determined by energy-weighted integrals over the photoabsorption cross
sections. In particular, Baldin's sum rule yields the following results for
proton and neutron \cite{Babusci:1997ij}:
\begin{eqnarray}
(\alpha_{E1}+\beta_{M1})_p & = & 13.69 \pm 0.14\, ,\nonumber \\
(\alpha_{E1}+\beta_{M1})_n & = & 14.40 \pm 0.66\, .\label{eq:4.1.7a}
\end{eqnarray}
\indent The T-matrix for general scattering angles is described by the 6 L'vov
amplitudes $A_i(\nu,t)$. These amplitudes have no kinematical constraints, are
symmetrical under crossing, and contain both the pole terms of
(Fig.~\ref{fig:RCS_graph}~a, b, and f) and an integral over the excitation
spectrum, which we call the dispersive amplitude,
\begin{equation}
A_i(\nu,t) = A_i^\text{pole}(\nu,t) \, + \, A_i^\text{disp}(\nu,t)\, .
\label{eq:4.1.7}
\end{equation}
The polarizabilities are determined by the dispersive amplitudes at $\nu=t=0$,
that is, at the threshold for RCS. This defines 6 real numbers $a_i
=A_i^\text{disp}(0,0)$, from which we can derive 2 scalar and 4 vector (or
spin) polarizabilities  by linear combinations. In the scalar sector, we find
the familiar electric ($\alpha_{E1}$) and magnetic ($\beta_{M1}$)
polarizabilities, which appear as $\alpha_{E1} + \beta_{M1}$ for forward  and
$\alpha_{E1} - \beta_{M1}$ for backward Compton scattering. The physical
content of the 4 vector polarizabilities is best described in a multipole
notation. Since the initial and the final states contain a nucleon in its
ground state with total spin $J=\textstyle{\frac{1}{2}}$, the transition
operator must have even parity and angular momentum 0 or 1. The electric
polarizability describes the absorption of an electric dipole photon followed
by the emission of a photon with the same multipolarity, that is $\alpha_{E1}
\sim [ E_1 \times E_1]^{[0 ]}$, and in the same way we find $\beta_{M1} \sim [
M_1 \times M_1]^{[0 ]}$ with the multipoles coupled to 0 (scalar
polarizabilities). In the spin-dependent sector there are 4 polarizabilities at
the lowest order: $\gamma_{E1E1}$, $\gamma_{M1M1}$, $\gamma_{M1E2}$, and
$\gamma_{E1M2}$. In this case the multipolarities are coupled to 1 (vector
polarizabilities). As an example, $\gamma_{M1E2} \sim [ M_1 \times E_2 ]^{[ 1
]}$ defines a spin polarizability with an electric quadrupole absorption
followed by a magnetic dipole emission. It is useful to define the forward
($\theta = 0$) and backward ($\theta = \pi$) polarizabilities also in the
spin-dependent sector,
\begin{eqnarray}
\gamma_0 &=& - \gamma_{E1E1} - \gamma_{M1M1} - \gamma_{M1E2} - \gamma_{E1M2}\,,
\label{eq:4.1.8} \\
\gamma_\pi &=&- \gamma_{E1E1} + \gamma_{M1M1} + \gamma_{M1E2} - \gamma_{E1M2}
\, .\label{eq:4.1.9}
\end{eqnarray}
It is of course possible to define higher polarizabilities related to the $\nu$
and $t$ derivatives of $A_i^\text{disp}(\nu,t)$ taken at $\nu=t=0$.  An often
discussed example is the electric quadrupole polarizability, which appears
among the terms of ${\cal{O}}(\nu^4)$ in Eq.~(\ref{eq:4.1.5}).

\subsubsection{Theoretical developments}
\label{sec:IV.1.2}
In a nonrelativistic approach like the constituent quark model (CQM), the
scalar dipole polarizabilities can be expressed by
\begin{eqnarray}
\alpha_{E1} &=& 2\alpha_\text{em} \sum_{n \ne 0} \frac{|\langle n|
d_z|0 \rangle|^2} {E_n-E_0} + \Delta\alpha_{E1}\,, \label{eq:4.1.14}\\
\beta_{M1} &=& 2\alpha_\text{em} \sum_{n\ne0} \frac{|\langle n| \mu_z|0
\rangle|^2} {E_n-E_0}+ \Delta\beta_{M1} \,,\label{eq:4.1.15}
\end{eqnarray}
where $\vec {d} = \sum \vec {d}_q = \sum e_q \vec {r}_q$ and $\vec {\mu} = \sum
\vec {\mu}_q $ are sums over the electric and magnetic dipole operators of the
constituents. For simplicity the quark masses may be taken as
$m_q={\textstyle{\frac{1}{3}}}M$, and the quark charges $e_q$ are in units of
$e$. Clearly the first terms on the rhs of the above equations are positive,
because the excitation energy $E_n-E_0$ is positive. The second terms describe
recoil and retardation, $\Delta\alpha_{E1}= \alpha_\text{em} \langle 0|\,\sum
e_q\, \vec {r}\,_q^2\,|0\rangle/(3M)$ and $\Delta\beta_{M1} =
-\alpha_\text{em}\langle 0|\,\vec {d}\,^2+\sum \vec{d}\,_q^2\,|0\rangle/(2M)$.
These are small corrections in atomic physics but quite sizeable for the quark
dynamics of the nucleon. They turn out positive for $\alpha_{E1}$ but negative
for $\beta_{M1}$. The leading term of the magnetic polarizability describes the
paramagnetism, mainly by a quark spin-flip transition from the nucleon to the
$\Delta$~(1232), while the sub-leading term represents Langevin's diamagnetism.
The simple CQM with an oscillator potential connects the rms radius $\langle
r^2\rangle^{1/2}$ with the oscillator frequency, $\omega_0=3/(M \langle
r^2\rangle)$, and yields $ \alpha_{E1} = 2\alpha_\text{em} / (M\,\omega_0^2) +
{\mathcal{O}}(M^{-2})$. However, this model is not able to describe both size
and excitation energy. If we use the proper size, $\alpha_{E1}$ is grossly
overestimated, whereas a fit to the excitation energy of the dominant dipole
mode N$^{\ast}$~(1520) leads to a value much below the experiment. For the
magnetic polarizability, the $M1$ transition to the $\Delta$~(1232) yields a
large paramagnetic value, $\beta^{\Delta}_{M1}\approx12$, which is somewhat
reduced by the subleading diamagnetic terms. It was therefore early recognized
that a complete picture of the nucleon must also include the
pion cloud \cite{Weiner:1985ih}.\\

Systematic calculations of pion cloud effects became possible with the
development of chiral perturbation theory (ChPT), an expansion in the external
momenta and the pion or quark mass ($p$ expansion). The first calculation of
Compton scattering in that scheme was performed by \textcite{Bernard:1991rq}.
Keeping only the leading term in $1 / m_\pi$, they found the following
remarkable relation at ${\mathcal {O}}(p^3)$:
\begin{equation} \label{eq:4.1.16}
\alpha_{E1} = 10\beta_{M1} = \frac {5\alpha_\text{em}g_A^2}{96\pi
f_{\pi}^2m_{\pi}} = 12.2\,, \end{equation}
with $f_{\pi}\approx 93$~MeV the pion decay constant and $g_A\approx1.26$ the
axial coupling constant. The calculation was later repeated in heavy-baryon
ChPT, which allows for a consistent chiral power counting, and extended to
${\mathcal{O}}(p^4)$ yielding $\alpha_{E1}^p = 10.5\pm2.0$ and $\beta_{M1}^p =
3.5\pm3.6$ \cite{Bernard:1993ry,Bernard:1993bg}. The error bars for these
values indicate that several low-energy constants appear at this order, which
were determined by resonance saturation, that is by use of phenomenological
information about resonances and vector mesons. Since the $\Delta (1232)$ is
close in energy and very important for photoabsorption, it has been proposed to
include this resonance dynamically. This leads to an additional expansion
parameter, the N$\Delta$ mass splitting ($\varepsilon$ expansion).
Unfortunately, the ``dynamical'' $\Delta$ increases the polarizabilities to
values far above the data, $\alpha_{E1}^p =16.4$ and $ \beta_{M1}^p = 9.1$
\cite{Hemmert:1997tj}. This can be changed by introducing large low-energy
constants within a higher-order calculation, however at the expense of losing
the predictive power.\\

The spin polarizabilities have been calculated to $\mathcal {O}(p^3)$ in both
relativistic ChPT \cite{Bernard:1995dp} and heavy-baryon ChPT
\cite{Hemmert:1997tj}. As an example we give the predictions of the latter
reference:
\begin{eqnarray}
\gamma_0 &=& 4.6-2.4-0.2+0=2.0\,,\label{eq:4.1.16a} \\
\gamma_\pi &=&4.6+2.4-0.2-43.5=-36.7 \, .\label{eq:4.1.16b}
\end{eqnarray}
the 4 separate contributions referring to N$\pi$-loops, $\Delta$-poles,
$\Delta\pi$-loops, and the pion pole, in order. As is obvious from these
results, the $\pi^0$ pole dominates the backward spin polarizability but does
not contribute in the forward direction. Independent calculations of the
forward spin polarizability to $\mathcal {O}(p^4)$ resulted in $\gamma_0 =
-3.9$ \cite{Birse:2000ve,Ji:1999sv}, which indicates a slow convergence of the
expansion.\\

Because a reliable data analysis is based on dispersion relations (DRs), we
recall some pertinent features of this technique in the following. The
invariant amplitudes $A_i$ are free of kinematical singularities and
constraints, they also obey the crossing symmetry and gauge invariance. 
Assuming further analyticity and an appropriate high-energy behavior, these 
amplitudes fulfill unsubtracted DRs at fixed $t$,
\begin{equation}\label{eq:4.1.17}
{\rm {Re}}~A_i(\nu, t) = A_i^\text{pole}(\nu, t) +\frac {2}{\pi} \, \mathcal
{P} \int_{\nu_{0}}^{\infty} d\nu' \; \frac{\nu' \, {\rm {Im}}~A_i(\nu',t)} {
\nu'^2 - \nu^2}\,,
\end{equation}
where $A_i^\text{pole}$ is the nucleon pole term and $\mathcal {P}$ denotes the
principal value integral. The latter can be calculated if the absorptive part
of the amplitude, Im~$A_i$, is known to a sufficient accuracy. Because of the
energy-weighting, the pion production near threshold and the mesonic decay of
the low-lying resonances yield the biggest contributions to the integral.  With
the existing information on these processes and some reasonable assumptions on
the lesser known higher part of the spectrum, the integrand can be constructed
up to cm energies $W \approx 2$~GeV. However, Regge theory predicts that the
amplitudes $A_1$ and $A_2$ do not drop sufficiently fast to warrant a
convergence of the integral. This behavior is mainly due to fixed poles in the
$t$-channel. In particular the $t$-channel exchange of pions and $\sigma$
mesons leads to the bad convergence for $A_2$ and $A_1$, respectively. The
latter meson has a mass of about 600~MeV and a very large width, it models
correlations in the two-pion channel with spin zero and positive parity. In
order to obtain useful results for these two amplitudes,
\textcite{L'vov:1996xd} proposed to close the contour integral in the complex
plane by a semi-circle of finite radius $\nu_{max}$, and to replace the
contribution from the semi-circle by a number of energy independent poles in
the $t$ channel. This procedure is relatively safe for $A_2$ because the
$\pi^0$ pole or triangle anomaly is well established by both experiment and
theory. However, it introduces a considerable model-dependence for $A_1$.\\

In order to avoid the convergence problem and the phenomenology necessary to
determine the asymptotic contributions, it was suggested to subtract the DRs at
$\nu=0$ \cite{Drechsel:1999rf}. This subtraction improves the convergence by
two additional powers of $\nu'$ in the denominator of the dispersion integrals,
Eq.~(\ref{eq:4.1.17}). The subtraction functions $A_i(\nu=0, t)$ can be
obtained from subtracted DRs in $t$ with the imaginary part of the amplitude
$\gamma\gamma\rightarrow\pi\pi\rightarrow N\bar{N}$ as input. In a first step,
a unitarized amplitude for the $\gamma\gamma\rightarrow\pi\pi$ subprocess is
constructed from the available experimental data. This information is then
combined with the $\pi\pi\rightarrow N\bar{N}$ amplitudes determined by
analytical continuation of $\pi N$ scattering amplitudes \cite{Hoehler:1983dd}.
Once the $t$ dependence of the subtraction functions $A_i(0,t)$ is known, the
subtraction constants $a_i=A_i(0, 0)$ have to be fixed. Although all 6
subtraction constants $a_1$ to $ a_6$ could be used as fit parameters, it is
sufficient to fit $a_1$ and $a_2$, or equivalently ($\alpha_{E1} -
\beta_{M1}$) and $\gamma_\pi$ to the data. The remaining 4 subtraction
constants can be calculated through an unsubtracted dispersion integral. Yet
another method are hyperbolic (fixed-angle) DRs, which improve the convergence
for large values of $t$ or backward scattering
angles \cite{Bernabeu:1974zu,Holstein:1994tw,L'vov:1999dd}.
\textcite{Holstein:1994tw} investigated backward DRs in order to get rigorous
bounds for the backward scalar polarizability of the proton. The important
finding was that the phenomenological $\sigma$ meson can be replaced by
experimental information on the $\pi\pi$ continuum. The results of a more
recent analysis are $(\alpha_{E1}-\beta_{M1})_s = -5.6$ and
$(\alpha_{E1}-\beta_{M1})_t=16.5$, leading to a total value of about 10.9 in
good agreement with the data  \cite{Drechsel:2002ar}. The importance of the
$t$-channel contribution has also been found in  a careful analysis of the new
experimental data by \textcite{Schumacher:2007xr} who obtained $(\alpha_{E1} -
\beta_{M1})_t=15.2$. We conclude that the polarizability of the nucleon is
largely determined by the subprocess $\gamma + \gamma \rightarrow \pi + \pi$,
and therefore intertwined with correlations of the two-pion system and the
polarizability of the pion.
\subsubsection{RCS data and extraction of the proton polarizabilities}
\label{sec:IV.1.3}
\begin{figure}[]
\begin{center}
\includegraphics[width=\columnwidth,angle=0]{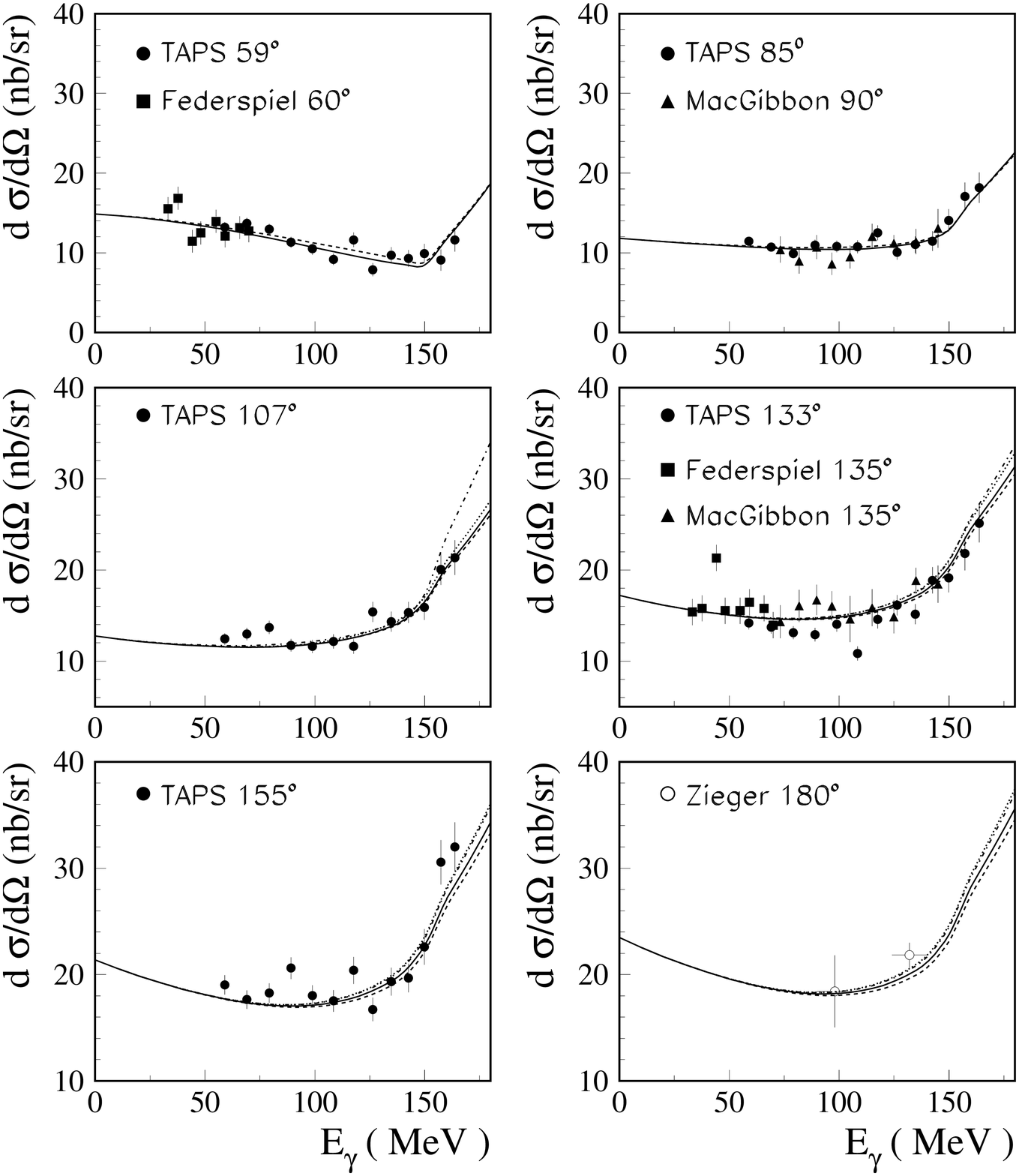}
\end{center}
\caption{Differential cross section for Compton scattering off the proton as a
function of the lab photon energy $E_\gamma$ and at different scattering angles
$\theta_{\text{lab}}$. Solid lines: fixed-$t$ subtracted DRs, dashed lines:
fixed-$t$ unsubtracted DRs, dotted lines: hyperbolic subtracted DRs,
dashed-dotted lines: hyperbolic unsubtracted DRs. All results are shown for
fixed values of $\alpha_{E1}^p+\beta_{M1}^p=13.8$,
$\alpha_{E1}^p-\beta_{M1}^p=10$, and $\gamma_{\pi}^p=-37.$ The data are
from \textcite{OlmosdeLeon:2001zn}~(full circles),
\textcite{Federspiel:1991yd}~(squares),
\textcite{MacGibbon:1995in}~(triangles), and \textcite{Zieger:1992jq}~(open
circles). Figure from \textcite{Drechsel:2002ar}.} 
\label{fig:RCS_olmos_DR}
\end{figure}

\begin{figure}[]
\begin{center}
\includegraphics[width=0.65\columnwidth,angle=0]{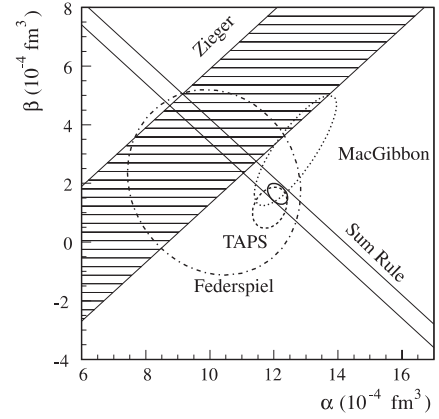}
\end{center}
\caption{Contour plot of $\chi^2 + 1$ for different measurements of
$\alpha_{E1}^p$ and $\beta_{M1}^p$. The 3 dashed ellipses are obtained from
\textcite{OlmosdeLeon:2001zn}, \textcite{Federspiel:1991yd}, and
\textcite{MacGibbon:1995in} as indicated. The dashed area corresponds to the
measurement of \textcite{Zieger:1992jq} at $\theta = 180^{\circ}$, the area
between the other two straight lines to the Baldin sum rule. The small ellipse
drawn with the solid line is a common fit to all the above data. Figure from 
\textcite{OlmosdeLeon:2001zn}.}
\label{fig:RCS_contour}
\end{figure}

The pioneering experiment in Compton scattering off the proton was performed by
\textcite{Gol'danski:1960dd}. They obtained an electric polarizability
$\alpha_{E1}^p = (9\pm 2)$, with a large uncertainty in the normalization of
the cross section giving rise to an additional systematical error of $\pm5$. In
a later experiment \textcite{Baranov:1975ju} used bremsstrahlung providing
photons with energies up to 100~MeV. The data obtained by these authors were
later reevaluated by DRs with the result $\alpha_{E1}^p \approx 12$ and
$\beta_{M1}^p \approx -6$. This outcome was much to the surprise of everybody,
because one expected a large paramagnetic effect of at least
$\beta_\text{para}^p \approx 10$ from the quark spin alignment in the $N
\rightarrow \Delta(1232)$ transition. The first modern experiments were
performed at Illinois \cite{Federspiel:1991yd}, followed by the work of the
Saskatoon group \cite{Hallin:1993ft,MacGibbon:1995in}. With tagged photons at
70~MeV$\leq E_{\gamma}\leq 100$~MeV and untagged photons for the higher
energies, the latter group obtained $\alpha_{E1}^p=\left(12.1 \,\pm\,0.8
\,\pm\, 0.5 \right)$ and $\beta_{M1}^p = \left(2.1 \,\mp\,0.8 \,\mp\,0.5
\right)$. New precision measurements at MAMI \cite{OlmosdeLeon:2001zn} have
been performed with tagged photons and the photon detector TAPS. The  measured
differential cross sections from various laboratories are shown in
Fig.~\ref{fig:RCS_olmos_DR} as a function of the photon lab energy and at
different scattering angles. The data have been compared to the results from
four different types of DRs. The figure shows that the differences among the
predicted results are hardly visible, except for the unsubtracted hyperbolic DR
at $\theta_{\text{lab}}=107^{\circ}$, because this angle is too much forward
for this DR.
A fit to all modern low-energy data constrained by Baldin's sum rule of
Eq.~(\ref{eq:4.1.7a}) yields \cite{OlmosdeLeon:2001zn}
\begin{eqnarray}\label{eq:4.1.20}
\alpha_{E1}^p & = & 12.1  \pm 0.3 \mp 0.4 \pm 0.3\, , \\
\beta_{M1}^p & = & 1.6 \pm 0.4 \pm 0.4 \pm 0.4\, , \label{eq:4.1.21}
\end{eqnarray}
in units of $10^{-4}~\rm{fm}^3$ and with errors denoting the statistical,
systematical, and model-dependent errors, in order. This new global average
confirms, beyond any doubt, the dominance of the electric polarizability
$\alpha_{E1}^p$ and the tiny value of the magnetic polarizability
$\beta_{M1}^p$, which has to come about by a cancelation between the large
paramagnetic s-channel contribution of the $N\Delta$ spin-flip transition and
the somewhat smaller diamagnetic t-channel contribution of the ``pion cloud''.
The huge improvement by the new data is seen in Fig.~\ref{fig:RCS_contour},
which displays the error ellipses in the $\alpha_{E1}^p$ - $\beta_{M1}^p$ plane
as obtained from  recent experiments. For further details of the experiments
and their interpretation, see the review of \textcite{Schumacher:2005an}.\\

Much less is known about the spin sector, except for the forward and backward
spin polarizabilities. The most recent values are
\begin{eqnarray}\label{eq:4.1.22}
\gamma_0 &=& -0.94\pm 0.15 \, , \\
\gamma_{\pi} &=& \left \{
\begin{array}{lll}
-36.1 \pm 2.2 \,\mbox{\cite{OlmosdeLeon:2001zn}}\\
-37.9 \pm 3.6 \, \mbox{\cite{Galler:2001ht}}\\
-38.7\pm 1.8\,\mbox{\cite{Schumacher:2005an}} \\
\end{array} \right \}\quad
\label{eq:4.1.23}
\end{eqnarray}
in units of $10^{-4}~{\rm{fm}}^4$. The small value for $\gamma_0$ in
Eq.~(\ref{eq:4.1.22}) was not measured by Compton scattering but has been
determined by a sum rule based on the helicity-dependent absorption cross
sections, see section~\ref{sec:VI.1}. The upper line in Eq.~(\ref{eq:4.1.23})
gives $\gamma_{\pi}$ as determined from low-energy data
\cite{OlmosdeLeon:2001zn}, in which case the error is dominated by the
statistical plus systematical uncertainty, whereas the middle line refers to
the work of \textcite{Galler:2001ht} who found that the model error prevails in
the $\Delta$ region, and the lower line gives the weighted average of several
MAMI results \cite{Schumacher:2005an}. For all the other spin and higher order
polarizabilities, both ChPT and DR predict small values that can not be
determined without dedicated polarization studies. A new generation of
experiments with polarized beams, polarized targets, and recoil polarimetry
holds the promise to disentangle all the scalar and vector polarizabilities of
the nucleon and to quantify the proton's full spin response to an external
electromagnetic field \cite{Babusci:1998ww,Hildebrandt:2003fm,Beane:2004ra}. 
The HIGS project \cite{Weller:2006dd} of a high-intensity beam with circularly
polarized photons in an energy range up to 140-160~MeV is ideally suited to
perform such experiments in the pion threshold region. Complementary
investigations are planned in the first resonance region using the Crystal Ball
detector at MAMI \cite{Arends:2007dd}. We strongly believe that only a
combination of these experimental projects will provide the ``sharp knife'' to
extract the spin polarizabilities in an unambiguous way \cite{Pasquini:2007hf}.
\subsubsection{RCS data and extraction of the neutron polarizabilities}
\label{sec:IV.1.4}
The experimental situation concerning the polarizabilities of the neutron is
still quite unsatisfactory. The electric polarizability $\alpha_{E1}^n$ can in
principle be measured by scattering low-energy neutrons on the Coulomb field of
a heavy nucleus, whereas the magnetic polarizability $\beta_{M1}^n$ remains
essentially unconstrained by such an experiment. This technique seemed to be
very promising until the beginning of the 1990's, when
\textcite{Schmiedmayer:1991dd} obtained the value
$\alpha_{E1}^n=12.6\pm1.5\,(\rm{stat})\pm2.0\,(\rm{syst})$ by scattering
neutrons with energies 50~eV$\le E_n\le 50$~keV off a $^{208}$Pb target.
Shortly later, however, \textcite{Nikolenko:1992dd} argued that the errors were
underestimated by a factor 5. These findings were confirmed by a similar
experiment \cite{Koester:1995nx} resulting in $\alpha_{E1}^n=0\pm5$, and by
\textcite{Enik:1997cw} who obtained $7\lesssim\alpha_{E1}^n\lesssim19$ after a
further analysis of the systematic errors.\\

The neutron polarizabilities can also be measured by quasi-free Compton
scattering off a bound neutron and elastic scattering on a deuteron. The former
experiment was performed by \textcite{Rose:1990eu}. Interpreted in conjunction
with Baldin's sum rule, the result was $0 <\alpha_{E1}^n < 14$ with a mean
value $\alpha_{E1}^n\approx10.7$. The large error bar arises from the fact that
the Thomson amplitude vanishes for a neutral particle, and therefore also the
interference between this term and the leading non-Born amplitude is absent. It
was therefore proposed to repeat such an experiment at higher energies and
backward angles for which the sensitivity to $\alpha_{E1}^n - \beta_{M1}^n$
is highest. Because the data analysis is very sensitive to final-state
interactions and two-body currents, it was suggested to measure the
polarizabilities of the bound proton at the same time. The proton values
obtained by \textcite{Wissmann:1999dd} were quite promising,
$\alpha_{E1}^p-\beta_{M1}^p=10.3\pm1.7\ ({\rm{stat + syst}})\pm1.1$~(mod). The
experiment was then extended to the neutron by the CATS/SENECA Collaboration
\cite{Kossert:2002jc}. Data were collected for both deuterium and hydrogen
targets and analyzed by \textcite{Levchuk:1999zy}. The agreement between the
polarizabilities of free and bound protons was still quite satisfactory, and
the final result for the (bound) neutron was
\begin{equation} \label{eq:4.1.24}
\alpha_{E1}^n-\beta_{M1}^n = 9.8\pm3.6\,
(\rm{stat})^{+2.1}_{-1.1}\,(\rm{syst})\pm2.2\,(\rm{mod})\, .
\end{equation}
This value is compatible with an earlier datum of the Saskatoon group
\cite{Kolb:2000ix}, obtained at similar energies and angles but with a much
larger error bar, $\alpha_{E1}^n-\beta_{M1}^n\approx12$. The comparison between
proton and neutron demonstrates that there is no significant isovector
contribution to the scalar polarizabilities of the nucleon. Unfortunately, the
experimental data from elastic photon scattering off a deuteron are much more
prone to model errors. Such experiments have been performed at SAL
\cite{Hornidge:1999xs} and at MAX-lab \cite{Lundin:2002jy}, and within the
formalism of \textcite{Levchuk:1999zy} the following results have been
obtained:
\begin{eqnarray}\label{eq:4.1.25}
\alpha_{E1}^n-\beta_{M1}^n = \left \{\begin{array}{ll}
-4.8 \pm 3.9 \;\mbox{\cite{Hornidge:1999xs}} \\
+2.3 \pm 3.4 \; \mbox{\cite{Lundin:2002jy}}
\end{array}\right \}.
\end{eqnarray}
Altogether these numbers speak for a very small value of the backward scalar
polarizability, which is difficult to understand on theoretical grounds. The
quasi-free Compton scattering experiments off a bound neutron have also
provided a first glimpse at the backward spin polarizability of the neutron.
Whereas the large pion-pole contribution is negative for the proton, it carries
a positive sign for the neutron. The dispersive contributions, on the other
hand, are positive for both nucleons. As a result, we expect a large positive
number for $\gamma_{\pi}^n$. This is consistent with the value
\begin{equation} \label{eq:4.1.26}
\gamma_{\pi}^n = 58.6 \pm 4.0 \,,
\end{equation}
obtained from a fit to quasi-free Compton scattering off a bound
neutron \cite{Schumacher:2005an}.
%
\subsection{Generalized polarizability of the nucleon at $Q^2>0$}
\label{sec:IV.2}
\begin{figure}[]
\begin{center}
\includegraphics[width=0.6\columnwidth,angle=0.]{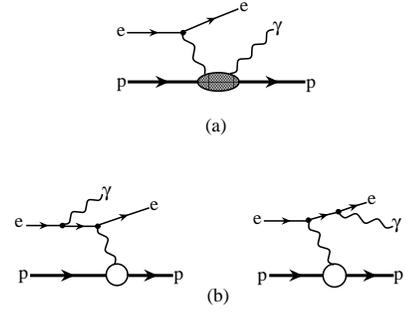}
\end{center}
\caption[]{Contributions to the reaction $ e+p \rightarrow e'+p'+\gamma$, (a):
virtual Compton scattering on the proton, (b): Bethe-Heitler process. The blob
in diagram (a) represents both nucleon intermediate states (Born terms) and
excited states of the nucleon (non-Born terms).\label{fig:VCS_graph}}
\end{figure}

Virtual Compton scattering is formally obtained from real Compton scattering by
replacing the incident real photon with a virtual photon $\gamma^{\ast}$. It is
realized by a subprocess of the reaction $e+p \rightarrow e'+p'+\gamma$. As
displayed in Fig.~\ref{fig:VCS_graph}, the real photon can be emitted by either
the electron or the proton. The former process is called Bethe-Heitler (BH)
scattering that can be calculated from QCD, whereas the latter process is
referred to as virtual Compton scattering (VCS). Because the two processes lead
to the same final state, the amplitudes add coherently,
\begin{equation}\label{eq:4.2.1}
\mathcal {T}^{ee'\gamma}=\mathcal {T}^{\text{BH}}+\mathcal {T}^{\text{VCS}}.
\end{equation}
The VCS amplitude $\cal {T}^{\text{VCS}}$ can be further decomposed into a Born
and a non-Born contribution. For the Born contribution, the nucleon remains
always in its ground state, and therefore this amplitude can be calculated once
the (ground state) form factors of the nucleon are known. The non-Born term
contains all the contributions with excited intermediate states, that is nucleon
resonances, pion-nucleon scattering states, and so on. The physics interest is,
of course, in the non-Born amplitude, because it contains the information on
the nucleon's internal structure in the form of generalized polarizabilities
(GPs). These GPs depend on the virtuality $Q^2$ transferred by the virtual
photon. The physics of VCS is visualized best if we consider the time-reversed
version: As in RCS the real photon plays the role of a quasi-static
electromagnetic field that induces a polarization of the charges, currents, and
magnetizations whose spatial distributions are resolved by the virtual photon
through variation of $Q^2$. To lowest order in the energy, VCS is determined by
6 independent GPs, which can be determined by measuring the interference
between the Bethe-Heitler and VCS amplitudes by means of angular distributions
\cite{Guichon:1995pu} and double-polarization asymmetries
\cite{Vanderhaeghen:1997bx}. A word of caution for the reader familiar with the
formalism of meson electroproduction: the cross section for the reaction $e+N
\rightarrow e'+N'+\gamma$ does not take the form of Eq.~(\ref{eq:2.23}), which
is based on particle production from the nucleon only. Instead, the final-state
photon can be emitted from both the electron and the nucleon.
The pioneering VCS experiment was made at MAMI by \textcite{Roche:2000ng}, and
a first  double-polarization experiment is underway. Further experiments have
recently been performed at MIT/Bates for very small $ Q^2 $
\cite{Bourgeois:2006js} and at JLab for large $ Q^2 $
\cite{Laveissiere:2004nf}. From such experiments we find different spatial
distributions for the diamagnetism and paramagnetism in the nucleon.
Furthermore, the planned double-polarization experiments will give a direct
comparison with the spin polarizabilities predicted by ChPT, which are free of
low-energy constants at the leading order.
\subsubsection{Kinematics and invariant amplitudes}
\label{sec:IV.2.1}
In the following we only consider the subprocess
\begin{equation}\label{eq:4.2.2}
\gamma^{\ast} (q)+ N(p) \rightarrow  \gamma (q') + N(p')\, .
\end{equation}
Because the space-like virtual photon has a ``mass'' $q^2=-Q^2$, the kinematic
relations change with regard to the real photon case, Eq.~(\ref{eq:4.1.1}), in
particular
\begin{eqnarray}\label{eq:4.2.3}
&&s + t + u = 2 M^2 - Q^2 \;,\\
&&\nu = \frac {s - u}{4 M} = E_{\gamma} + \frac { t - Q^2}{4 M} \,,\label{eq:4.2.4}\\
&&t =2 E'_{\gamma} \, (\cos\,\theta_{\text{lab}}\,\sqrt {E_{\gamma}^2 + Q^2}
-E_{\gamma}) - Q^2 \,, \label{eq:4.2.5}
\end{eqnarray}
with $\theta_{\text{lab}}$ the lab scattering angle, and $E'_\gamma$ and
$E_\gamma$ the lab energies of the real and virtual photon, respectively. In
the following we choose $\nu$, $t$, and $Q^2$ as the
independent variables. \\

The VCS Compton tensor is constructed as for RCS, Eq.~(\ref{eq:4.1.2}), except
that the polarization four-vector of the virtual photon has 3 independent
components, that is the helicities $\lambda=\pm 1$ (transverse polarization)
and  $\lambda = 0$ (longitudinal polarization). The VCS tensor $\tilde {H}^{\mu
\nu}$ can be expanded in a basis of 12 independent tensors with amplitudes
depending on 3 variables,
\begin{equation}
\tilde {H}^{\mu \nu} = \sum_{i = 1}^{12} \;\tilde {\mathcal {M}}_i^{\mu \nu} \,
F_i(\nu, t,Q^2)\, . \label{eq:4.2.6}
\end{equation}
The number 12 is given by the possible choices for the helicities in the
initial and final states, namely $3 \times 2 \times 2 \times 2$ divided by two
because of parity invariance. The same consideration yields 8 for RCS, but this
number is further reduced to 6 independent combinations by time-reversal, which
of course does not apply for VCS. It is possible to find a special tensor basis
such that each term is gauge invariant, even under crossing and free of
kinematical singularities and constraints \cite{Drechsel:1997xv}. Furthermore,
only 6 amplitudes contribute for $Q^2 \rightarrow 0$, because 4 tensor
structures and 2 amplitudes vanish in this limit. The result are 6 relations
between the VCS amplitudes $F_i$ and the RCS amplitudes $A_i$.
\subsubsection{Generalized polarizabilities}
\label{sec:IV.2.2}
If the emitted photons have small energies, the VCS experiments can be analyzed
in terms of a low-energy expansion (LEX) as proposed by
\textcite{Guichon:1995pu}. In this approximation the non-Born part of the
amplitudes is expanded in $E'_{\gamma}$, and only the linear term is kept. This
reduces the multipolarities of the emitted photon to electric and magnetic
dipole radiation. Furthermore, the GPs are given by linear combinations of the
amplitudes at threshold ($\nu = 0, t = -Q^2$), which contains the definition of
the polarizabilities for RCS in the limit $Q^2 \rightarrow 0$. Let us next
discuss the multipole decomposition of the non-Born VCS tensor $\tilde {H}^{\mu
\nu}_{\text{nB}}$ of Eq.~(\ref{eq:4.2.6}) at small real photon energy, $q'
\rightarrow 0$, but for arbitrary three-momentum $q \,\equiv\, |\vec
{q}_{\text{cm}}\,| $ of the virtual photon. For this purpose we denote the GPs
by $P^{(\mathcal{M}'\,\mathcal{L}', \, \mathcal{M}
\,\mathcal{L})\,\mathcal{S}}$ \cite{Guichon:1995pu}. In this notation,
$\mathcal{L}$ refers to the angular momentum and $\mathcal{M}$ to the electric
$(E)$, magnetic $(M)$, or longitudinal $(L)$ nature of the virtual photon, with
the primed variables denoting the real photons. Furthermore, the quantum number
$\mathcal{S}$ differentiates between the spin-flip ($\mathcal{S}=1$) and non
spin-flip ($\mathcal{S}=0$) character of the hadronic transition. Within the
LEX we may use the dipole approximation, $\mathcal{L}' = 1$. With this
assumption, the conservation of angular momentum and parity restricts the
number of GPs to 10 \cite{Guichon:1995pu}. Four more constraints are provided
by nucleon crossing combined with charge conjugation symmetry, which leaves 6
independent GPs \cite{Drechsel:1997xv},
\begin{eqnarray}
&& P^{(L1,L1)0}(q), \quad P^{(M1,M1)0}(q)\,, \nonumber \\
&& P^{(L1,L1)1}(q), \quad P^{(M1,M1)1}(q)\,, \label{eq:4.2.7} \\
&& P^{(M1,L2)1}(q), \quad P^{(L1,M2)1}(q)\,.  \nonumber
\end{eqnarray}
We note that the transverse electric multipoles have been eliminated from the
above equations, because they differ from the longitudinal multipoles only by
terms of higher order in $q$. In the limit $q\rightarrow 0$ one finds the
following relations between the VCS and RCS polarizabilities
\cite{Drechsel:1997xv}~:
\begin{eqnarray}
P^{(L1,L1)0} & \rightarrow &  -\frac{\sqrt{2}}{\sqrt{3}\,\alpha_\text{em}} \; \alpha_{E1}\, ,\nonumber \\
P^{(M1,M1)0} & \rightarrow &  -\frac{\sqrt{8}}{\sqrt{3}\,\alpha_\text{em}} \; \beta _{M1} \, , \nonumber \\
P^{(L1,L1)1} & \rightarrow &  0 \, , \; \;  P^{(M1,M1)1} \rightarrow  0 \, , \label{eq:4.2.8} \\
P^{(L1,M2)1} & \rightarrow &  -\frac{\sqrt{2}}{3\alpha_\text{em}} \; \gamma_3  \, , \nonumber \\
P^{(M1,L2)1} & \rightarrow &  -\frac{2 \sqrt{2}}{3 \sqrt{3}\,\alpha_\text{em}}
\; (\gamma_2 + \gamma_4)  \, , \nonumber
\end{eqnarray}
In order to connect the scalar VCS and RCS polarizabilities, we introduce the
definitions
\begin{eqnarray}
\alpha_{E1}(Q^2)&=& -{\sqrt{\frac{3}{2}}}\,\alpha_\text{em} \; P^{(L1 \,, L1) \,0}\,(Q^2)\,,\label{eq:4.2.9} \\
\beta_{M1}(Q^2)&=& -{\sqrt{\frac{3}{8}}}\,\alpha_\text{em}\; P^{(M1 \,,
M1)\,0}\,(Q^2) \,,\label{eq:4.2.10}
\end{eqnarray}
with $\alpha_{E1}(0)=\alpha_{E1}$ and $\beta_{M1}(0)=\beta_{M1}$ as measured by
RCS according to Eqs.~(\ref{eq:4.1.20}) and (\ref{eq:4.1.21}).

\subsubsection{Theoretical developments}
\label{sec:IV.2.3}
For the given tensor basis, the associated non-Born VCS amplitudes
$F_i^{\text{nB}}$, ($i$ = 1,...,12) are free of kinematical singularities and
constraints, and even under crossing. Assuming further an appropriate analytic
and high-energy behavior, these amplitudes fulfill unsubtracted dispersion
relations in the variable $\nu$ and at fixed $t$ and $Q^2$,
\begin{eqnarray}
\mathrm{Re} F_i^{\text{nB}}(Q^2, \nu, t) = F_i^{\text{pole}}(Q^2,
\nu, t)- F_i^{\text{B}}(Q^2, \nu, t)\nonumber \\
+\frac {2} {\pi} \, \mathcal {P} \int_{\nu_{0}}^{\infty} d\nu' \, \frac{\nu' \,
\mathrm{Im} F_i(Q^2, \nu',t)} {\nu'^2 - \nu^2}\, . \label{eq:4.2.11}
\end{eqnarray}
We recall that the Born amplitudes $F_i^{\text{B}}$ are given by diagrams with
nucleons in the intermediate state, whereas the pole amplitudes
$F_i^{\text{pole}}$ are obtained from the Born amplitudes at the pole position,
that is, with all numerators evaluated at the pole. Furthermore, Im~$F_i$ are
the discontinuities across the $s$-channel cuts, starting at the
pion production threshold $\nu_{0} = m_\pi + (2 m_{\pi}^2 + t + Q^2)/(4 M)$.\\

Besides the absorptive singularities due to physical intermediate states, one
might wonder whether additional singularities like anomalous thresholds can
contribute to the dispersion integrals. The latter arise when a hadron is a
loosely bound system of other hadronic constituents which can go on-shell, thus
leading to so-called triangular singularities. However, it was shown that
within the strong confinement of QCD, the quark-gluon structure of hadrons does
not give rise to additional anomalous thresholds
\cite{Jaffe:1991ib,Oehme:1995xe}, and that possible quark singularities turn
into hadron singularities as described through an effective field theory.
Therefore, the only anomalous thresholds arise for hadrons which are loosely
bound systems of other hadrons, as for example the $\Sigma$ particle in terms
of a $\Lambda$-$\pi$ system. Such anomalous thresholds are absent for the
nucleon, and therefore the imaginary parts in Eq.~(\ref{eq:4.2.11}) are only
given by absorptive effects due to $\pi N$, $\pi \pi N$, and heavier hadronic
states. Of course, Eq.~(\ref{eq:4.2.11}) is only valid if the amplitudes drop
fast enough such that the integrals converge. The high-energy behavior of the
amplitudes $F_i$ was investigated by \textcite{Pasquini:2001yy} in the Regge
limit ($\nu \rightarrow \infty$, $t$ and $Q^2$ fixed). As in
subsection~\ref{sec:IV.1.2}, the dispersion integrals diverge for two
amplitudes, $F_1$ and $F_5$ in our notation. These amplitudes are dominated by
the $t$-channel exchange of $\sigma$ and $\pi^0$ mesons, respectively. As long
as we are interested in the energy region up to the $\Delta(1232)$, we may
saturate the $s$-channel dispersion integral by the $\pi N$ contribution,
choosing $\nu_{\text{max}} \approx 1.5$~GeV as upper limit of integration. The
asymptotic contribution to $F_5$ is saturated by the pion pole and therefore
independent of $\nu$,
\begin{equation}\label{eq:4.2.12}
F^{\text{as}}_{5}(\nu, t, Q^2) = - \frac { g_{\pi NN}} {M e^2} \,
\frac{F_{\pi^0 \gamma \gamma}(Q^2)} {t - m_{\pi}^2} \, ,
\end{equation}
with a monopole form factor from the $\pi^0 \rightarrow \gamma +\gamma$ decay.
As a result $\alpha^{\text{as}}_{E1} (Q^2) \sim F^{\text{as}}_5(0, -Q^2, Q^2)$
has a dipole form, with $\alpha^{\text{as}}_{E1}(0)$ known from RCS. Although
the pion pole contribution is certainly dominant, there may be other effects
such as more-pion and heavier intermediate states. In view of our still limited
knowledge on these reactions, we simply parameterize the $Q^2$ dependence in a
dipole form with a parameter $\Lambda_{\alpha}$,
\begin{equation}\label{eq:4.2.14}
\alpha^{\text{as}}_{E1} (Q^2)=
\frac{\alpha^{\text{as}}_{E1}(0)}{(1+Q^2/\Lambda_{\alpha}^2)^2}\,.
\end{equation}
In the same spirit we also estimate the contribution of the $\sigma$ meson by a
dispersion relation in $t$ at $\nu=0$,
\begin{equation}\label{eq:4.2.15}
\bar F_1^{\text{as}} (Q^2)\,=\, \frac{1}{\pi}\int^\infty_{4m_{\pi}^2} dt'\,
\frac{\mathrm{Im}_t F_1 (0, t', Q^2)}{t'+Q^2}\,,
\end{equation}
with $\mathrm{Im}_t F_1$  as determined from the $t$-channel reaction $\gamma +
\gamma^{\ast} \rightarrow \pi + \pi \rightarrow N \bar {N}$. The result of this
calculation can also be parameterized by a dipole form \cite{Pasquini:2001yy}:
\begin{equation}\label{eq:4.2.16}
\beta^{\text{as}}_{M1} (Q^2)\approx
\frac{\beta^{\text{as}}_{M1}(0)}{(1+Q^2/\Lambda_{\beta}^2)^2}\,,
\end{equation}
with $\Lambda_{\beta} \approx 0.4$~GeV and $\beta^{\text{as}}(0)$ known from
RCS. We note that $\Lambda_{\beta}$ is small compared to the parameter
$\Lambda_D = 0.84$~GeV of Eq.~(\ref{eq:3.1}), which gives the scale of the
nucleon's magnetic form factor, that is, the asymptotic diamagnetic
polarization is related to surface phenomena as expected from the pion cloud.
\subsubsection{Experiments and data analysis}
\label{sec:IV.2.4}
At small 3-momentum  $q'$ of the emitted real photon, the measured cross
section can be analyzed through the LEX of \textcite{Guichon:1995pu}. This
expansion is based on a low-energy theorem (LET) stating that the radiative
amplitude for point-like particles diverges like $1/q'$ for $q' \rightarrow 0$,
whereas the dispersive amplitude vanishes like $q'$ in that limit.  As a
consequence the spin-averaged (unpolarized) square of the matrix element takes
the form
\begin{equation}\label{eq:4.2.17}
|\mathcal {M}|^2 = \frac{\mathcal {A}_{-2}}{(q')^2} +\frac{\mathcal
{A}_{-1}}{q'} +\mathcal {A}_0 +\mathcal {O}(q') \,,
\end{equation}
with coefficients $\mathcal {A}_{-2}$ and $\mathcal {A}_{-1}$ that are fully
described by the Bethe-Heitler (BH) and Born terms, which can be calculated
from QED once the proton (ground state) form factors are known. The next order
term $\mathcal {A}_0$ contains contributions from the BH and Born terms but
also an interference between the $\mathcal {O}(1/q')$ contribution of BH plus
Born amplitudes and the leading term of the non-Born amplitude $\mathcal
{O}(q')$, which is proportional to the GPs. This interference term can be
expressed by the structure functions $P_{LL}(q)$, $P_{TT}(q)$, and $P_{LT}(q)$
 \cite{Guichon:1995pu},
\begin{multline}\label{eq:4.2.18}
\mathcal {A}_0^{\text{exp}} - \mathcal {A}_0^{\text{BH+B}}  =  2 K_2
\bigg \{ v_1 \left[\varepsilon P_{LL}(q) - P_{TT}(q)\right] \\
 +  (v_2-\frac{\tilde
{q}_0}{q}v_3)\sqrt {2\varepsilon (1+\varepsilon)} P_{LT}(q) \bigg \} \,,
\end{multline}
with $K_2$, $v_1$ ,$v_2$, and $v_3$ kinematical functions depending on
$\varepsilon$, $q$, and the polar and azimuthal cm angles, $\theta_{\text{cm}}$
and $\phi_{\text{cm}}$, respectively. Furthermore, $\tilde {q}_0$ is the cm
energy of the virtual photon in the limit $q' \rightarrow 0$. The 3 structure
functions of Eq.~(\ref{eq:4.2.18}) can be expressed by the GPs as follows
\cite{Guichon:1995pu,Guichon:1998xv}:
\begin{eqnarray}
P_{LL} &=& - 2\sqrt{6} \, M  G_E  P^{( {L1,L1} )0} \,,
\label{eq:4.2.19} \\
P_{TT}&=& - 3  G_M \frac{q^2}{\tilde {q}_0} \left( P^{(M1,M1)1}-\sqrt{2}\,
\tilde {q}_0  P^{(L1,M2)1} \right) \,,
\nonumber \\
P_{LT}&=& \sqrt{\frac{3}{2}}\, \frac{M q}{Q} G_E  P^{(M1,M1)0}+\frac{3}{2}
\frac{Q q}{\tilde {q}_0}G_M P^{(L1,L1)1}\,, \nonumber
\end{eqnarray}
with $G_E$ and $G_M$ the electric and magnetic nucleon
form factors.\\

\begin{figure}[]
\begin{center}
\includegraphics[width=1.0\columnwidth,angle=0.]{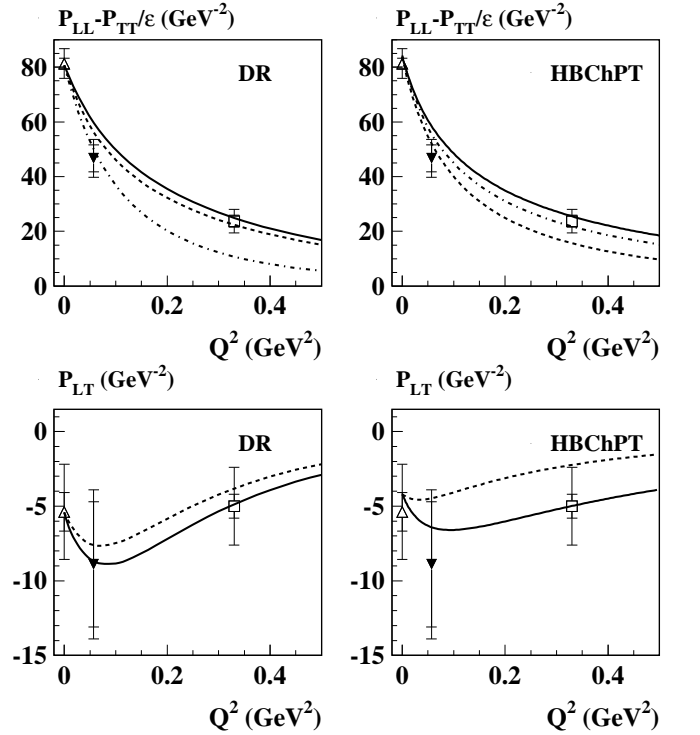}
\end{center}
\caption{Comparison between the unpolarized VCS structure functions calculated
by dispersion relations \cite{Pasquini:2001yy} (left column) and HBChPT
\cite{Hemmert:1996gr,Hemmert:1999pz} at $\mathcal {O}(p^3)$ (right column).
Upper row: Result for $P_{LL} - P_{TT}/\varepsilon$ with $\varepsilon=0.62$
(solid lines) and $\varepsilon=0.9$ (dashed-dotted lines) compared to the
result for $\alpha_{E1}$ only and $\varepsilon=0.62$ (dashed lines). The
dispersive results for $\varepsilon=0.62$ and $\varepsilon=0.9$ are obtained
with $\Lambda_{\alpha}=1.79$~GeV and $\Lambda_{\alpha}=0.7$~GeV, respectively.
Lower row: Results for $P_{LT}$ (solid line) compared to the result for
$\beta_{M1}$ only (dashed line). The data are from
\textcite{OlmosdeLeon:2001zn} (open triangles), \textcite{Bourgeois:2006js}
(full triangles), and \textcite{Roche:2000ng}
(squares). Figure from \textcite{Drechsel:2002ar} updated by B.~Pasquini.}
\label{fig:P_LL_TT_LT_low}
\end{figure}

In Fig.~\ref{fig:P_LL_TT_LT_low} we compare the measured response functions to
the predictions of DR (left column) and HBChPT (right column). The response
function $P_{LL} - P_{TT}/\varepsilon$ is displayed in the upper panels of
this figure. According to Eqs.~(\ref{eq:4.2.9}) and (\ref{eq:4.2.19}), 
$P_{LL}$ is directly proportional to the scalar GP $\alpha_{E1}(Q^2)$, whereas
$P_{TT}$ contains only spin GPs. As discussed in subsection~\ref{sec:IV.1},
the dispersive and asymptotic contributions to $\alpha_{E1}$ have the same
sign at the real photon point, which leads to a large total value. However,
$\alpha_{E1}(Q^2)$ drops rapidly as function of $Q^2$. The difference between
the solid and the dashed lines is due to the spin GPs whose importance rises
with $Q^2$. There is a general agreement between the results from DR
\cite{Pasquini:2001yy} and the HBChPT \cite{Hemmert:1997at,Hemmert:1999pz},
however the spin GPs turn out much larger in the latter approach. The lower 
row of Fig.~\ref{fig:P_LL_TT_LT_low} gives the same comparison for the
response function $P_{LT}$, which contains both the scalar magnetic
polarizability $P^{(M 1, M 1)0} \sim -\beta_{M1}(Q^2)$ and the spin GP 
$P^{(L 1, L 1)1}$. As shown in subsection~\ref{sec:IV.1}, $\beta_{M1}(0)$ 
is the sum of a large dispersive (paramagnetic) contribution, which is 
dominated by $\Delta(1232)$ excitation, and a somewhat smaller asymptotic 
(diamagnetic) contribution with opposite sign. Moreover, we expect that the 
diamagnetic contribution is largely due to pionic degrees of freedom, and 
therefore of longer range in r-space than the paramagnetic component. This
expectation is corroborated by the minimum of $P_{LT}$ at $Q^2\approx
0.05$~GeV$^2$: the (positive) diamagnetic component of $P_{LT}$ decreases
faster in $Q^2$-space than the (negative) paramagnetic term, and therefore
$P_{LT}$ decreases over a small $Q^2$ region to the minimum, from whereon the
form factor effects lead to a rapid approach towards zero. Although the full
results of DR and HBChPT agree qualitatively, there is again a large
difference in the spin-dependent sector.\\

\begin{figure}[]
\begin{center}
\includegraphics[width=0.8\columnwidth,angle=0.]{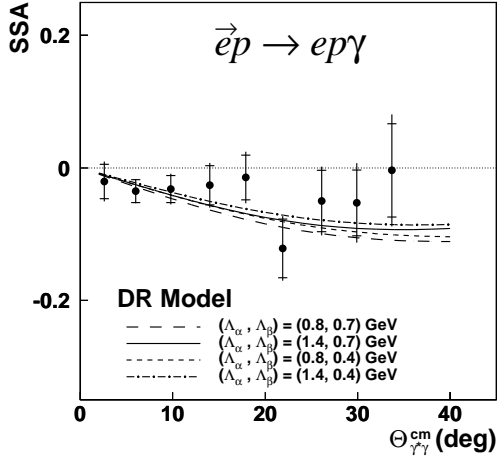}
\end{center}
\caption{Beam single spin asymmetry (SSA) for VCS as function of the photon
scattering angle. The dispersive predictions are shown for different pairs of
($\Lambda_\alpha$, $\Lambda_\beta$) given in GeV. Solid line: (1.4, 0.7),
dashed-dotted line: (1.4, 0.4), long-dashed line: (0.8, 0.7), short-dashed
line: (0.8, 0.4). Figure from \textcite{Bensafa:2006wr}.}
\label{fig:VCS_SSA}
\end{figure}

In the region between pion threshold and $\Delta$-resonance peak, the
sensitivity to the GPs is much enhanced, because the contributions of the GPs
interfere with the rapidly rising amplitude of the $\Delta$-resonance
excitation. It is of course not possible to extend the LEX to these energies,
but the dispersive approach is expected to give a reasonable frame to extract
the GPs. When crossing the pion threshold, the VCS amplitude also acquires an
imaginary part due to the opening of the $\pi N$ channel. As an interesting
result, single polarization observables appear above pion threshold. A
particularly relevant observable is the electron single-spin asymmetry (SSA),
which is obtained by flipping the electron beam helicity \cite{Guichon:1998xv}.
The main source of the SSA is an interference between the (real) Bethe-Heitler
and Born amplitudes and the imaginary part of the VCS amplitude. Because the
SSA vanishes in-plane, its measurement requires an out-of-plane experiment.
Such an experiment has recently been performed at MAMI \cite{Bensafa:2006wr}.
The measured asymmetry at $W$=1.19~GeV and $Q^2=0.35$~GeV$^2$ is displayed in
Fig.~\ref{fig:VCS_SSA} and compared to the predictions of dispersion theory.
The figure shows a rather weak dependence of the asymmetry on variations of the
GPs. Therefore, a measurement of the SSA provides an excellent cross-check of
the dispersive input, i.e., the imaginary parts of the $\pi~N$ multipoles, in
particular by studies of the $\Delta$ region by VCS and pion electroproduction
in parallel.\\

\begin{figure}[]
\begin{center}
\includegraphics[width=1.0\columnwidth,angle=0.]{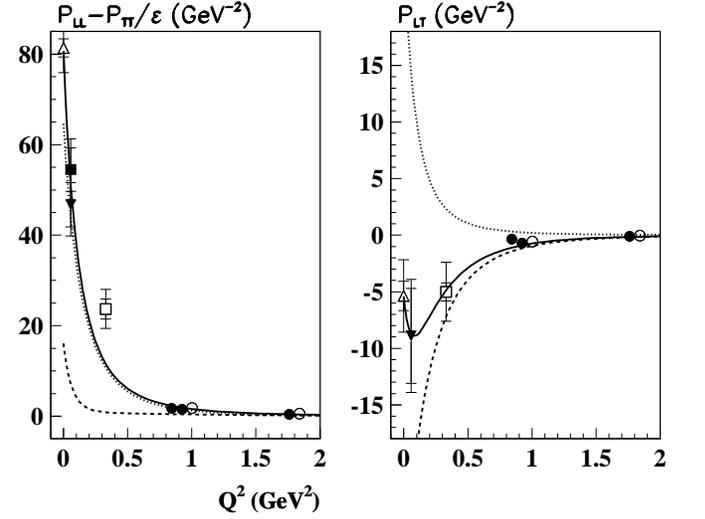}
\end{center}
\caption{Results for $P_{LL}-P_{TT}/\varepsilon$ (left panel) and $P_{LT}$ 
(right panel). Dashed lines: dispersive $\pi N$ contributions. Dotted lines: 
asymptotic contributions with $\Lambda_\alpha = 0.71$~GeV and $\Lambda_\beta = 
0.51$~GeV. Solid lines: full results. The data are from 
\textcite{OlmosdeLeon:2001zn} (open triangles), \textcite{Bourgeois:2006js} 
(full triangles), \textcite{Roche:2000ng} (squares), and 
\textcite{Laveissiere:2004nf} as obtained by the LEX (open circles) and 
DR (full circles) analysis. The inner error bars describe the statistical
error, the outer error bars include systematical errors. Figure from
\textcite{Drechsel:2002ar}, updated by B.~Pasquini.}
\label{fig:LL_TT_LT_hi}
\end{figure}

\begin{figure}[]
\begin{center}
\includegraphics[width=0.95\columnwidth,angle=0.]{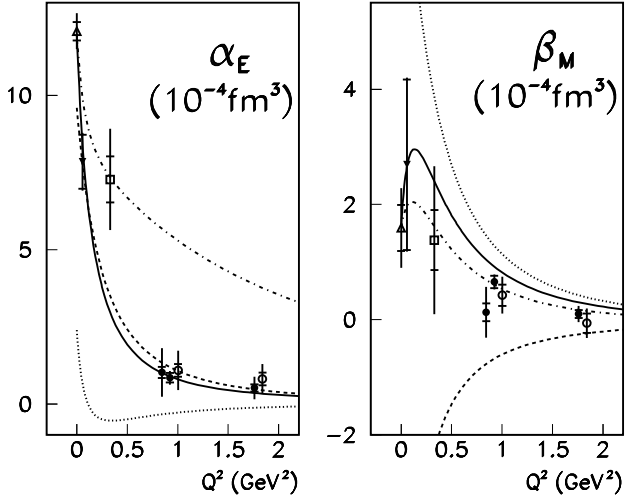}
\end{center}
\caption{Left panel: the electric GP $\alpha_{E1}(Q^2)$ obtained from the DR
formalism with $\Lambda_\alpha = 0.7$~GeV (solid line) and $\Lambda_\alpha =
1.79$~GeV (dashed-dotted line). The solid line is the sum of the asymptotic
(dashed line) and dispersive (dotted) contributions. Right panel: the magnetic
GP $\beta_{M1}(Q^2)$  obtained from the DR formalism with $\Lambda_\beta =
0.51$~GeV (solid line) and $\Lambda_\beta = 0.63$~GeV (dashed-dotted line). The
dashed-dotted line is the sum of the asymptotic (dashed line) and dispersive
(dotted line) contributions. The data are from \textcite{OlmosdeLeon:2001zn}
(open triangles), \textcite{Bourgeois:2006js} (full triangles),
\textcite{Roche:2000ng} (squares), and \textcite{Laveissiere:2004nf} as
obtained by the LEX (open circles) and DR (full circles) analysis. The inner
error bars describe the statistical error, the outer error bars also include
systematical errors. Figure from \textcite{Laveissiere:2004nf}.}
\label{fig:VCS_alpha-beta}
\end{figure}

At larger virtuality, the VCS process has been investigated by the Hall A
Collaboration at JLab, and data have been obtained at $Q^2 = 0.92$~GeV$^2$ and
$Q^2 = 1.76$~GeV$^2$ \cite{Laveissiere:2004nf}. A reasonable description of
these data is obtained by the values $\Lambda_\alpha =0.71$~GeV and
$\Lambda_\beta =0.51$~GeV shown by the solid lines in
Fig.~\ref{fig:LL_TT_LT_hi}. We note that $\Lambda_{\alpha} $ corresponds to a
spatial range comparable to the nucleon's charge distribution. However, the
best fit value for $\Lambda_{\beta} $ is substantially lower, indicating that
the diamagnetism is related to pion cloud effects at distances above 1~fm.
Subtracting the spin-dependent terms according to the dispersion predictions,
we obtain the $Q^2$ dependence of the scalar GPs shown in
Fig.~\ref{fig:VCS_alpha-beta}. It is obvious that the electric GP
$\alpha_{E1}(Q^2)$ is dominated by the asymptotic term, which however can not
be described by a single dipole form over the full $Q^2$ range. The magnetic GP
$\beta_{M1}(Q^2)$ clearly shows a characteristic maximum at $Q^2~\approx
0.05$~GeV$^2$, which comes about by cancelation between the positive
paramagnetic $\Delta$ contribution and the negative diamagnetic contribution of
the $t$-channel $\pi\pi$ exchange. By Fourier transforming the GPs
$\alpha_{E1}(Q^2)$  and $\beta_{M1}(Q^2)$ in the Breit frame, one obtains the
spatial distribution of the induced electric polarization and magnetization of
the nucleon \cite{L'vov:2001fz}. The emerging picture is as expected from a
classical interpretation of diamagnetism. Due to the external magnetic field,
pionic currents start circulating in the nucleon and give rise to an induced
magnetization opposite to the applied field. At distances $r \gtrsim 1/m_\pi$,
the diamagnetic effect dominates and the Fourier transform $\beta_{M1}(r)$
takes negative values, whereas the paramagnetic contributions prevail at the
smaller distances giving rise to positive values of $\beta_{M1}(r)$ in the
interior of the nucleon. As the momentum transfer $Q^2$ increases, the negative
contribution due to the long-range pion cloud vanishes fast and hence
$\beta_{M1}(Q^2)$ increases. This nicely explains the positive slope of
$\beta_{M1}(Q^2)$ at $Q^2=0$ and the maximum at $Q^2 \approx 0.05~\text{GeV}^2$
as indicated by the experimental data.\\

According to Eqs.~(\ref{eq:4.2.18}) and (\ref{eq:4.2.19}), the unpolarized VCS
experiment gives access to only 3 combinations of the 6 GPs. As was shown by
\textcite{Vanderhaeghen:1997bx}, it takes experiments with polarized lepton
beams and polarized targets or recoil nucleons to measure the remaining 3 GPs.
These double-polarization observables require measuring the cross sections for
a definite electron helicity $h$ and recoil (or target) proton spin orientation
parallel and opposite to a specified axis. As shown for the unpolarized cross
section by Eq.~(\ref{eq:4.2.17}), also the polarized squared amplitude has a
low-energy expansion, and again the GPs  are obtained from the term $\mathcal
{O}~(1)$. This term contains the structure functions $P_{LT}^z(q)$,
$P_{LT}^{'z}(q)$, and $P_{LT}^{'\perp}(q)$, which are related to the spin GPs
by~\cite{Vanderhaeghen:1997bx}
\begin{eqnarray}
P_{LT}^z &=& \frac{3 Q q}{2 \tilde {q}_0} G_M P^{(L1,L1)1}
-\frac{3 M q}{Q} G_E P^{(M1,M1)1}\,,\label{eq:4.2.21}\\
P_{LT}^{'z} &=& -\frac{3}{2}\, Q G_M P^{(L1,L1)1}
+ \frac{3 Mq^2}{Q \tilde {q}_0} \, G_E P^{(M1,M1)1}\,,\nonumber\\
P_{LT}^{'\perp} &=& \frac{3 q Q}{2 \tilde {q}_0} \,G_M \left(P^{(L1,L1)1}-
\sqrt{\frac{3}{2}} \, \tilde {q}_0 P^{(M1,L2)1}\right).\nonumber
\end{eqnarray}
While $P_{LT}^z$ and $P_{LT}^{'z}$ can be accessed by in-plane kinematics,
$P_{LT}^{'\perp}$ requires an out-of-plane measurement.\\

\begin{figure}[]
\begin{center}
\includegraphics[width=0.95\columnwidth,angle=0.]{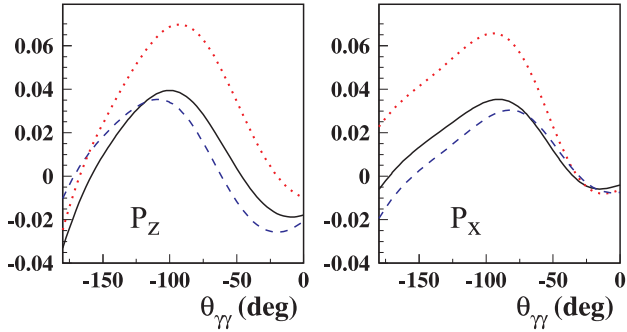}
\end{center}
\caption{Predicted double-polarization asymmetries as function of the
photon scattering angle $\theta_{\gamma \gamma}$ and for the following
fixed kinematic values: $q=600~{\text {MeV}}$, $q'=111.5~{\text {MeV}}$,
$\varepsilon = 0.62$, and $\Phi = 0^{\circ}$. In order to highlight the
model dependence, the (known) Bethe-Heitler and Born contributions to
the asymmetry have been subtracted. Solid lines: Results of dispersion
relations for $\Lambda_\alpha$ = 1~GeV and $\Lambda_\beta$ = 0.6~GeV
\cite{Drechsel:2002ar}, dotted lines: predictions of HBChPT at $\mathcal
{O}(p^3)$ \cite{Hemmert:1999pz}, and dashed lines: HBChPT
at $\mathcal {O}(p^4)$ \cite{Kao:2004us}. See text for further
explanations.}
\label{fig:VCS_double-pol}
\end{figure}
\begin{figure}[]
\begin{center}
\includegraphics[width=1.0\columnwidth,angle=0.]{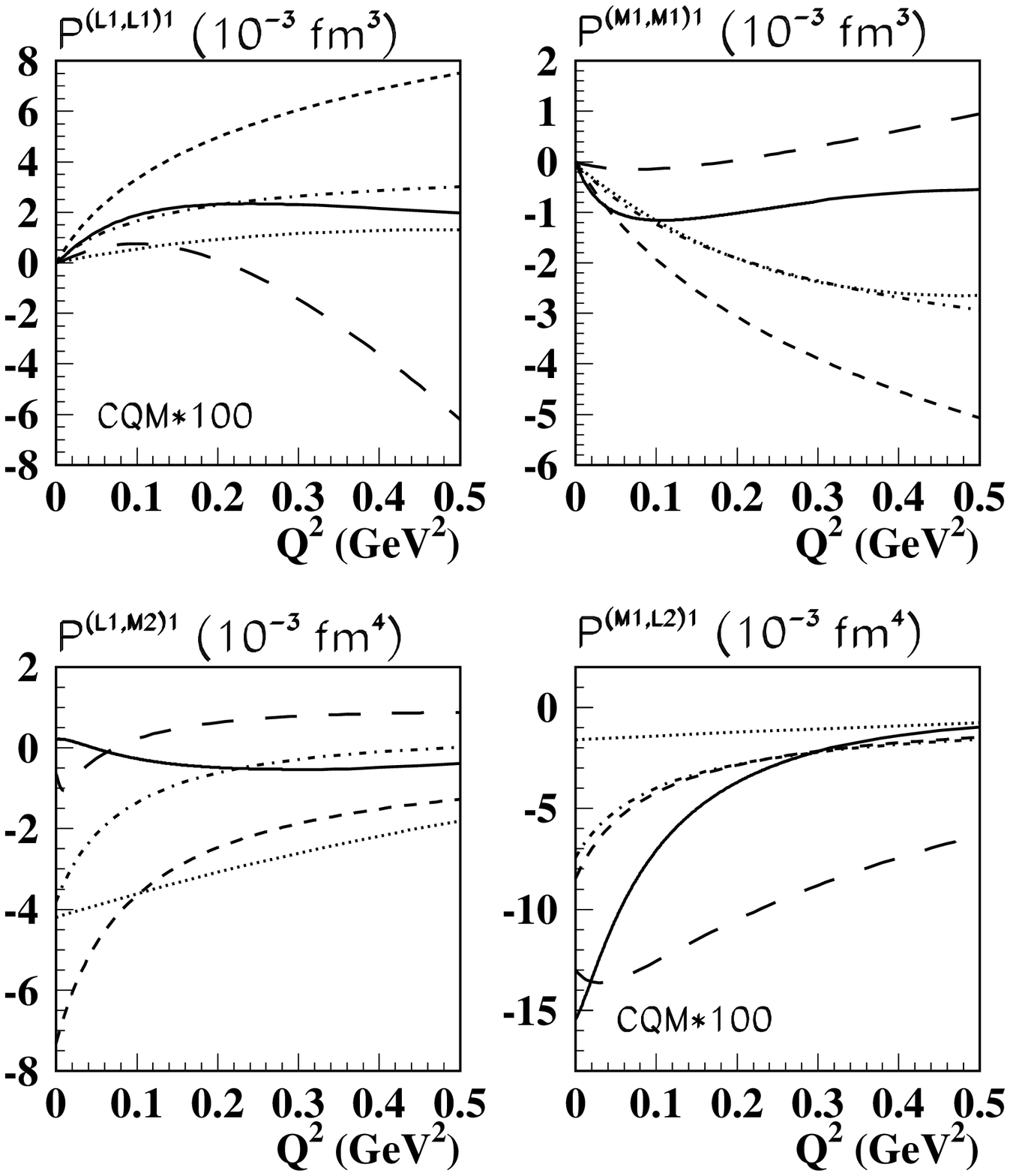}
\end{center}
\caption{The spin-flip GPs (without the $\pi^0$-pole contribution) from several
calculations. Solid lines: dispersive $\pi N$ contribution
\cite{Pasquini:2001yy}, short-dashed lines: ${\mathcal O}(p^3)$ HBChPT
\cite{Hemmert:1999pz}, long-dashed lines: ${\mathcal O}(p^4)$ HBChPT
\cite{Kao:2004us}, dashed-dotted lines: linear $\sigma$ model
\cite{Metz:1996fn}, dotted lines: non-relativistic CQM \cite{Pasquini:2000ue}.
For visibility, the tiny CQM results for $P^{(L 1,L 1)1}$ and $P^{(M 1,L 2)1}$
are multiplied by a factor 100. Figure from \textcite{Drechsel:2002ar} updated
by B.~Pasquini.} 
\label{fig:VCS_gamma_i}
\end{figure}

In Fig.~\ref{fig:VCS_double-pol} we compare the results of DR and HBChPT for
the double-polarization observables, with longitudinally polarized electrons
and recoil proton polarization either along the virtual photon direction
($z$-direction) or in the reaction plane and perpendicular to the virtual
photon ($x$-direction). The large but well-known asymmetries from the
Bethe-Heitler and Born terms have been subtracted in this figure in order to
highlight the differences between DR \cite{Pasquini:2001yy} and HBChPT at
${\mathcal O}(p^3)$ \cite{Hemmert:1999pz}. We note that the latter approach
yields significantly larger effects due to higher predicted values for the spin
GPs. Although these double polarization observables are tough to measure, a
first test experiment is underway at MAMI. Contrary to the scalar
polarizabilities, the spin-flip GPs are still unknown territory. In
Fig.~\ref{fig:VCS_gamma_i} we compare the dispersive results for the spin-flip
GPs with the predictions of the non-relativistic constituent quark model
\cite{Pasquini:2000ue}, the HBChPT to ${\mathcal O}(p^3)$
\cite{Hemmert:1999pz,Hemmert:1997at} and ${\mathcal O}(p^4)$ \cite{Kao:2002cn},
and the linear $\sigma$-model \cite{Metz:1997fr}. We refrain from commenting on
the theoretical predictions which clearly open a wide range of values for the
spin polarizabilities. An absolute must for further progress are dedicated
experiments with a large sensitivity to the spin-dependent GPs. Such
experiments are (I) unpolarized VCS with variation of the transverse photon
polarization $\epsilon$ in order to separate the response functions $P_{LL}$
and $P_{TT}$ and (II) double-polarization experiments as discussed above.
%
\subsection{Polarizability of mesons}
\label{sec:IV.3}
Mesons are systems of a quark and an anti-quark and, therefore, theoretically
simpler to describe than baryons. Let us set the scene with the classical
picture of two charges bound in a quark-antiquark potential in the presence of
an additional static electric field. For this system we can derive the
following relation \cite{Walcher:2006ju}:
\begin{equation}
\alpha^{\pi^+}_{E1} = \frac{\alpha_{em}}{\alpha_{q{\bar{q}}}} 4 \pi R^3 \zeta^2
\,, \label{eq:4.3.1}
\end{equation}
with $\zeta = 1/6$ the effective charge of the system and $\alpha_{q\bar{q}}
\approx 5$ derived from the heavy quark potential, of course somewhat outside
its applicability. Furthermore, $R$ is a characteristic dimension of the
system, for example the equivalent charge radius. With the pion rms radius of
Eq.~(\ref{eq:3.16}), we obtain $R =\sqrt{(5/3) \langle r^2 \rangle_{\pi}} =
0.86$~fm, and as a result $\alpha^{\pi^+}_{E1} = 3.2$, here and in the
following in units of $10^{-4}~\text{fm}^3$. Comparing these numbers with the
results of subsection~\ref{sec:IV.1}, we find that the pion is a ``dielectric
medium'' with $\varepsilon \approx 1.001$, i.e., even more rigid than the
nucleon.\\

The pion polarizabilities have been calculated in ChPT at the two-loop order,
${\mathcal{O}}(p^6).$ Contrary to the situation of the nucleon, no ``matter
fields'' with their own mass scale are present, and therefore the calculations
can be performed in the original formulation of ChPT
\cite{Gasser:1983yg,Gasser:1984gg}. This makes the following predictions for
the polarizabilities a very significant test of this theory
\cite{Gasser:2006qa}:
\begin{eqnarray}
\alpha^{\pi^+}_{E1} + \beta^{\pi^+}_{M1} &=& 0.16 \pm 0.1 \,,\label{eq:4.3.2}\\
\alpha^{\pi^+}_{E1} - \beta^{\pi^+}_{M1} &=& 5.7 \pm 1.0\,.\label{eq:4.3.3}
\end{eqnarray}
The very small value predicted by Eq.~(\ref{eq:4.3.2}), that is Baldin's sum
rule applied to the pion, makes a measurement of this observable close to
impossible. The experiments are therefore analyzed
with the constraint $\alpha_{E1}^{\pi^+} \approx -\beta_{M1}^{\pi^+}$.\\

Unfortunately, the experimental situation is rather contradictory, see
\textcite{Ahrens:2004mg} and \textcite{Gasser:2006qa} for recent reviews of the
data and further references to the experiments. There exist basically three
different methods to measure $\alpha_{E1}$: (I) the reactions $e^+e^-
\rightleftarrows \gamma \gamma \rightleftarrows \pi^+\pi^-$, (II) the Primakov
effect of scattering a relativistic pion in the Coulomb field of a heavy
nucleus, and (III) the radiative pion photoproduction, $p(\gamma, \gamma' \pi^+
n)$, which contains Compton scattering on a (bound) pion as a subprocess. The
latter reaction was recently investigated at the Mainz Microtron MAMI, by use
of a kinematically optimized set-up consisting of the backward photon detector
TAPS, a forward $\pi^+$ detector, and a neutron detector realized by a large
scintillator wall of dimensions 3x3x0.5~m$^3$. The largest error of this
measurement is due to the systematic error of the neutron efficiency. The final
result of the experiment is \cite{Ahrens:2004mg}
\begin{equation}
\alpha^{\pi^+}_{E1} - \beta^{\pi^+}_{M1} = 11.6 \pm 1.5_{\text{stat}} \pm
3.0_{\text{syst}} \pm 0.5_{\text{mod}}\,, \label{eq:4.3.4}
\end{equation}
which is at variance with the prediction of \textcite{Gasser:2006qa} by two
standard deviations. In view of the theoretical uncertainties from the fact
that the Compton scattering is off a bound pion, the deviation from theory is
an open problem. In particular we point out that the model error in
Eq.~(\ref{eq:4.3.4}) is estimated by comparing the analysis with 2 specific
models. This does not exclude that a wider range of models will lead to a
larger model error. Considering the fact that the scattering is off a
``constituent'' pion in the nucleon, we may attribute the deviation to binding
effects. For example, as suggested by Eq.~(\ref{eq:4.3.1}), an increase of the
bound pion radius by 20~\% would give a hand-waving explanation for the
experimental data. Because the pion polarizability is extremely important for
our understanding of QCD in the confinement region, it is prerequisite to check
the given arguments by a full-fledged ChPT calculation of the reaction
$p(\gamma,\gamma' \pi^+n)$ .\\

The second method to determine the polarizability, the Primakov effect, has
been studied at Serpukhov with the result
\begin{equation}
\alpha^{\pi^+}_{E1} - \beta^{\pi^+}_{M1} = 13.6 \pm 2.8_{\text{stat}} \pm
2.4_{\text{syst}}\,, \label{eq:_4.3.5}
\end{equation}
in agreement with the value from MAMI. Recently, also the COMPASS Collaboration
at CERN has investigated this reaction. However, at this time the analysis is
still in a too preliminary stage to include the result. Unfortunately, the
reactions $e^+e^- \rightleftarrows \gamma \gamma \rightleftarrows \pi^+\pi^-$
have led to even more contradictory results in the range $4.4 \leq
\alpha^{\pi^+}_{E1}\leq 52.6$, as listed in the work of
\textcite{Gasser:2006qa}. In conclusion one has to wait for an improved
analysis and possibly also independent experimental efforts before final
conclusions can be drawn.


\section{Excitation spectrum of the Nucleon}
\label{sec:V}
\subsection{Threshold production of mesons}
\label{sec:V.1}
As outlined in subsection~\ref{sec:II.4}, the threshold photoproduction of
mesons provides a significant test of our theoretical understanding, because
only few partial waves contribute and, therefore, all relevant multipoles can
be directly determined by the experiment. The case of neutral pion
photoproduction on the proton, $\gamma(q) + p(p_1) \rightarrow \pi^0(k) +
p(p_2)$ is of particular interest. For the s-wave threshold multipole $E_{0+}$
of this reaction, several authors had derived a low energy theorem (LET) based
on current algebra and PCAC \cite{DeBaenst:1971hp,Vainshtein:1972ih}. According
to the theorem, the leading terms of the threshold multipole were directly
determined by the Born diagrams, evaluated with the pseudovector pion-nucleon
interaction. However, this prediction had to be revised in the light of
surprising experimental evidence. The reason for the discrepancy between the
theorem and the data was first explained in the framework of ChPT by pion-loop
corrections. An expansion in the mass ratio $\mu=m_{\pi}/M\approx1/7$ yielded
the result \cite{Bernard:1991rt}
\begin{equation}
E_{0+} (\pi^0 p) = \frac{eg_{\pi N}}{8\pi m_{\pi}} \left \{ \mu-\mu^2\,
\frac{3+\kappa_p}{2}- \mu^2\, \frac{M^2}{16f^2_{\pi}} +\ ...\right \}\ ,
\label{eq:5.1}
\end{equation}
where $g_{\pi N}$ is the pion-nucleon coupling constant and $f_{\pi} \approx
93$~MeV the pion decay constant. We observe that $E_{0+}(\pi^0 p)$ is
proportional to $\mu$, which suppresses this reaction relative to charged pion
production. The first and the second term on the rhs of Eq.~(\ref{eq:5.1}) are
identical to the ``LET'' of \textcite{DeBaenst:1971hp} and
\textcite{Vainshtein:1972ih}, which however has to be corrected by the third
term on the rhs. Although this loop correction is formally of higher order in
$\mu$, its numerical value
is of the same size as the leading term.\\

\begin{figure}[]
\begin{center}
\includegraphics[width=0.8\columnwidth,angle=0.]{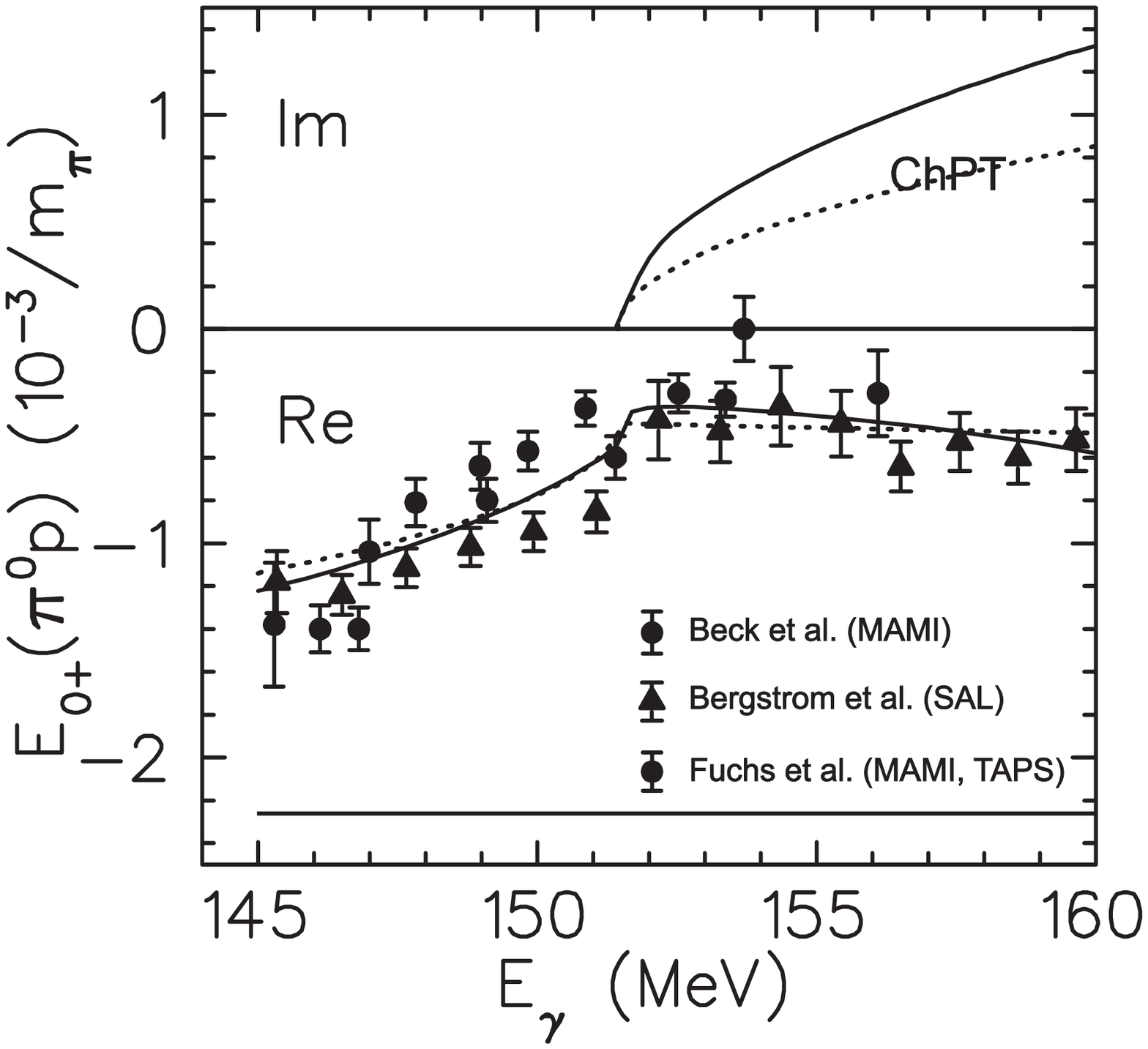}
\end{center}
\caption{The real (Re) and imaginary (Im) parts of the s-wave amplitude
$E_{0+}$ for $\pi^0$ photoproduction at threshold energies. The MAMI data of
\textcite{Beck:1990da} and \textcite{Fuchs:1996ja} are represented by circles,
the SAL data of \textcite{Bergstrom:1996fq} by triangles. Dashed lines:
predictions of ChPT at ${\cal{O}}$($p^3$) \cite{Bernard:1994gm,Bernard:1995cj},
solid lines: results from dispersion relations \cite{Hanstein:1996bd}. The
solid horizontal line at about -2.2 shows the prediction of
\textcite{DeBaenst:1971hp} and \textcite{Vainshtein:1972ih}. Figure by
courtesy of R.~Beck.}
\label{fig:E0+_pi0p}
\end{figure}
The energy dependence of $E_{0+}(\pi^0 p)$ is shown in Fig.~\ref{fig:E0+_pi0p}.
The discrepancy between predictions of \textcite{DeBaenst:1971hp} and
\textcite{Vainshtein:1972ih} and the experimental data obtained at the Mainz
Microtron MAMI and at SAL (Saskatoon) is apparent. Furthermore, the real part
of the amplitude shows a characteristic ``Wigner cusp'' at the threshold for
charged pion production, which lies about 5~MeV above the $\pi^0$ threshold.
This cusp in the real part is related to the sharp rise of the imaginary part
at the second threshold. The physical picture behind the large loop correction
is based on (I) the high production rate of charged pions and (II) the
charge-exchange scattering between the nucleon and the slow $\pi^+$ in the
intermediate state, which leaves a $\pi^0$ in the final state. However, the
direct experimental determination of the imaginary part will require
double-polarization experiments with linearly polarized photons and polarized
targets. The excellent agreement between ChPT and the data for $E_{0+}$ is
somewhat flawed by the fact that higher order diagrams are sizeable, that is,
the perturbative series converges slowly and low-energy constants appearing at
the higher orders reduce the predictive
power.\\

\begin{figure}[]
\begin{center}
\includegraphics[width=1.0\columnwidth,angle=0.]{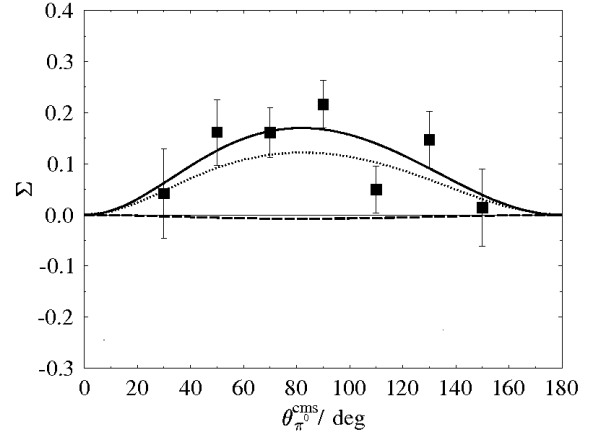}
\end{center}
\caption{The angular distribution of the photon asymmetry $\Sigma$ for the
reaction $p(\vec{\gamma},\pi^0)p$. Dashed line: results of dispersion
relations \cite{Hanstein:1996bd}, dotted line: prediction of ChPT
 \cite{Bernard:1994gm}, full line: empirical fit to the data with
Eqs.~(\ref{eq:2.20}) and (\ref{eq:2.21}). Figure from 
\textcite{Schmidt:2001vg}.} 
\label{fig:assym_pi0_photo_prod}
\end{figure}

For a more quantitative presentation of the results, the $E_{0+}$ amplitude was
parameterized as the sum of a direct and a charge-exchange
term \cite{Bernstein:1996vf},
\begin{eqnarray}
E_{0+} (\pi^0 p) & = &A_0(q) +i\, a_{\pi^+\pi^0}\, A_+ k_{\pi^+}\\
& =& A_0(q) + i\,\beta \,k_{\pi^+}\,,
\label{eq:5.2}
\end{eqnarray}
with $A_0(q)=a_0 + a_1 (q - q_{\text{thr}})$ and $A_+$ describing the neutral
and charged pion production in the absence of the charge exchange reaction,
$a_{\pi^+\pi^0}$ the scattering length for charge exchange, and $k_{\pi^+}$ the
momentum of the charged pion appearing in the intermediate state. This leaves
the 3 fit parameters $a_0$, $a_1$, and $\beta$ in oder to determine the s wave.
As discussed in subsection~\ref{sec:II.4}, the unpolarized cross section gives
information on only 2 of the 3 p-wave amplitudes. A complete experiment
therefore requires measuring a further observable, e.g., the photon asymmetry
$\Sigma$ of Eq.~(\ref{eq:2.21}). The result obtained at MAMI is depicted in
Fig.~\ref{fig:assym_pi0_photo_prod}.\\

The data of \textcite{Schmidt:2001vg} for the s- and p-wave amplitudes are
compared to the results of ChPT and dispersion relations in
Table~\ref{tab:thresh_mult}. The predictions of ChPT for the p waves are in
good agreement with the data, whereas $P_3$ comes out much too small in
dispersion theory. Since the latter approach is mainly based on input from the
imaginary parts of the multipoles $M_{1+}$ and $M_{1-}$ in the resonance
region, this failure may indicate that even the structure of the low-lying
resonances $\Delta(1232)$ and $N^{\ast}(1440)$ is not yet completely
unraveled.\\

\begin{table}[h]
\begin{tabular}{c|c|c|c}
\hline
&\cite{Schmidt:2001vg} & ChPT & DR\\
\hline
$ E_{0+}(\pi^0)$ & $-1.23\pm0.08\pm0.03$  & -1.16 & -1.22\\
$ E_{0+}(\pi^+)$ & $-0.45\pm0.07\pm0.02 $  & -0.43 & -0.56\\
$\beta$ & $2.43\pm0.28\pm1.0$ & 2.78& 3.6\\
$ P_1$ & $ 9.46\pm0.05\pm0.28$ & $9.14\pm0.5$& 9.55\\
$ P_2$ & $-9.5\pm0.09\pm0.28 $&  $-9.7\pm0.5$ & -10.37\\
$ P_3$ & $11.32\pm0.11\pm0.34$ & $10.36$ & 9.27\\
$ P_{23}$ & $ 10.45\pm 0.07$ & 11.07 & 9.84\\
\hline
\end{tabular}
\caption{Experimental results of \textcite{Schmidt:2001vg} for $E_{0+}(\pi^0p)$
at the $\pi^0$ and $\pi^+$thresholds in units of $10^{-3}/m_{\pi^+}$, four
combinations of the (reduced) $P$-wave amplitudes in units $k \cdot
10^{-3}/m_{\pi^+}^2$, and the parameter $\beta$ (with statistical and
systematic errors, in order) compared to the predictions of ChPT
\cite{Bernard:1994gm,Bernard:1995cj} and dispersion relations (DR)
\cite{Hanstein:1996bd}. For the definition of the reduced $P$-wave amplitudes
see section~\ref{sec:II.4}.} 
\label{tab:thresh_mult}
\end{table}
\begin{figure}[]
\begin{center}
\includegraphics[width=0.9\columnwidth,angle=0.]{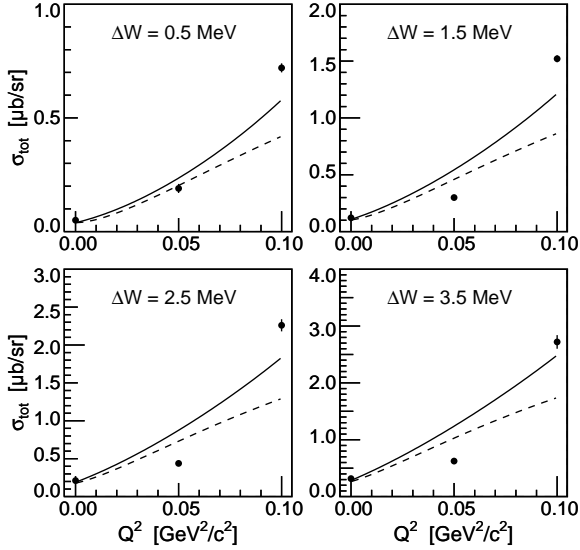}
\end{center}
\caption{The total cross section for the $p(e,e'p)\pi^0$ reaction as function
of  $Q^2$ for several values of the excitation energy $\Delta W = W
-M-m_{\pi^0}$. The data at $Q^2 = 0.05$~GeV$^2$ and $0.10$~GeV$^2$ are
from \textcite{Merkel:2001qg} and \textcite{Distler:1998ae}, respectively. The
solid line represents the prediction of ChPT \cite{Bernard:1996bi} and the
dashed line the result of the phenomenological model MAID 
\cite{Drechsel:1998hk}. Figure from \textcite{Merkel:2001qg}.} 
\label{fig:sigma_0_small_Delta_W}
\end{figure}
\begin{figure*}[]
\parbox{\columnwidth}
{\center
\includegraphics[width=0.8\columnwidth]{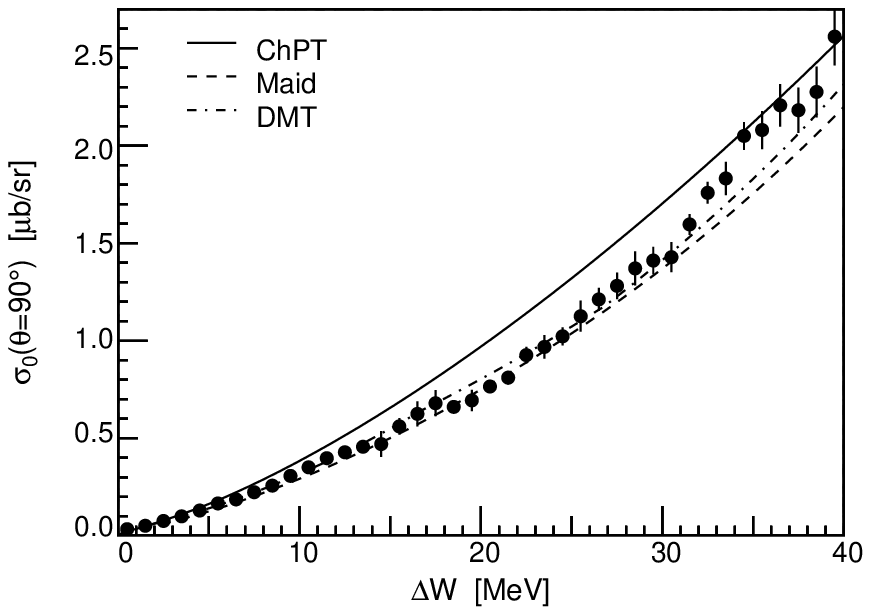}\\[5mm]
\includegraphics[width=0.8\columnwidth]{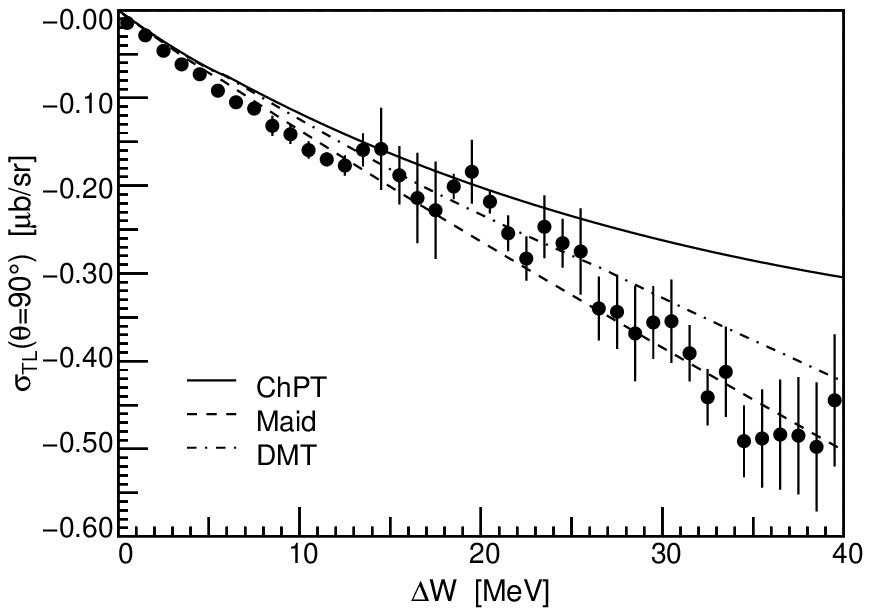}\\[5mm]
} \hfill
\parbox{\columnwidth}
{\center
\includegraphics[width=0.8\columnwidth]{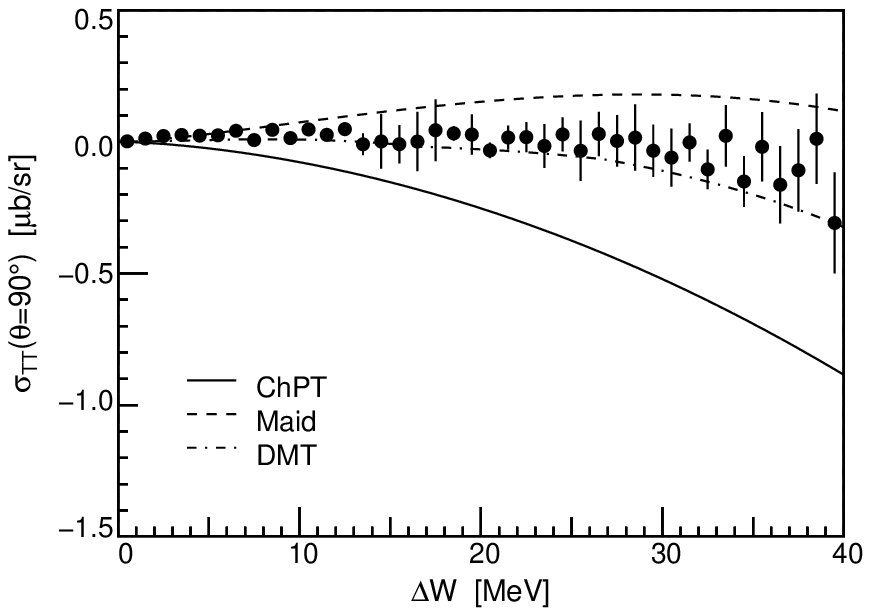}\\[5mm]
\includegraphics[width=0.8\columnwidth]{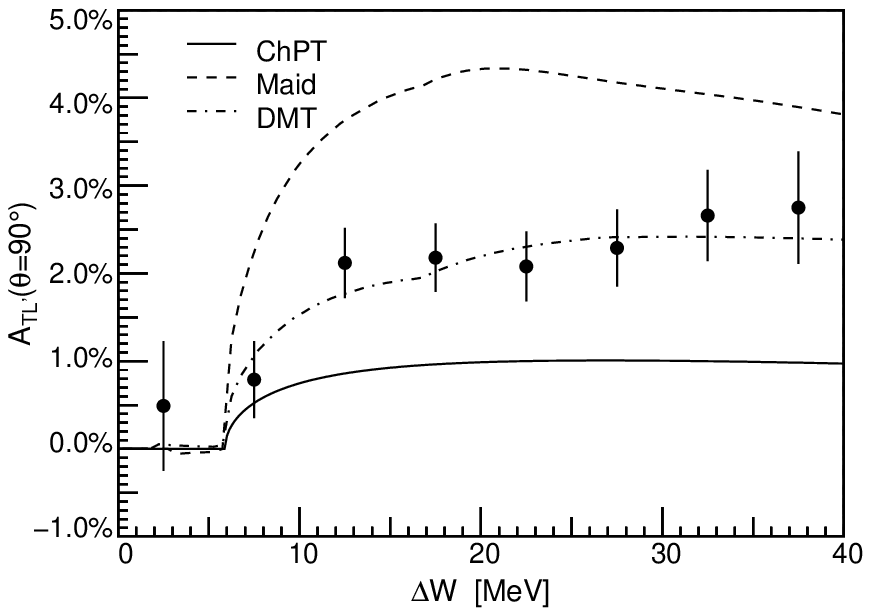}\\[5mm]
} \caption{The separated cross sections $\sigma_0$, $\sigma_{LT}$, and
$\sigma_{TT}$ as well as the beam helicity asymmetry $A_{LT}'$ at
$\theta_{\pi}^{\ast}=90^\circ$. Solid line: HBChPT \cite{Bernard:1996bi},
dashed line: MAID \cite{Drechsel:1998hk}, dashed-dotted line: DMT model
\cite{Kamalov:2001qg}. Figure from \textcite{Weis:2007kf}.}
\label{fig:structure_functions_thr}
\end{figure*}
The great success of ChPT for photoproduction at threshold was a strong
motivation to extend the experimental program to electroproduction. Because the
virtual photon has an additional longitudinal component, 3 more partial waves
appear to leading order: the longitudinal s-wave amplitude $L_{0+}$ and the
p-wave amplitudes $L_{1+}$ and $L_{1-}$, describing the excitation of the
$\Delta (1232)$) and $N^{\ast}(1440)$, respectively. Moreover, all the
amplitudes are functions of $Q^2$, that is, they probe the spatial distribution
of pion production on the nucleon. The first investigations performed at NIKHEF
\cite{vandenBrink:1995uk} and MAMI \cite{Distler:1998ae} for $Q^2=
0.10~\text{GeV}^2$ provided another confirmation of ChPT although at the
expense of 2 new low-energy constants, which were fitted to the data. In order
to further check this agreement, data were also taken at the lower momentum
transfer $Q^2 = 0.05~\text{GeV}^2$ \cite{Merkel:2001qg}. The total cross
section obtained by these measurements is compared with the predictions of ChPT
in Fig.~\ref{fig:sigma_0_small_Delta_W}, which shows the total cross sections
near threshold  as function of $Q^2$. The comparison was made on the basis of
the total cross sections in order to eliminate all possible systematic and
model errors connected with the separation in longitudinal and transverse
parts. It is apparent that the $Q^2$ dependence of the data can not not be
fully described by theory. We consider this an important issue that deserves
further investigations.\\

We conclude this subsection by presenting some recent results of
\textcite{Weis:2007kf}. Whereas former experiments were only sensitive to the
real part of the amplitudes, these authors also determined the fifth structure
function ($LT'$) given in Eq.~(\ref{eq:2.23}). This function can only be
measured with polarized electrons and out-of-plane, i.e., for finite values of
the pion azimuthal angle with regard to the electron scattering plane.
Furthermore, its multipole decomposition is of the form
Im\,$(L_{0+}^{\ast}\,M_{1+}+ \cdots)$, i.e., this function contains information
on the phase of the s-wave amplitude. With the shorthand notation $d\sigma_i /
d\Omega_{\pi}^{\ast}\equiv \sigma_i$, \textcite{Weis:2007kf} separated the
partial cross sections $\sigma_0=\sigma_T + \varepsilon \sigma_L$,
$\sigma_{TT}$, and $\sigma_{LT}$ as well as the beam asymmetry $A_{LT}'$
corresponding to $\sigma_{LT}'$. The result is displayed in
Fig.~\ref{fig:structure_functions_thr}. We observe that only the dynamical
Dubna-Mainz-Taipei (DMT) model \cite{Kamalov:2001qg} is able to fully describe
the experiment, in particular its prediction for the helicity asymmetry is
right on top of the data. Such dynamical models start from a description of the
pion-nucleon scattering phases by a quasi-potential, which serves as input for
an integral equation to account for multiple scattering. In this sense the
model contains the loop corrections to an arbitrary number of rescattering
processes, and is therefore perfectly unitary, albeit on a phenomenological
basis that may violate gauge invariance to some extent.
\subsection{Nucleon resonances and meson production}
\label{sec:V.2}
\begin{figure}[]
\begin{center}
\includegraphics[width=6.3cm,angle=90.]{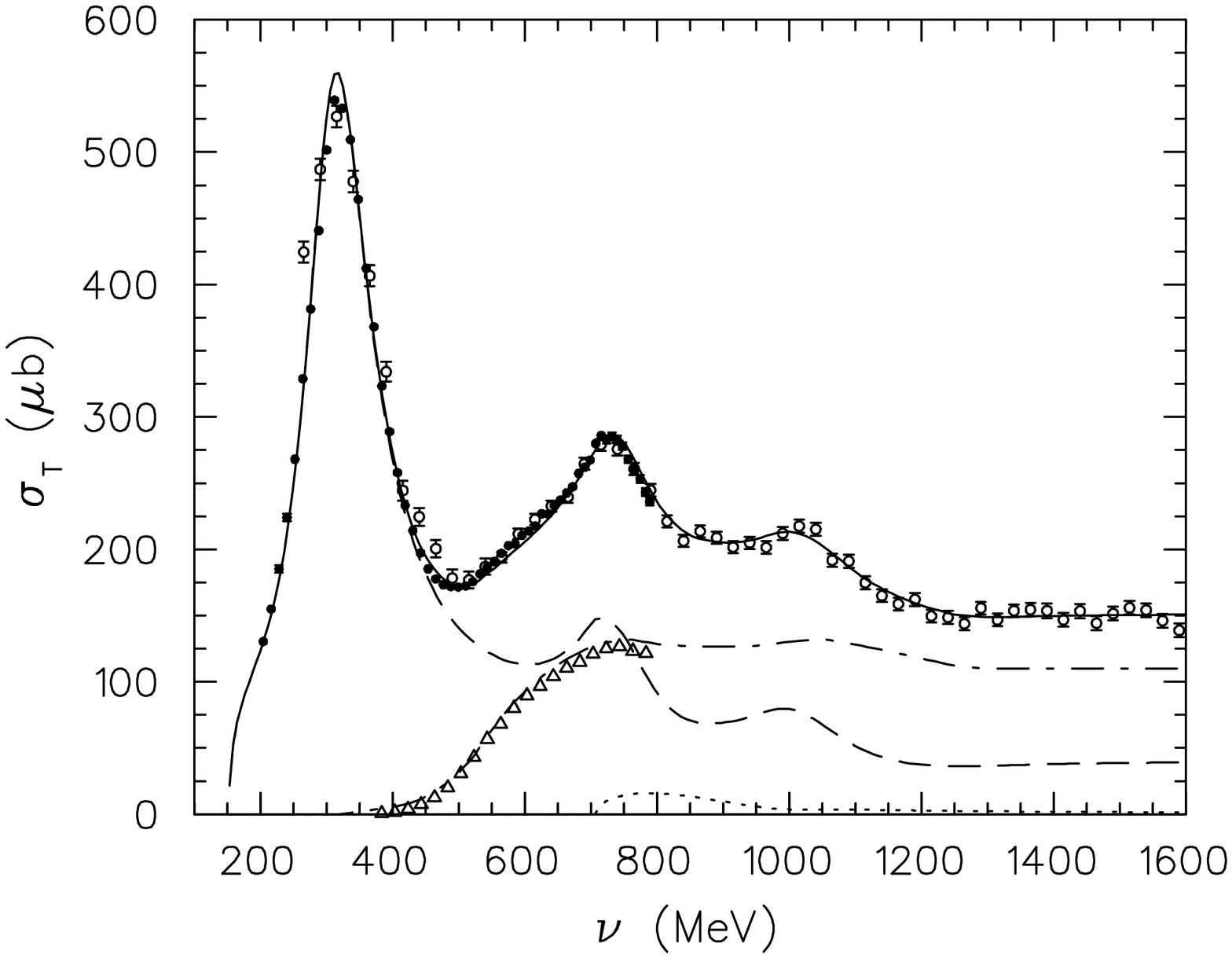}
\end{center}
\caption{The total photoabsorption cross section $\sigma_T$ on the proton as
function of the photon lab energy $\nu$ . The various lines represent the MAID
results \cite{Drechsel:1998hk} for the total cross section (solid line),
one-pion channels (dashed line), more-pion channels (dashed-dotted line), and 
$\eta$ channel (dotted line). Full circles: total cross section from MAMI
\cite{MacCormick:1996jz}, open circles: data from Daresbury
\cite{Armstrong:1971ns}, open triangles: two-pion production
\cite{Braghieri:1994rf}. Figure by courtesy of J.~Ahrens.}
\label{fig:SUM_sigmaT}
\end{figure}

\begin{figure}[]
\begin{center}
\includegraphics[width=0.9\columnwidth,angle=0.]{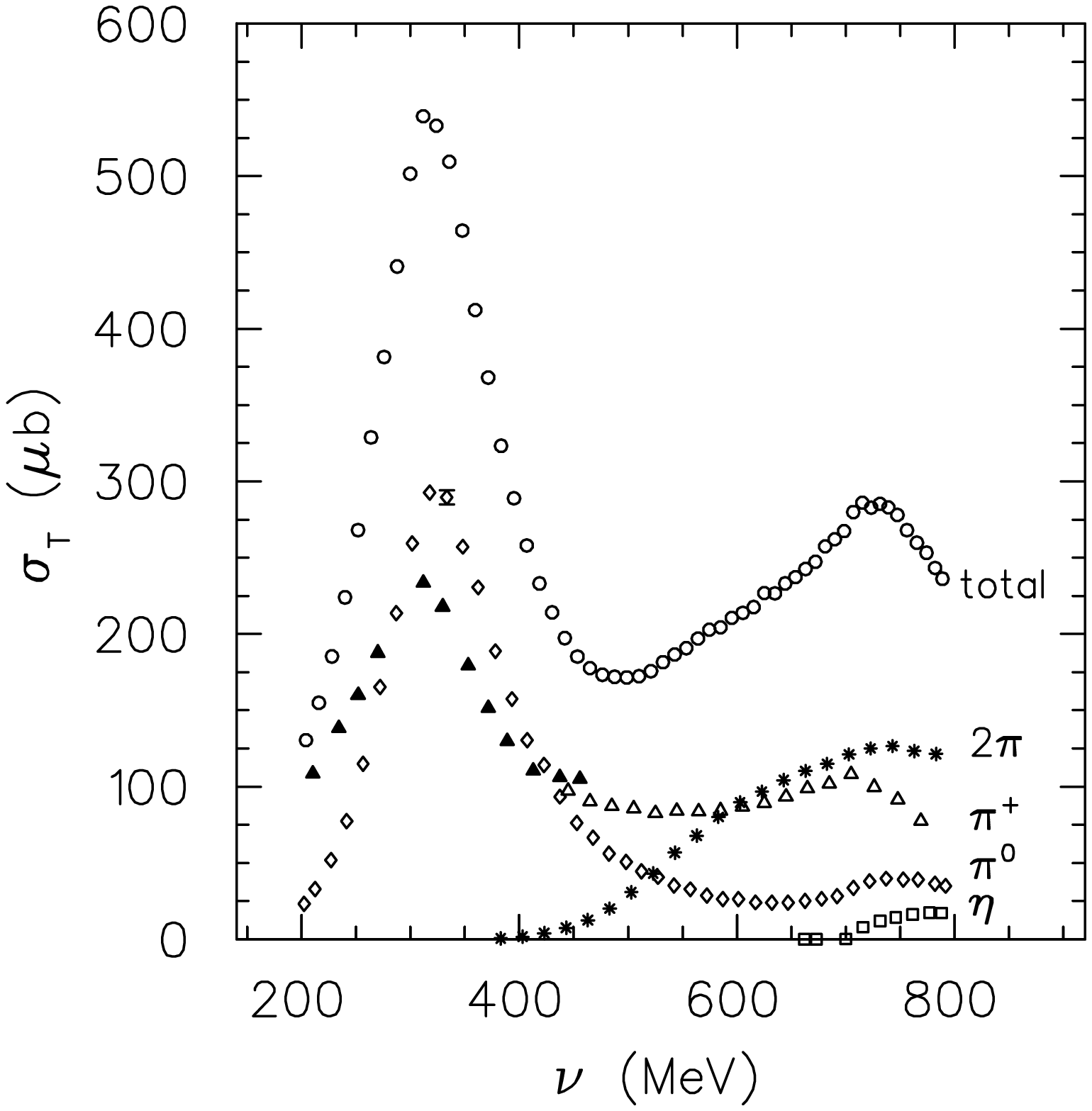}
\end{center}
\caption{The partial and total cross sections for the absorption of photons on
the proton as a function of the photon lab energy $\nu$ obtained at MAMI. Open
circles: total cross section as in Fig.~\ref{fig:SUM_sigmaT}
 \cite{MacCormick:1996jz}, solid and open triangles: $n\pi^+$ decay channel
 \cite{MacCormick:1996jz}, open diamonds: $p\pi^0$ decay channel
 \cite{MacCormick:1996jz}, asterisks: two-pion production
 \cite{Braghieri:1994rf}, open squares: $p\eta$ channel
 \cite{Krusche:1995nv}. Figure by courtesy of J.~Ahrens.}
\label{fig:SUM_channels}
\end{figure}

The total photoabsorption cross section $\sigma_T$ for the proton is displayed
in Fig.~\ref{fig:SUM_sigmaT}. It clearly exhibits 3 broad resonance structures
on top of a strong background. These structures correspond, in order, to
magnetic dipole $(M1)$ excitation of the $\Delta (1232)$ resonance, electric
dipole $(E1)$ strength near the resonances $N^{\ast}(1520)$ and
$N^{\ast}(1535)$, and electric quadrupole $(E2)$ strength near the
$N^{\ast}(1680)$. The figure also shows the contributions of the most important
channels: the one-pion channels dominate up to $\nu\approx500$~MeV, the
two-pion branching ratio becomes comparable in the second resonance region at
$\nu\approx700$~MeV, and the large $\eta$ branch of the resonance
$N^{\ast}(1535)$ is hidden in the background. Because the nucleon resonances
lie above the one-pion threshold, any separation in a continuous background and
discrete resonances is necessarily model-dependent. In particular, because
background and resonance contributions interfere, a careful analysis of the
partial waves and their relative phases is mandatory.
Figure~\ref{fig:SUM_channels} depicts the partial cross sections for the
different decay channels investigated so far. The rapid increase of the
two-pion contribution between 400 and 600~MeV is clearly seen, from whereon it
provides more than half of the total absorption. Another interesting feature is
the dominance of charged pion production both below and above the
$\Delta(1232)$ resonance. Finally, one notes the small $\eta$ decay branch,
which corresponds to s-wave $\eta$ production mediated by
the $S_{11}(1535)$.\\

In the past, the primary method to unravel the nucleon resonance spectrum were
experiments with strong interactions, in particular pion-nucleon scattering.
These data have been systematically studied by the Karlsruhe group
\cite{Hoehler:1983dd,Koch:1985bp} and the GWU Collaboration using the code SAID
\cite{Arndt:2002xv,Arndt:2003if,Arndt:2006bf}. A summary of the known
spectroscopic information on nucleon resonances is given by the Particle Data
Group (PDG) \cite{Yao:2006px}. It is the objective of these studies to
determine the relevant characteristics of the resonances, their pole positions,
widths, decay channels, and branching ratios. In a first step, the full data
base is fitted within the framework of a partial wave analysis. If a particular
partial wave shows some rapid increase of the scattering phase over a limited
energy region, the fit is then repeated with a form containing both a smooth
background and a resonance form, mostly of the Breit-Wigner shape. Because of
the strong decay channels and large resonance widths of typically 100~MeV and
more, the ideal resonance form is only realized for the first resonance, the
$\Delta(1232)$: the pion-nucleon scattering phase $\delta_{33}(W)$ goes through
$90^{\circ}$, the real part of the multipole vanishes, and the imaginary part
has a maximum near $W=M_R\approx 1232~$MeV. Because of inelastic channels,
overlapping resonances, and energy-dependent backgrounds, these conditions are
not fulfilled by the higher resonances. A mere ``bump'' in a partial wave is
not necessarily a resonance, it may also originate from the opening of a new
channel, which usually produces an asymmetric resonance shape. The
``speed-plot'' technique is particularly useful to probe the resonance
structure. It requires the derivative of the partial wave amplitude with regard
to the energy $W$, which is then compared to the corresponding derivative of an
ideal Breit-Wigner resonance. In this way one determines the pole position and
the residue of the multipole in the complex energy plane, which are unique
characteristics of a resonance.\\

\begin{figure}[]
\begin{center}
\includegraphics[width=0.9\columnwidth,angle=0.]{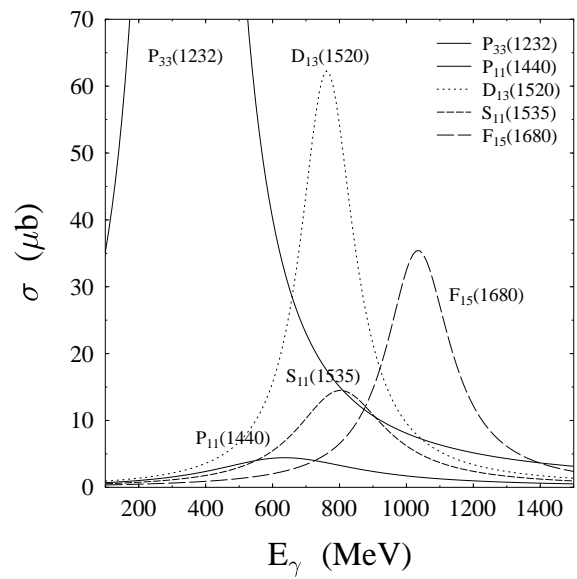}
\end{center}
\caption{Resonance contributions to the total photoabsorption cross section of
the proton. The calculations are from the unitary isobar model MAID2003
\cite{Drechsel:1998hk}. Figure by courtesy of R.~Beck and L.~Tiator.}
\label{fig:resonances}
\end{figure}
Because the cross sections for electroexcitation are suppressed by the fine
structure constant $\alpha_{\text{em}}$, systematic studies of the resonance
structure were only possible after the advent of the new electron accelerators
providing high duty-cycle and flux density of photons and electrons. The
virtues of these investigations are obvious: the selectivity through transverse
electric and magnetic as well as longitudinal photon fields allows for the
separation of multipoles and investigations of the spin degrees of freedom, the
study of specific final states with different mesons by coincidence
experiments, and the possibility to resolve the spatial distribution of charge
and current densities by varying the virtuality of the photon.
Figure~\ref{fig:resonances} sets the scene by showing the individual
contributions of the main nucleon resonances to the total photoabsorption cross
section. The resonances in the figure are labeled in the spectroscopic
notation, see subsection~\ref{sec:II.5}. As an example, the
$\Delta(1232)=P_{33}(1232)$ has the pion in a p wave and both isospin and spin
are 3/2. As is evident from Fig.~\ref{fig:resonances}, the $\Delta(1232)$ is
the dominant feature in the resonance spectrum. It contributes more than
400~$\mu$b at the maximum of the absorption cross section. Already in the
second resonance region, there are several overlapping resonances, the small
Roper resonance $P_{11}(1440)$ with the same quantum numbers as the nucleon,
as well as the relatively strong $D_{13}(1520)$ and the weaker $S_{11}(1535)$,
both mainly excited by electric dipole radiation, all on top of a large
background (not shown in the figure!) and on the tails of the neighboring 
resonances. In the third resonance region, the figure shows only the dominant 
$F_{15}(1680)$ resonance with a concentration of the electric quadrupole
strength. The number of known (and unknown) resonances increases in the higher
part of the spectrum, see \textcite{Yao:2006px}. However, somewhere above 
$W \approx 1.5$~GeV the notion of ``resonances'' becomes problematic and has 
to be replaced by a continuum which, of course, can be expanded in partial 
waves as discussed in subsection~\ref{sec:II.5}. The problem is clearly not on 
the experimental side, but in the modeling of the spectrum, which for 
$W\geq1.5$~GeV should be based on coupled channel calculations. As has been 
mentioned before, the most valid technique is the ``speed plot'' 
\cite{Hoehler:1983dd}, which gives information about the resonance position 
in the complex energy plane. However, the distinction of ``resonances'' and 
``background'' becomes increasingly difficult, and looses its meaning for 
energies somewhere above 2~GeV. The ``missing resonances'' issue of the 
constituent quark model (CQM) that predicts more resonances than have been 
observed \cite{Isgur:1979be,Metsch:2003ix}, may disappear in view of the 
``resonance-background'' problem \cite{Thoma:2005vk}.\\

As in the case of the elastic form factors discussed in section~\ref{sec:III},
the $Q^2$ dependence of the multipole transitions provides information on the
spatial distribution of these observables. Let us discuss this issue for the
$\Delta(1232)$. The 3 (real) transition form factors $G_M^{\ast}$,
$G_E^{\ast}$, and $G_C^{\ast}$ are related to the (complex) partial wave
amplitudes according to Eq.~(\ref{eq:2.25}). The magnetic dipole transition
$M_{1+}$ dominates, the electric and Coulomb quadrupole transitions $E_{1+}$
and $S_{1+}$ are much smaller. A finite value of the quadrupole moment requires
that the wave functions of either the nucleon or the $\Delta(1232)$, or most
likely both are deformed. In the CQM, such a deformation follows from the
tensor force contained in the color-hyperfine interaction among the quarks
\cite{DeRujula:1975ge,Isgur:1981yz,Capstick:1989ck}. However, typical CQM
calculations \cite{Capstick:1994ne,DeSanctis:2005vq} underestimate the electric
and Coulomb quadrupole amplitudes $E_{1+}$ and $S_{1+}$. In models with pionic
degrees of freedom, the deformation arises naturally from the spin-dependent
coupling of the pion to the quarks. The pions have been introduced in several
approaches, such as chiral bag models
\cite{Vento:1980mu,Bermuth:1988ms,Lu:1996rj}, dynamical models 
\cite{Sato:2000jf,Kamalov:1999hs}, and effective field theories
\cite{Pascalutsa:2005ts,Pascalutsa:2005vq,Gail:2005gz}. Although differing
considerably in the details, all these models describe the experimental data
reasonably well. In particular the chiral effective field theories (EFTs) are
based on a systematic expansion in terms of the external momenta, the pion
mass, and the N$\Delta$ mass splitting. Contrary to the dynamical models they
are gauge and Lorentz invariant, however, being based on a perturbative
expansion, the unitarity condition is only approximately fulfilled.\\

\begin{figure*}[]
\parbox{2.0\columnwidth}
{\center
\includegraphics[width=1.5\columnwidth,angle=0.]{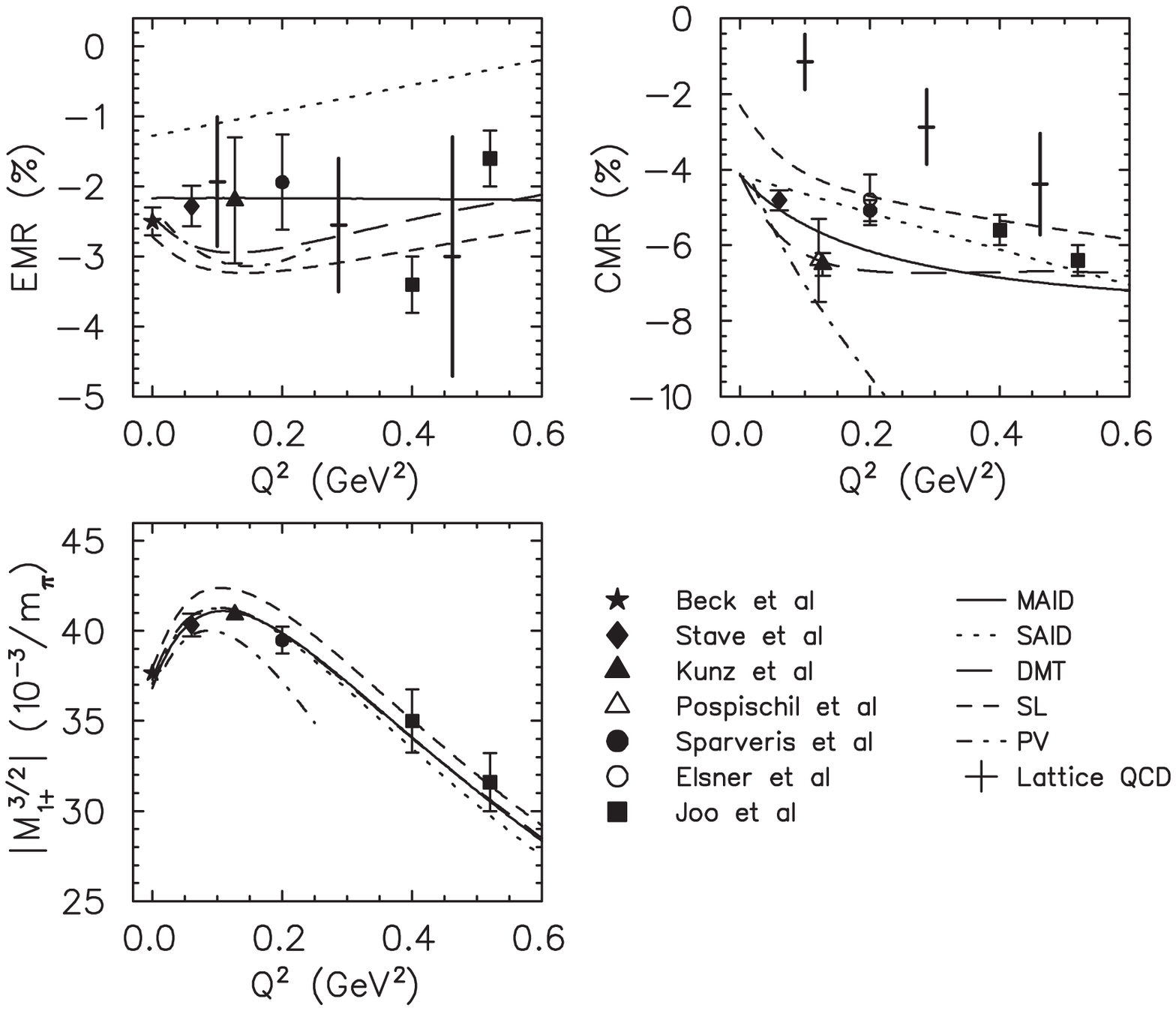}
} \caption{The ratios EMR=$R_{EM}$ and CMR=$R_{EM}$ together with the amplitude
$M^{3/2}_{1+}$ of the N$\Delta(1232)$ transition as function of $Q^2$. The data
are from \textcite{Beck:1999ge} (asterisks), \textcite{Stave:2006ea}
(diamonds), \textcite{Kunz:2003we} (solid triangles),
\textcite{Pospischil:2000ad} (open triangles), \textcite{Sparveris:2006uk}
(solid circles), \textcite{Elsner:2005cz} (open circles), and
\textcite{Joo:2004mi} (squares). The theoretical predictions are represented by
solid lines: MAID2007 \cite{Drechsel:2007if}, dotted lines: SAID
\cite{Arndt:2002xv}, long-dashed lines: DMT \cite{Kamalov:1999hs}, short-dashed
lines: Sato-Lee model \cite{Sato:2000jf}, dot-dashed lines: ChEFT
\cite{Pascalutsa:2005ts,Pascalutsa:2005vq}, and crosses: predictions with error
bars from Lattice QCD \cite{Alexandrou:2004xn}. Figure adapted from
\textcite{Sparveris:2006uk} by L.~Tiator.} \label{fig:EMR-CMR-M1+}
\end{figure*}
Of course, the deformation can not be measured for the nucleon ground state,
because the intrinsic static quadrupole moment $Q_0$ and the observed
quadrupole moment are related by $Q^{\text{obs}}_J=[3 \langle M \rangle^2 -
J(J+1)]/[J(2J-1)]\,Q_0$. This rule allows one to observe the quadrupole
moment of the $\Delta(1232)$ but, of course, the highly unstable
$\Delta(1232)$ can not serve as a static target. Therefore,
the deformation issue can only be accessed through the electromagnetic
transitions from the nucleon to the $\Delta(1232)$. In the following we discuss
the ratios of the multipole transitions at the resonance position, $R_{EM}$ and
$R_{SM}$ as defined by Eq.~(\ref{eq:2.28}). We repeat that the resonance
position of the $\Delta(1232)$ is uniquely defined within the validity of the
Fermi-Watson theorem, that is, (I) the real parts of all 3 N$\Delta$ amplitudes
vanish at $W=M_{\Delta}$, because all the  amplitudes carry the same phase,
namely the pion-nucleon phase shift in the partial wave with $\ell=1$ and
$I\, = \,J = \,\frac {3}{2}$, $\delta_{33}(W)$.\\

Many experimental and theoretical investigations have been devoted to the
N$\Delta$ transition with real photons, for a summary of this work see
\textcite{Beck:2006ye}. The most precise value for the ratio of the multipoles,
\begin{equation}
R_{EM}= (-2.4 \pm 0.16_{\text{stat}}\pm 0.24_{\text{syst}})\%\,,
\end{equation}
was obtained with linearly polarized photons \cite{Beck:1999ge,Leukel:2001rr}.
These authors also studied neutral and charged pion production on the proton,
which is necessary to isolate the isospin $\frac{3}{2}$ amplitude relevant for
the N$\Delta$ transition. At finite $Q^2$ the same physics questions were
addressed by pion electroproduction, $e+p\rightarrow e'+p'+\pi^0$, measuring
the scattered electron in coincidence with the recoil proton detected in a
high-resolution magnetic spectrometer. The small solid angle of these
instruments does not limit the accuracy since the protons are focused by the
relativistic boost along $\vec{q}$. Such experiments were first performed at
MIT/Bates \cite{Mertz:1999hp} and then extended by \textcite{Sparveris:2004jn}
to out-of-plane angles with the OOPS spectrometer. This work was continued in
the framework of the A1 Collaboration at MAMI/Mainz
\cite{Sparveris:2006uk,Elsner:2005cz,Stave:2006ea}. In particular,
 \textcite{Stave:2006ea} also measured the structure function
$\sigma'_{LT}$ by use of polarized electrons and out-of-plane proton detection.
The experimental results for the $\Delta$ multipoles in the low-$Q^2$ region
are compared to several model calculations in Fig.~\ref{fig:EMR-CMR-M1+}. We
observe that the leading multipole $M_{1+}$ is described quite well by the
models. It is also also worthwhile mentioning that the dynamic models ascribe a
third of the magnetic dipole strength to the pion cloud. The predictions
scatter much more with regard to the quadrupole strength as shown by the ratios
$R_{EM}$ and $R_{SM}$. Concerning the origin of ``deformation'', both the
dynamic models \cite{Sato:2000jf,Kamalov:1999hs} and effective field theory
\cite{Pascalutsa:2005ts,Pascalutsa:2005vq,Gail:2005gz} agree that the
multipoles $E_{1+}$ and $S_{1+}$ are essentially due to the pion cloud. For a
detailed comparison of these models in the $\Delta$ region, see
\textcite{Drechsel:2006hc}.\\

In many of the mentioned contributions, the authors have pointed out a
considerable model-dependence of the analysis. It is therefore a substantial
progress in this field that \textcite{Kelly:2005jy} performed a series of
double-polarization experiments at $Q^2=1$~GeV$^2$ near the $\Delta$ region.
Altogether they extracted 16 of the 18 independent response functions, most of
them for the first time. As mentioned before, the experimental cross section is
obtained by summing bilinear products of multipole amplitudes, $\sum_{\ell,
\ell'} {\mathcal {M}}^{\ast}_{\ell}{\mathcal {M}}_{\ell'}$. Whereas the
unpolarized response functions or cross sections, in shorthand $\sigma_T$,
$\sigma_L$, $\sigma_{TT}$, and $\sigma_{LT}$ are obtained from the real parts
of these products, many of the polarized cross sections are given by the
imaginary parts, the first example being the ``fifth structure function''
$\sigma_{LT}'$ measured with polarized beams. Because the N$\Delta$ multipoles
carry the same phase, their product can only contribute to responses built from
the real parts. On the other hand, the responses containing the imaginary parts
yield information on the interference between the N$\Delta$ and background
multipoles. The findings of \textcite{Kelly:2005jy} can be summarized as
follows: (I) response functions governed by real parts are in general agreement
with recent model calculations, (II) response functions determined by imaginary
parts may differ substantially from the experiment and among the calculations,
(III) the multipole analysis yields better results than a (truncated) Legendre
series, and (IV) the model builders should go back to the drawing board to get
better control of the non-resonant background, in
particular at the larger virtualities $Q^2$.\\

As mentioned above, the typical CQM calculations underestimate the $\Delta$
multipoles, in particular the electric and longitudinal ones. Therefore, any
successful description of these observables needs a pion cloud, at least if one
insists on a reasonable size of the quark bag. Such descriptions are the chiral
bag models, the dynamical models, and effective field theories. Taking all
facts together, one is again forced to accept the dominant role that pions play
for the structure of the nucleon. Of special interest is the recent work to
solve QCD on the space-time lattice. Since these calculations can not yet be
performed at the small (current) quark masses corresponding to the physical
pion mass, one uses very large quark masses leading to pion masses $m_{\pi}
\gtrsim 300$~MeV. The results are then extrapolated to the physical pion mass
by extrapolating functions, ideally as derived from chiral effective field
theories. Such procedure is not undisputed because the chiral expansion is
hardly valid at pion masses much larger than the physical mass. However, the
chiral extrapolation shows that unexpected phenomena occur when the pion mass
is lowered from a few hundred MeV to its physical value: near the pion mass
value for which the $\Delta$ resonance can decay, the chiral extrapolation
becomes non-analytic, which leads to a kink and strong curvature
in the extrapolation formula \cite{Pascalutsa:2006up}.\\

\begin{figure}[]
\begin{center}
\includegraphics[width=0.8\columnwidth,angle=0.]{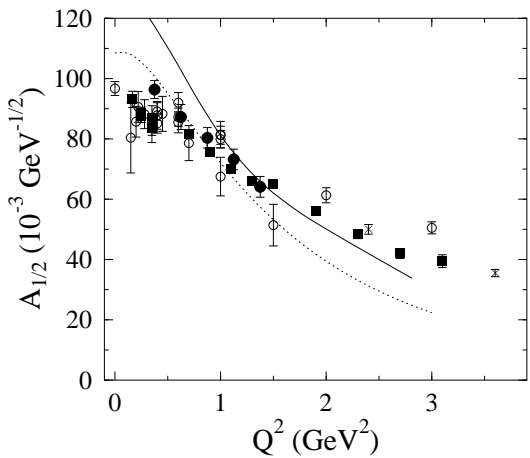}
\end{center}
\caption{The helicity amplitude $A_{1/2}$ for the $S_{11}(1535)$ in the
$p(e,e'p)\eta$ reaction. The data are represented by filled circles:
\textcite{Denizli:2007tq}, crosses: \textcite{Armstrong:1998wg}, and open
circles: earlier publications as specified by \textcite{Denizli:2007tq}. The
predictions are based on quark models and given by the solid
line~\cite{Capstick:1994ne} and the dotted line~\cite{Aiello:1998xq}. Figure
from \textcite{Denizli:2007tq}.}
\label{fig:A_12_S_11}
\end{figure}

As a further instructive example we present some results for the
electroproduction of $\eta$ mesons, which are mainly produced by the decay of
the $S_{11}(1535)$ resonance with $J=I=\frac{1}{2},\,\ell=0$. As is evident
from Fig.~\ref{fig:SUM_channels}, the $S_{11}(1535)$ is buried under the total
cross section. However it is clearly seen in the $\eta$ channel, because this
resonance has an $\eta$ branching ratio of 45-60\% compared to a few per cent
for other excitations of the nucleon \cite{Vrana:1999nt}. Eta photoproduction
experiments at threshold show a strong increase of the cross section in the
range of the resonance and an s-wave angular distribution in agreement with the
given assignment \cite{Krusche:1995nv}. However, the shape of the resonance is
very asymmetric, and also the speed-plot analysis does not yield satisfactory
solutions \cite{Hoehler:1994dd}. Because of this unusual behavior several
alternative interpretations have been given, for example in terms of a
K$\Sigma$ molecular state \cite{Kaiser:1996js}. Also the helicity amplitude
$A_{1/2}$ of this resonance has an unusually soft form factor, as displayed in
Fig.~\ref{fig:A_12_S_11} for the $p(e,e'p)\eta$ reaction measured at the
Jefferson Lab \cite{Denizli:2007tq}.
It is striking how flat this transition form factor stays compared to the
typical dipole form for other form factors of the nucleon. We further notice
that the constituent quark model calculations, relativistic or not, can not
fully explain the slow falloff with $Q^2$. It is also tempting to identify the
structure near $Q^2 \approx 0.2~\text{GeV}^2$ with a meson cloud effect as
discussed in section~\ref{sec:III.1}.\\

\begin{figure}[]
{\centering
\parbox{0.8\columnwidth}{\raggedleft
\includegraphics[width=0.8\columnwidth,angle=0.]{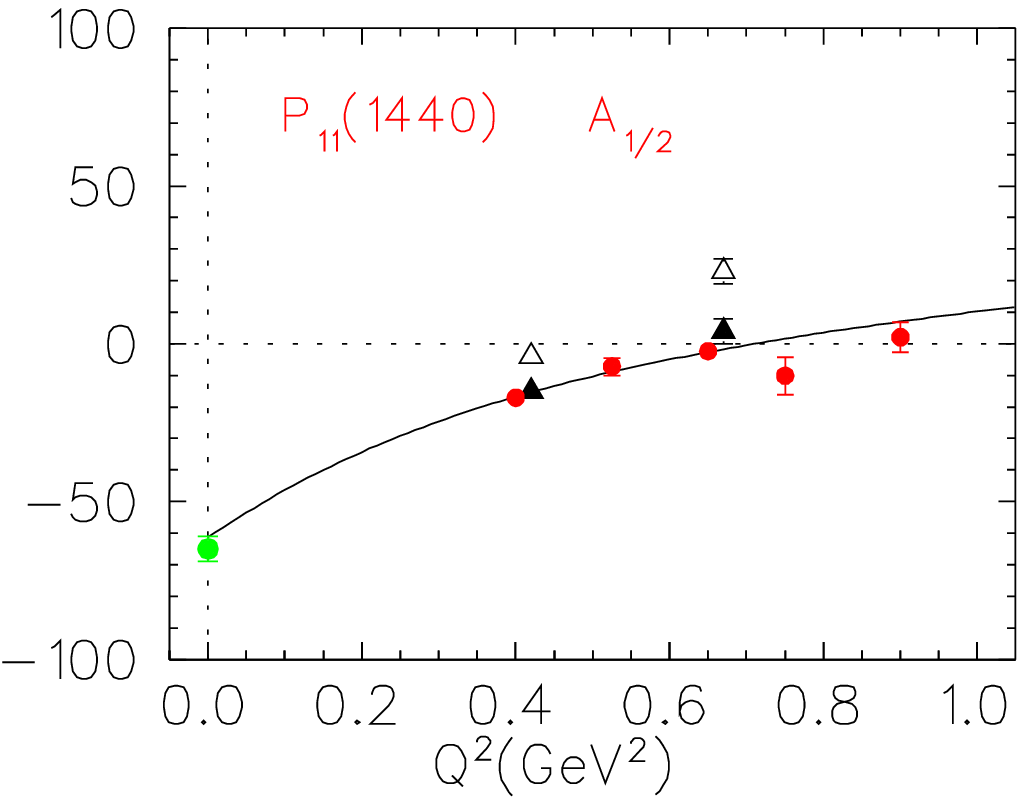}\\
\includegraphics[width=0.77\columnwidth,angle=0.]{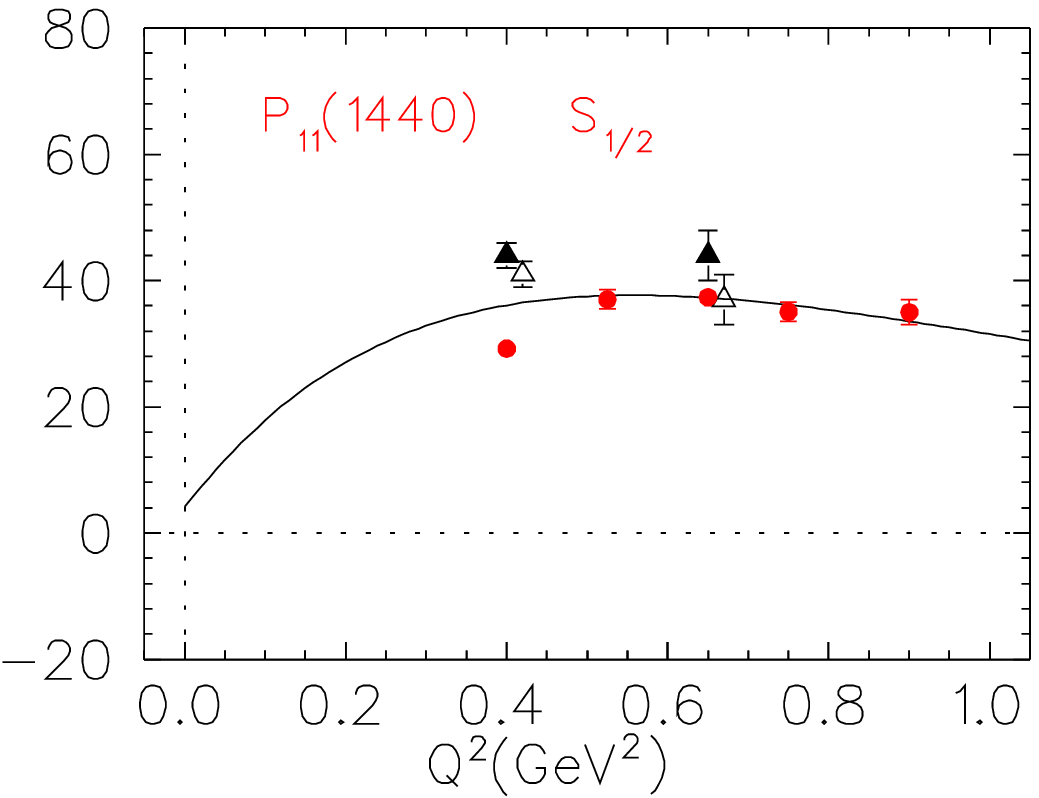}}
} \caption{The helicity amplitudes $A_{1/2}$ and $S_{1/2}$ of the
$P_{11}(1440)$ as function of $Q^2$. The triangles are from the analysis of
\textcite{Aznauryan:2004jd} based on data of the CLAS Collaboration at JLab and
obtained from dispersion relations (open triangles) and the unitary isobar
model (filled triangles). The solid line is the global solution of
MAID2007 \cite{Drechsel:2007if}, a fit to the full data basis, and the filled
circles are local fits to the data in the particular energy region. The data
point at $Q^2=0$ is from the PDG \cite{Yao:2006px}. Figures from 
\cite{Drechsel:2007if}.} 
\label{fig:Roper}
\end{figure}

Another small but interesting resonance in the second resonance region is the
Roper resonance $P_{11}(1440)$ with the same quantum numbers as the nucleon.
The CQM describes this resonance by a radial excitation of the nucleon, and
therefore it should be sensitive to the radial form of the bag potential.
However, its mass is much lower than expected in simple quark models.
\textcite{li:1991yb} suggested that the Roper could also be a quark-gluon
hybrid state, which, however, can not be excited by the longitudinal current or
Coulomb field. Recent data on the electroproduction of this resonance are shown
in Fig.~\ref{fig:Roper} and compared with the single-energy and global
solutions of MAID2007. We note that the amplitude $A_{1/2}$ (magnetic dipole
transition) has a zero crossing at small momentum transfer. Furthermore, the
amplitude $S_{1/2}$ (Coulomb monopole transition) rises to quite large values,
which rules out the quark-gluon hybrid model.\\

In section~\ref{sec:VI} we also discuss the helicity structure of the second
and third resonance regions with regard to several sum rules for real and
virtual photons. However, with increasing excitation energy it becomes more and
more difficult to isolate individual resonances by inclusive cross sections. It
will therefore take a full-fledged program with polarized beams and targets as
well as recoil polarization to analyze the higher mass region and to identify
resonance structures on top of the large background.

\section{Sum rules} \label{sec:VI}
The persisting problem of gaining a quantitative understanding of nucleon
structure is one of the reasons why we are interested in sum rules. Being based
on quite general principles like causality, unitarity, Lorentz and gauge
invariance, sum rules should be valid for every model or theory respecting
these principles and having a ``reasonable'' high-energy behavior. Therefore,
the agreement or disagreement between theoretical predictions and the sum rule
provides invaluable information on the quality of the approximations involved
and whether or not the relevant degrees of freedom have been included.
Specifically, if we compare a sum rule value with accurate experimental data up
to a certain maximum energy, we learn whether the physics responsible for the
sum rule is provided by the phenomena up to that energy, or whether possibly
new degrees of freedom come into the game. As an example, Baldin's sum rule
relates the forward scalar polarizability to an energy-weighted integral over
the total photoabsorption cross section and thus allows for an independent
check of the results from Compton scattering. Moreover, recent
double-polarization experiments have determined the helicity structure of this
cross section for the proton. From these data, the forward spin polarizability
has been obtained, and the Gerasimov-Drell-Hearn (GDH) sum rule has been
verified within the experimental error bars of less than 10~\%. Data have also
been taken for the neutron and are under evaluation. With some caveat in mind,
these sum rules can be generalized to the scattering of virtual photons. The
integrands of the respective integrals are related to the electroexcitation
cross sections and, in the limit of deep inelastic scattering (DIS), to the
nucleon structure functions. The resulting generalized integrals and
polarizabilities depend on the photon's 4-momentum and therefore contain
information on the spatial distribution of these observables.
\subsection{Sum rules for real photons}
\label{sec:VI.1}
\subsubsection{Forward dispersion relations and sum rules}
\label{sec:VI.1.1}
\begin{figure}[]
\begin{center}
\includegraphics[width=0.8\columnwidth]{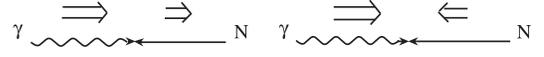}
\end{center}
\vspace{0.5cm} \caption{Spins and helicities in the reaction
$\vec{\gamma}+\vec{N}\rightarrow N^{\ast}$. The open arrows denote the
projections of the spin on the photon momentum, $S_z$, the helicities $h$ are
the projections on the respective particle momentum, and the photon has
right-handed helicity, $\lambda=+1$.\\ Left:
$N(S_z=1/2,\ h=-1/2)\rightarrow N^{\ast}(S_z=h=3/2)$,\\
right: $N(S_z=-1/2,\ h=1/2)\rightarrow N^{\ast}(S_z=h=1/2)$.}
\label{fig:SUM_helicity}
\end{figure}

The forward scattering amplitudes $f$ and $g$ of Eqs.~(\ref{eq:4.1.4b}) -
(\ref{eq:4.1.6}) can be determined by scattering circularly polarized photons
(helicity $\lambda=\pm 1$) off nucleon targets that are polarized along or
opposite to the photon momentum $\vec{q}$ as shown schematically in
Fig.~\ref{fig:SUM_helicity}. If the spins are parallel, the helicity of the
intermediate hadronic state takes the value $\textstyle{\frac {3}{2}}$. Since
this requires a total spin $J \ge \textstyle{\frac {3}{2}}$, the transition can
only take place on a correlated 3-quark system. For opposite spins, on the
other hand, the helicity is conserved and the scattering can also take place on
an individual quark. Denoting the Compton scattering amplitudes for these two
experiments by $T_{3/2}$ and $T_{1/2}$, we find $f(\nu)=\textstyle{\frac
{1}{2}}(T_{1/2}+T_{3/2})$ and $g(\nu)=\textstyle{\frac
{1}{2}}(T_{1/2}-T_{3/2})$. Furthermore, the total absorption cross section is
given by the spin average over the helicity cross sections,
\begin{equation}\label{eq:6.1.3}
\sigma_T=\frac{1}{2}\,(\sigma_{1/2}+\sigma_{3/2})\ ,
\end{equation}
and the helicity-dependent cross section by the helicity difference,
\begin{equation}\label{eq:6.1.4}
\sigma_{TT} = \frac{1}{2}\,(\sigma_{1/2}-\sigma_{3/2})\ .
\end{equation}
Based on unitarity and causality, the optical theorem relates the absorption
cross section to the imaginary parts of the respective forward scattering
amplitudes,
\begin{eqnarray}
{\rm {Im}}~f(\nu) & = & \frac{\nu}{8\pi} (\sigma_{1/2}(\nu)+\sigma_{3/2}(\nu))
=
\frac{\nu}{4\pi}\,\sigma_T (\nu)\, , \nonumber \\
{\rm {Im}}~g(\nu) & = & \frac{\nu}{8\pi} (\sigma_{1/2}(\nu)-\sigma_{3/2}(\nu))
= \frac{\nu}{4\pi}\,\sigma_{TT} (\nu) . \qquad \label{eq:6.1.5}
\end{eqnarray}
Because of the smallness of the fine structure constant $\alpha_\text{em}$, all
higher order electromagnetic processes are at the per cent level. We may
therefore neglect the absorption below the threshold for pion production,
$\nu_0 =m_{\pi}(1+m_{\pi}/2M)\approx150$~MeV, thereby assuming that the
scattering amplitude is real in this region. In the next step, one has to study
the high-energy behavior of the absorption cross sections. As predicted by
Regge theory and also seen by the data, the total absorption cross section
increases at the highest energies reached by the experiment. Although this
increase is not expected to continue forever, we can not expect that the
unsubtracted dispersion integral converges, and therefore we subtract $f(\nu)$
at $\nu=0$ and identify  $f(0)$ with the classical Thomson amplitude. By use of
the crossing relation and the optical theorem, the subtracted dispersion
relation takes the form
\begin{equation}\label{eq:6.1.6}
f(\nu) = f(0)+\frac{\nu^2}{2\pi^2}\,{\mathcal{P}}\,
\int_{\nu_0}^{\infty}\,\frac{\sigma_T(\nu')} {\nu'^2-\nu^2}\,d\nu'\ .
\end{equation}
For the odd function $g(\nu)$ we assume the existence of an unsubtracted
dispersion relation,
\begin{equation}\label{eq:6.1.7}
g(\nu) = \frac{\nu}{4\pi^2}\,{\mathcal{P}}\,
\int_{\nu_0}^{\infty}\,\frac{\sigma_{1/2}(\nu')-\sigma_{3/2}(\nu')}
{\nu'^2-\nu^2}\,\nu'd\nu'\ .
\end{equation}
If these dispersion integrals exist, they can be expanded as a Taylor series in
$\nu^2$, which converges for $|\nu|< \nu_0$. Comparing these power series to
the low energy theorems of Eqs.~(\ref{eq:4.1.5}) and (\ref{eq:4.1.6}), we
obtain the sum rule of \textcite{Baldin:1960dd},
\begin{equation}\label{eq:6.1.10}
\alpha_{E1} + \beta_{M1} = \frac{1}{2\pi^2}\,
\int_{\nu_0}^{\infty}\,\frac{\sigma_T(\nu')}{\nu'^2} \,d\nu'\ ,
\end{equation}
the sum rule of \textcite{Gerasimov:1965dd,Gerasimov:1965dda}
and \textcite{Drell:1966jv},
\begin{equation}\label{eq:6.1.11}
\frac{\pi e^2\kappa^2_N}{2M^2}= \int_{\nu_0}^{\infty}\,\frac{\sigma_{3/2}(\nu')
-\sigma_{1/2}(\nu')}{\nu'}\,d\nu' \, \equiv I_{\text{GDH}} \, ,
\end{equation}
and the forward spin polarizability \cite{Gell-Mann:1954db},
\begin{equation}\label{eq:6.1.12}
\gamma_0= \,-\,\frac{1}{4\pi^2}\,\int_{\nu_0}^{\infty}\,
\frac{\sigma_{3/2}(\nu')-\sigma_{1/2}(\nu')} {\nu'^3}\,d\nu'\ .
\end{equation}
\subsubsection{Photoabsorption cross sections for the proton}
\label{sec:VI.1.2}
The total photoabsorption cross section $\sigma_T$ in the resonance region of
the proton is shown by Fig.~\ref{fig:SUM_sigmaT}. This figure displays 3
resonance peaks on top of a large background. Above the resonance region,
$\sigma_T$ is slowly decreasing towards a minimum of about 115~$\mu$b at
$W\approx 10 $~GeV. At the highest energies, $W \approx 200$~GeV (corresponding
to $\nu \simeq 2 \cdot 10^4$~GeV), the experiments show a slow increase with
energy of the form $\sigma_T \sim W^{0.2}$ \cite{Derrick:1994dt,Aid:1995bz}, in
accordance with Regge parametrizations through a soft pomeron exchange
mechanism \cite{Cudell:1999tx}. Given this information, the rhs of
Eq.~(\ref{eq:6.1.10}) can be constructed, with the most recent numerical result
given by Eq.~(\ref{eq:4.1.7a}). In this way Baldin's sum rule provides a rather
precise value for the sum of the 2 scalar polarizabilities, which serves as an
important constraint for the analysis of
Compton scattering.\\

\begin{figure}[]
\begin{center}
\includegraphics[width=0.75\columnwidth,angle=90]{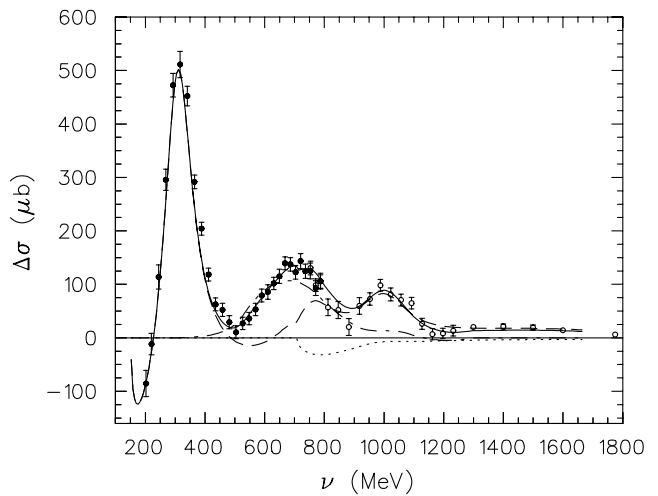}
\end{center}
\caption{The helicity difference $\Delta\sigma = \sigma_{3/2}-\sigma_{1/2}$ for
the proton as function of the photon energy $\nu$. The experimental data are
from MAMI \cite{Ahrens:2000bc, Ahrens:2001qt} (full circles) and
ELSA \cite{Dutz:2003mm, Dutz:2005ns} (open circles). The various lines
represent MAID results \cite{Drechsel:1998hk} for the total helicity difference
(solid line), one-pion channels (dashed line), more-pion channels
(dashed-dotted line), and $\eta$ channel (dotted line). Figure from
\textcite{Drechsel:2004ki}.} 
\label{fig:SUM_sigmaTT}
\end{figure}

During the past years also the helicity difference $\sigma_{TT}$ has been
measured. The pioneering experiment was carried out by the GDH Collaboration at
MAMI for photon energies between 200 and 800~MeV
\cite{Ahrens:2000bc,Ahrens:2001qt}, and then extended into the energy range up
to 3~GeV at ELSA  \cite{Dutz:2003mm,Dutz:2005ns}. These data allow us to verify
the GDH sum rule for the proton within an accuracy of less than 10~\%. Because
the integral for the forward spin polarizability converges much better, it is
essentially saturated by the MAMI data at 800~MeV. As shown in
Fig.~\ref{fig:SUM_sigmaTT}, the helicity difference fluctuates much more
strongly than the total cross section $\sigma_T$. The threshold region is
dominated by s-wave pion production, i.e., intermediate states with spin
$\textstyle{\frac {1}{2}}$ that can only contribute to the cross section
$\sigma_{1/2}$. In the region of the $\Delta (1232)$ with spin
$J=\textstyle{\frac {3}{2}}$, both helicity cross sections contribute, but
since the transition is essentially $M1$, we find the ratio
$\sigma_{3/2}/\sigma_{1/2}\approx3$, and therefore the helicity difference
becomes large and positive. The figure also shows that $\sigma_{3/2}$ dominates
in the second and third resonance regions. It was in fact an early success of
the quark model to understand this feature by a cancelation of the spin and
convection currents for $\sigma_{1/2}$. The data at the higher energies
indicate a fourth resonance region (1800~MeV$<W<2000$~MeV) followed by a
continuing decrease of $\Delta\sigma$ with a cross-over to negative values at
$\nu\gtrsim 2.0$~GeV, as predicted by an extrapolation of data from deep
inelastic scattering~\cite{Bianchi:1999qs,Simula:2001iy}. At high $\nu$, above
the resonance region, one usually invokes Regge phenomenology to argue that the
integral converges \cite{Bass:1999fh}. In particular, one obtains for the
isovector channel $\sigma_{1/2} - \sigma_{3/2} \to \nu^{\alpha_V - 1}$ at large
$\nu$, with $-0.5 \lesssim \alpha_V \lesssim 0$ being the intercept of the
$a_1(1260)$ meson Regge trajectory. For the isoscalar channel, Regge theory
predicts a similar energy behavior with $\alpha_S \simeq - 0.5$, which is the
intercept of the isoscalar $f_1(1285)$ and $f_1(1420)$ Regge trajectories.
However, these ideas have still to be tested experimentally. We observe that
the large background of non-resonant photoproduction in $\sigma_T$
(Fig.~\ref{fig:SUM_sigmaT}) has almost disappeared in the helicity difference
$\Delta\sigma$ (Fig.~\ref{fig:SUM_sigmaTT}), i.e., the background is ``helicity
blind''. As a result the two helicity cross sections for real photons remain
large and nearly equal up to the highest energies, at values of
$\sigma_{1/2}\approx\sigma_{3/2}\approx120~\mu$b. We conclude that the real
photon is essentially absorbed by coherent processes, which require
interactions among the constituents such as gluon exchange between two quarks.
This behavior differs from DIS, which refers to incoherent scattering off the
constituents. As a consequence the ratio $\sigma_{3/2}/ \sigma_{1/2}$ tends to
zero with increasing virtuality $Q^2$, because the absorption on an individual
quark leads only to final states
with helicity $\textstyle{\frac {1}{2}}$.\\

\begin{figure}[h]
\begin{center}
\includegraphics[width=0.95\columnwidth,angle=0]{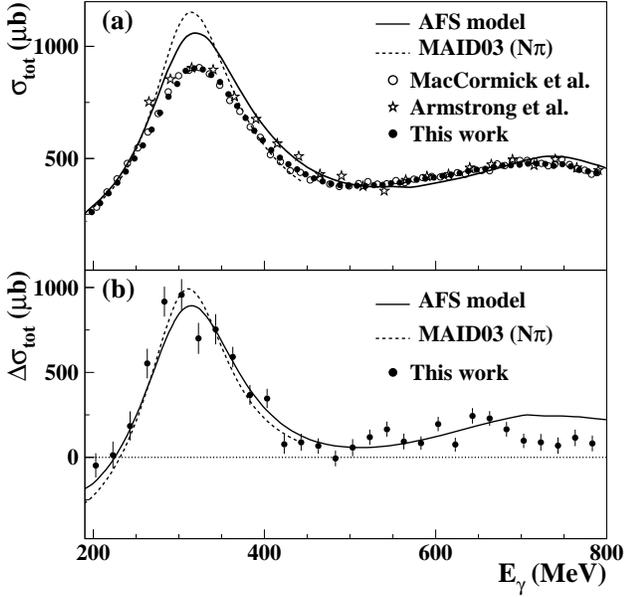}
\end{center}
\caption{The total photoabsorption cross section $\sigma_T$ (upper panel) and
the helicity difference $\Delta\sigma = \sigma_{3/2}-\sigma_{1/2}$ (lower
panel) for the deuteron. The experimental data are from
\textcite{Ahrens:2006yx} (full circles),  \textcite{MacCormick:1997ek} (open
circles), and \textcite{Armstrong:1971ns} (asterisks). The theoretical
predictions are from MAID03 (dashed lines) and \textcite{Arenhovel:2004ha}
(solid lines). Figure from \textcite{Ahrens:2006yx}.}
\label{fig:SUM_sigmaTT_deut}
\end{figure}
As shown in Fig.~\ref{fig:SUM_sigmaTT_deut}, the GDH Collaboration has also
measured the helicity difference for the deuteron \cite{Ahrens:2006yx}. The
upper panel of this figure displays the total cross section, which yields a
considerably smaller resonance peak than predicted for free nucleons (MAID) and
also by the dynamical model of \textcite{Arenhovel:2004ha}. However, the
measured helicity difference (lower panel) agrees quite well with both models.
The experiment has been continued to energies up to $W=1.8$~GeV at
ELSA \cite{Dutz:2005ns}, yielding positive values of typically 50~$\mu$b above
the resonance region. This contrasts the proton result, which turns out
negative in this region. Both experimental findings agree with the prediction
of Regge theory that the asymptotic tail of $\Delta\sigma$ should be positive
for the neutron and negative for the
proton \cite{Bianchi:1999qs,Simula:2001iy}.
\subsubsection{The GDH sum rule}
\label{sec:VI.1.3}
The GDH sum rule, Eq.~(\ref{eq:6.1.11}), relates the anomalous magnetic moment
(amm) of a particle to an energy-weighted integral over the helicity-dependent
photoabsorption cross sections. This relation bears out that a finite value of
the amm requires the existence of an excitation spectrum, and that both
phenomena are different aspects of a particle with intrinsic degrees of
freedom. A further property of composite objects is their spatial extension in
terms of size and shape, which reveals itself through the form factors measured
by elastic lepton scattering. In conclusion, the discovery of the proton's
large amm by \textcite{Stern:1933dd} marked the beginning of hadronic physics.
The experiment indicated that the proton was a microcosm in itself, and in this
sense the findings of Stern and collaborators were revolutionary. Today,
seventy-five years later, we are still struggling to describe the structure of
the strongly interacting particles in a quantitative way. At this point the
reader may well ask why should the GDH sum rule exist and what do we learn from
it. In fact it was pointed out many years ago that the GDH sum rule holds at
leading order in perturbation theory for the standard model of electroweak
interactions \cite{Altarelli:1972nc}. Later on this result was generalized to
any $2\rightarrow2$ process in supersymmetric and other field
theories \cite{Brodsky:1995fj}. The essential criterion is that these theories
start from point-like particles, and then the GDH sum rule should hold order by
order in the coupling constant. As an example, the GDH sum rule has been proven
in QED up to ${\mathcal{O}}(e^6)$ \cite{Dicus:2000cd}. In passing we note that
the GDH sum rule was also investigated in quantum gravity to one-loop
order \cite{Goldberg:1999gc}. The result was a violation of the sum rule, which
may however be due to our ignorance of quantum gravity in the strong coupling
(high energy) limit.\\

The amm of a particle is defined by the following relation between the total
magnetic moment $\vec{\mu}$ and the spin $\vec{S}$:
\begin{equation}\label{eq:6.1.13}
\vec{\mu} = \frac{e}{M}\,(Q+\kappa)\,\vec{S}\, ,
\end{equation}
with $eQ$ the charge and $M$ the mass of the particle. We also recall that the
ratio between the magnetic moment $\vec{\mu}$ and the orbital angular momentum
$\vec{L}$ of a uniformly charged rotating body is $eQ/2M$, whereas
Eq.~(\ref{eq:6.1.13}) yields $eQ/M$ as ratio between the ``normal'' magnetic
moment $\vec{\mu}$ and the spin $\vec{S}$, because of the gyromagnetic ratio
g=2 predicted by Dirac's equation for a spin $\frac{1}{2}$ particle. Contrary
to the conjecture by \textcite{Belinfante:1953dd} that g=1/S, the ``natural''
value of the gyromagnetic ratio is g=2 for every point particle, independent of
its spin. This is necessary if one insists on a well-behaved high-energy
scattering amplitude and a reliable perturbative expansion
\cite{Weinberg:1970dd, Ferrara:1992yc}, in the sense that any deviation from
this value must be related to finite size effects. Such spatially extended
phenomena, however, do not affect the high-energy limit of Compton scattering.
In particular \textcite{Brodsky:1968ea} verified the GDH sum rule for a
composite system of any spin on the basis of the spin-dependent interaction
currents associated with the cm motion. To further illustrate this point let us
consider the small amm of the electron, which can be evaluated in QED to 10
decimal places. Since the associated photon-electron loops are spread over a
spatial volume of about $10^6$~fm$^3$, a high-energy photon of a few hundred
MeV or a wavelength $\lambda\lesssim1$~fm will decouple from such a large
volume. Therefore the amm does not affect the high-energy limit and, along the
same lines, the amm of the electron does not keep us back from using the
electron as an ideal point particle to study the form factors of the
nucleon \cite{Drechsel:1991dd}.\\

The lhs of the GDH sum rule, Eq.~(\ref{eq:6.1.11}), yields
$I_{\text{GDH}}^p=205~\mu$b and $I_{\text{GDH}}^n=233~\mu$b for proton and
neutron, respectively. However, the first estimates based on the then existing
photoproduction data led to $261~\mu$b for the proton and $183~\mu$b for the
neutron \cite{Karliner:1973em}. Over the following years the predictions moved
even further away from the sum rule values in spite of an improving data basis,
simply because these data were not sensitive to the helicity difference of the
inclusive cross sections. Many explanations for an apparent violation of the
sum rule followed, but in view of the new experimental evidence we may safely
discard these ideas. Table~\ref{tab:GDH} summarizes our present knowledge on
the GDH integral and the forward spin polarizability of the proton. The
threshold contribution for $\nu \le 0.2$~GeV is evaluated by the MAID multipole
analysis of pion photoproduction \cite{Drechsel:1998hk}, with an error
estimated by comparing to the SAID analysis \cite{Arndt:2002xv}. The resonance
region up to $\nu=2.9$~GeV is determined by the experimental data taken at MAMI
\cite{Ahrens:2000bc, Ahrens:2001qt} and ELSA \cite{Dutz:2003mm,Dutz:2005ns},
and the asymptotic contribution is based on the Regge analysis of DIS
\cite{Bianchi:1999qs,Simula:2001iy}. Summing up these contributions, the GDH
sum rule value is obtained within the given error bars. Because of the
different energy weighting, the forward spin polarizability converges much
better and is therefore completely determined by the existing data.
\begin{table}[h]
\begin{tabular}{|l|cc|}
\hline
energy [GeV]& $I_{\text{GDH}}^p$ $[\mu$b]  & $\gamma_0^p$ $[10^{-4}$~fm$^4]$ \\
\hline
$\le$0.2&  $-28.5\pm2$  &  $0.95\pm0.05$\\
0.2-0.8 &  $226\pm5\pm12$  &  $-1.87\pm0.08\pm0.10$\\
0.8-2.9 &  $27.5\pm2.0\pm1.2$  & $-0.03$  \\
$\ge$2.9&  $-14\pm2$  &  $+0.01$ \\
\hline \hline
total   & $211\pm15$  &  $-0.94\pm0.15$ \\
\hline
sum rule&  204  &  -- \\
\hline
\end{tabular}
\caption{The contribution of various energy regions to the GDH integral
$I_{\text{GDH}}^p$ and the forward spin polarizability $\gamma_0^p$ of the
proton (see text for explanation). \label{tab:GDH}}
\end{table}
Because the helicity dependent cross section $\sigma_{TT}$ is strongly energy
dependent, the GDH integral is very sensitive to experimental errors. It is
therefore quite satisfying that all the decay channels were separately
identified in the range of 200~MeV $<\nu<$ 800~MeV. The one-pion channel opens
at $\nu_0\approx150$~MeV and dominates the cross section up to
$\nu\approx500$~MeV, except for small contributions due to radiative decay and
the onset of two-pion production. The helicity-dependent cross section for the
one-pion channel has the following multipole expansion \cite{Drechsel:1992pn}:
\begin{eqnarray}\
\sigma_{TT}^{\pi} & = & 4\pi \frac{k} {q}\, \bigg \{|E_{0+}|^2 -
|M_{1+}|^2 + 6\ \mbox{Re}\ (E_{1+}^{\ast}M_{1+})\nonumber \\
&& + 3|E_{1+}|^2 + |M_{1^-}|^2 - |E_{2-}|^2 \label{eq:6.1.14}\\
&& -6\ \mbox{Re}\,(E_{2-}^{\ast}M_{2-}) + 3|M_{2^-}|^2 \pm \ ... \bigg \}\,.
\nonumber
\end{eqnarray}
As we have seen, the threshold region was not covered by the experiment. The
dominant multipole in this region is $E_{0+}$, which corresponds to an electric
dipole (E1) transition leading to the production of (mostly charged) pions in
an s wave. This multipole is well described by pion photoproduction data at
threshold (see section~\ref{sec:V.1}), ChPT \cite{Bernard:1991rt}, dispersion
theory \cite{Hanstein:1996bd}, and phenomenological analysis
\cite{Drechsel:1998hk,Arndt:2002xv}. With increasing photon energy, the first
resonance becomes more and more dominant, mainly because of the magnetic dipole
transition to the $\Delta$(1232). Although the associated electric quadrupole
transition is strongly suppressed by the ratio $R_{EM}=E_{1^+}/M_{1^+}$, the
GDH experiment permits an independent measurement \cite{Ahrens:2004pf}, because
the product $E_{1+}^{\ast}M_{1+}$ appears with a factor~6 in
Eq.~(\ref{eq:6.1.14}). Altogether the MAMI data are in good agreement with the
multipole analysis in the first resonance region
\cite{Drechsel:1998hk,Arndt:2002xv}. However, even relatively small effects
count, because this region provides the lion's share to the sum rule, that is
about 175$~\mu$b between $\nu = 250$ and 450~MeV. In a similar way the
$N\rightarrow D_{13}\,(1520)$ transition was studied \cite{Ahrens:2002gu}. The
multipoles $E_{2^-}$ and $M_{2^-}$, E1 and M2 transitions, respectively, are
related to the helicity amplitudes of this resonance as follows:
\begin{equation}
A_{1/2}^{1520} \sim E_{2^-} - 3M_{2^-} \ ,\quad A_{3/2}^{1520} \sim \sqrt{3}
(E_{2^-}+M_{2^-})\ .
\end{equation}
The new data yield $A_{1/2}=-38\pm3$ and $A_{3/2}=147\pm10$, to be compared
with the listing of the PDG, $-24\pm9$ and $166\pm5$ \cite{Yao:2006px}, all in
units of $10^{-3}~\text{GeV}^{-1/2}$. The given examples demonstrate that
double-polarization experiments provide a very
sensitive tool to study resonance properties.\\

\begin{figure}[]
\begin{center}
\includegraphics[width=0.48\columnwidth,angle=90]{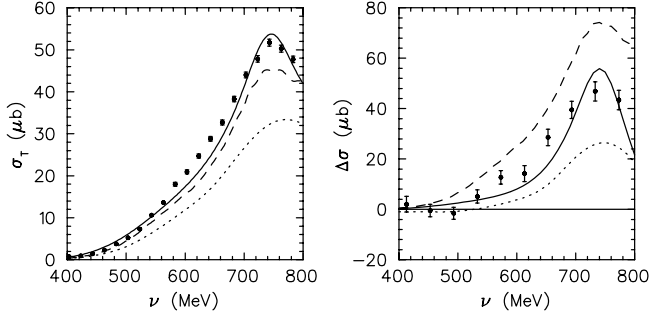}
\end{center}
\caption{The total cross section $\sigma_{T}$ and the helicity difference
$\Delta\sigma=\sigma_{3/2}-\sigma_{1/2}$ for the reaction $\vec{\gamma}
\vec{p}\rightarrow n\pi^+\pi^0$. The theoretical predictions are shown by solid
lines \cite{Hirata:2002tp}, dashed lines \cite{Nacher:2001yr}, and dotted
lines \cite{Holvoet:2001dd}. The data are from MAMI 
\cite{Ahrens:2003bp,Ahrens:2003na}. Figure completed by Lothar Tiator. 
\label{fig:SUM_two-pion}}
\end{figure}

Although the threshold for two-pion production lies already in the $\Delta$
region, these channels become important only for  $\nu \ge 500$~MeV. The
channels $n\pi^+\pi^0$, $p\pi^+\pi^-$, and $p\pi^0\pi^0$ were separately
analyzed at MAMI \cite{Ahrens:2003bp,Ahrens:2003na}. As an example,
Fig.~\ref{fig:SUM_two-pion} shows the cross sections for the reaction
$\vec{\gamma} \vec{p}\rightarrow n\pi^+\pi^0$, which also exhibits a clear
dominance of $\sigma_{3/2}$ over $\sigma_{1/2}$. The interesting and previously
unexpected feature is the peaking of the respective cross section at
$\nu\approx700$~MeV or $W\approx1480$~MeV, definitely below the positions of
the $D_{13}\,(1520)$ and $S_{11}\,(1535)$ resonances. This proves that two-pion
production can not be simply explained by a resonance driven mechanism as was
assumed in the earlier estimates for the sum rule. The $\eta$ channel provides
another interesting contribution to the sum rule. This channel is dominated by
the resonance $S_{11}\,(1535)$, which has an exceptionally large branching
ratio of about $50~\%$ for $\eta$ decay. Because this resonance has $\ell=0$,
it only contributes to the helicity cross section
$\sigma_{1/2}$ \cite{Ahrens:2003na}. For further information on the helicity
structure of the different channels we refer the reader to the review
by \textcite{Krusche:2003ik}.
The work of the GDH Collaboration at ELSA and MAMI has recently been summarized
by \textcite{Helbing:2006zp}. Several other ongoing activities or proposals for
future experiments have been reported at conferences by the LEGS Collaboration
at BNL, several groups at the GRAAL and Spring-8 facilities, the CLAS
Collaboration at JLab, and the E-159 collaboration at SLAC. Such further
experiments will be invaluable as independent tests of the sum rules, in
particular at the higher energies in order to probe the soft Regge physics in
the spin-dependent forward Compton amplitude.\\

\begin{table}[]
\begin{tabular}{|l|ccccccc|}
\hline
& $ e $ & $p$  & $n$  & $d$   &  $^3$He & $^1$H & C$_4$H$_9$OH \\
\colrule
$\mu$ & -1838 & 2.79 & -1.91 & 0.86 & -2.13 & -1836 & -2$\cdot10^4$  \\
$\kappa$ & $-1.2\cdot10^{-3}$ & 1.79 & -1.92 & -0.14 & -8.37 & -918 & -7$\cdot10^4$ \\
$I_{\text{GDH}}$  & 289 & 205 & 233 & 0.65 & 498 & $10^8$ & $10^9$ \\
\hline
\end{tabular}
\caption{The magnetic moment $\mu$ in units of the nuclear magneton $\mu_N$,
the amm $\kappa$, and the GDH sum rule $I_{\text{GDH}}$ in units of $\mu$b for
electrons, protons, neutrons, deuterons, and $^3$He nuclei as well as fully
polarized hydrogen atoms and butanol molecules.} \label{tab:SUM_kappa}
\end{table}
Table~\ref{tab:SUM_kappa} shows the amm for the nucleons and standard ``neutron
targets''. The most striking observation is the tiny value of $\kappa$ for the
deuteron, a loosely bound proton-neutron system, which has isospin $I=0$ and
spin $S=1$ and is essentially in a relative s state. The interplay between
nuclear and subnuclear degrees of freedom in the deuteron has been studied in
great detail by \textcite{Arenhovel:2004ha}. It turns out that the most
important nuclear channel is deuteron disintegration, $\gamma+d\rightarrow
p+n$, which yields a maximum value of $\sigma_P-\sigma_A\approx-1800~\mu$b at
$\nu\approx2.3$~MeV. This huge helicity asymmetry is due to the M1 transition
$^3S_1\rightarrow ^1S_0$, which changes the magnetic moments of proton and
neutron from parallel to antiparallel. Due to the energy denominator $\nu$ in
the GDH integral, it is precisely the small excitation energy of the weakly
bound deuteron that provides this large negative contribution to the GDH
integral. To the contrary, the $N\rightarrow\Delta$ transition aligns the quark
spins and peaks at $\nu\approx330$~MeV with a maximum value of
$\sigma_P-\sigma_A\approx1100~\mu$b for free nucleons. As a result, the large
negative contribution of deuteron break-up is canceled by large positive
contributions of the subnuclear degrees of freedom, and this happens to three
decimal places. The other neutron target is $^3$He, a system of two protons
with spins paired off and an ``active'' neutron, essentially again in s states
of relative motion. As a result we find ${\mu\,}_{^3\rm{He}}\approx\mu_n<0$,
whereas Eq.~(\ref{eq:6.1.13}) predicts a ``normal'' moment of $\frac
{2}{3}\mu_N$. Therefore, the amm of $^3\rm{He}$
has a large negative value, which leads to a large and positive GDH integral.\\

The recent experiments to determine the GDH sum rule for the neutron,
$I_{\text{GDH}}^n$, have been performed with a frozen-spin deuterated butanol
(C$_4$D$_9$OD) target \cite{Ahrens:2006yx} and a frozen-spin $^6$LiD
target \cite{Dutz:2005ns}. Setting aside the problems on the molecular and
atomic levels, these experiments provide polarized deuterons whose
helicity-dependent response has been measured in the energy region 200~MeV$<
\nu <$ 800~MeV. As is obvious from the above discussion, a quantitative
extraction of $I_{\text{GDH}}^n$ is necessarily model-dependent. Even the
presently most elaborate calculation of  \textcite{Arenhovel:2004ha} misses the
GDH sum rule for the deuteron by nearly 30~$\mu$b and overestimates the total
photoabsorption in the region of the $\Delta$(1232). It is therefore mandatory
to also measure the different decay channels in order to constrain the
theoretical analysis. Such experiments are in progress, and the first results
for the one-pion channels are already available \cite{Ahrens:2006yx}. To lowest
order we may assume that $I_{\text{GDH}}^p$+$I_{\text{GDH}}^n$ should be given
by the GDH integral for the deuteron, if extended from pion threshold to
infinity. On the basis of the present analysis we may conclude that more than
60\% of this contribution is due to $\pi^0$ production and another third from
two-pion channels with at least one charged pion. Furthermore,
Fig.~\ref{fig:SUM_sigmaTT_deut} shows a reasonable agreement between the data
and the MAID model in the $\Delta$(1232) resonance, in which region the model
yields quite similar values for proton and neutron. The additional sum rule
strength for the neutron should therefore come from energies above the
$\Delta$(1232). And indeed, the ELSA experiment \cite{Dutz:2005ns} yields a
contribution of about 34$\mu$b between 815 and 1825~MeV, contrary to earlier
estimates. Furthermore, the integrand remains positive at the highest energies,
which could indicate a further contribution of about 40$\mu$b on the basis of
Regge models. In conclusion, the present experiments confirm the GDH sum rules
for proton and neutron, however with a very large systematical error in the
latter case. It is also likely that the isovector combination $I_{\text{GDH}}^p
- I_{\text{GDH}}^n$ turns out negative, as required by the
sum rule prediction of $-28\mu$b.\\

The interplay of nuclear and subnuclear degrees of freedom is based on general
principles like low-energy theorems and dispersion relations. A complete answer
to the remaining questions calls for experiments covering both the nuclear and
the subnuclear energy range. It is therefore very promising that programs are 
being developed for energies between nuclear breakup and pion threshold at
TUNL/HI$\gamma$S (Duke) \cite{Weller:2002dd}, both for the nuclear physics
aspects by themselves and as a test of many-body calculations that are
inevitably required for further studies of the GDH sum rule of the neutron. We
conclude this subsection with a possibly academic but still interesting
question. Although we have talked about nucleons and nuclei as targets, we
could just as well think about projects to measure the GDH sum rule for atoms
or molecules. Table~\ref{tab:SUM_kappa} shows the amm $\kappa$ and the GDH sum
rule $I$ for several such systems. The comparison of the different hierarchies
is quite amusing. If one tried to reconstruct, for example, the amm of the
hydrogen atom by a GDH integral over its atomic spectrum in the eV or keV
region, one would find physics ``beyond'': the physics of $e^+e^-$ pair
production at the MeV scale (QED) and hadronic physics above pion threshold
(QCD).
\subsection{Sum rules for virtual photons}
\label{sec:VI.2}
\begin{figure}[]
\begin{center}
\includegraphics[width=0.5\columnwidth,angle=0]{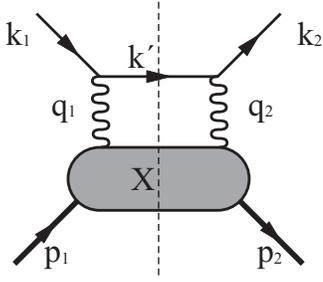}
\end{center}
\caption{The two-photon exchange diagram. The blob represents the possible
intermediate states of the hadronic system.} \label{fig:SUM_two_gamma_graph}
\end{figure}

Doubly-virtual Compton scattering (VVCS) offers a useful framework to study
generalized GDH integrals and polarizabilities \cite{Ji:1993mv}. This process
is based on the idea that an incident virtual photon with definite energy $\nu$
and virtuality $Q^2>0$ (space-like) hits a nucleon and excites this hadronic
system, which eventually decays into a nucleon and an outgoing virtual photon
with the same value of $Q^2$. Although this reaction can not be realized
experimentally, it is not merely a theoretical construct. As pointed out many
years ago, the imaginary part of this amplitude can be obtained by elastic
scattering of transversely polarized electrons off unpolarized
targets~\cite{DeRujula:1972te}. The asymmetry with regard to changing the
transverse polarization is parity conserving but time-reversal odd and
therefore vanishes in the one-photon exchange approximation. It appears only at
sub-leading order as the product of the (real) Born amplitude for one-photon
exchange and the imaginary part of the two-photon exchange diagram shown by
Fig.~\ref{fig:SUM_two_gamma_graph}. This imaginary part is related to the VVCS
tensor, however the experiment can only determine a weighted integral over a
range of virtualities $Q^2$, whereas the VVCS tensor refers to a fixed value of
$Q^2$. A pioneering experiment to measure this asymmetry has been performed at
MIT/Bates \cite{Wells:2000rx}, and more data are now available from MAMI
\cite{Maas:2005au,Maas:2004pd}. These measurements show that even for moderate
virtualities, $Q^2\approx 0.1$~GeV$^2$, the asymmetry is dominated by the
excited states of the nucleon, notably the $\Delta(1232)$ resonance. The same
physical effect can also be observed for other single-spin asymmetries, such as
the transverse target asymmetry or the normal recoil polarization for an
unpolarized electron beam.
\subsubsection{VVCS and nucleon structure functions}
\label{sec:VI.2.1}
The absorption of a virtual photon on a nucleon $N$ is described by inclusive
electroproduction, $e+N\rightarrow e'+X$. The cross section for this reaction
takes the form  \cite{Drechsel:2000ct},
\begin{eqnarray}\label{eq:6.2.1}
\frac{d\sigma}{d\Omega_2 d\epsilon_2}& =& \Gamma_V \sigma (\nu,Q^2) \,,\\
\sigma &=& \sigma_T+\varepsilon \sigma_L -hP_x
\sqrt{2\varepsilon(1-\varepsilon)}\, \sigma_{LT}\nonumber \\
&-& hP_z \sqrt{1-\epsilon^2}\, \sigma_{TT}\, . \label{eq:6.2.2}
\end{eqnarray}
Comparing this form with Eqs.~(\ref{eq:2.22}) and (\ref{eq:2.23}), we find that
the transverse ($\sigma_T$) and longitudinal ($\sigma_L$) absorption cross
sections are obtained by integration of the respective differential cross
sections over the angles of the emitted pion, and in general by integrating
over all angles and energies of the produced particles. The two (inclusive)
spin-flip cross sections, $\sigma_{TT}$ and $\sigma_{LT}$ can only be measured
with polarized electrons (helicity $h = \pm 1$) and target polarization in the
direction of the virtual photon momentum ($P_z$) and perpendicular to that
direction in the scattering plane ($P_x$), respectively, or equivalent recoil
polarizations. In order to avoid a possible misunderstanding, we note that
these inclusive spin-flip cross sections are not related to the differential
cross sections $d\sigma_{TT}/d\Omega_{\pi}^{\ast}$ and
$d\sigma_{LT}/d\Omega_{\pi}^{\ast}$ of Eq.~(\ref{eq:2.23}) but obtained from
double-polarization cross sections not shown in that equation. The following
discussion concentrates on the spin-flip cross sections, which are related to
the spin-dependent nucleon structure functions $g_1$ and $g_2$
by~\cite{Drechsel:2002ar}
\begin{eqnarray}\label{eq:6.2.3}
\sigma_{TT} &=& \frac{4\pi^2\alpha_\text{em}}{M K}(g_1-\gamma^2 g_2 )\\
\sigma_{LT} &=& \frac{4\pi^2\alpha_\text{em}}{M
K}\gamma\,\left(g_1+g_2\right)\, ,\label{eq:6.2.4}
\end{eqnarray}
with $\gamma = Q/\nu$ and $K$ the photon equivalent energy defined by
Eq.~(\ref{eq:2.24}). The nucleon structure functions are usually expressed as
functions of the Bjorken variable, $x=Q^2/(2M \nu)$, i.e.,
$g_{1,2}=g_{1,2}(x,Q^2)$.
\\

The VVCS amplitude for forward scattering of virtual photons generalizes
Eq.~(\ref{eq:4.1.4b}) by introducing an additional longitudinal polarization
vector $\hat{q}$,
\begin{eqnarray}
T(\nu,\,Q^2,\,\theta=0)&=& \vec
{\varepsilon}\,'^{\ast}\cdot \vec {\varepsilon}\,f_T(\nu,Q^2)+f_L(\nu,Q^2) \nonumber\\
&+& \;i\vec{\sigma}\cdot(\vec{\varepsilon}\,'^{\ast}\times\vec
{\varepsilon}\,)\, g_{TT}(\nu,\,Q^2) \label{eq:6.2.5} \\
& + &i (\vec{\varepsilon}\,'^{\ast}-\vec {\varepsilon}\,)\cdot(\vec{\sigma}
\times \hat{q})g_{LT}(\nu,Q^2) \ .\nonumber
\end{eqnarray}
Because the total amplitude is crossing-even, the amplitude $g_{TT}(\nu)$ is an
odd function of $\nu$ whereas $g_{LT}(\nu)$ is even. In order to set up
dispersion relations, we have to construct the imaginary parts of the
amplitudes with contributions from both elastic and inelastic scattering. The
elastic contributions are obtained from the direct and crossed Born diagrams
with nucleons in the intermediate states and expressed by the nucleon form
factors of section~\ref{sec:II.2},
\begin{eqnarray}
g_{TT}^{\text{el}}& = & -\frac{\alpha_\text{em} \nu}{2M^2} \bigg ( F_2^2 +
\frac{Q^2}{\nu^2-\nu_B^2+i \varepsilon} \, G_M^2 \bigg ) \, ,\label{eq:6.2.6} \\
g_{LT}^{\text{el}} & = & \frac{\alpha_\text{em} Q}{2M^2} \bigg ( F_1F_2 -
\frac{Q^2}{\nu^2-\nu_B^2+i \varepsilon} \,G_EG_M \bigg ) \, ,\nonumber
\end{eqnarray}
with $\nu_B=Q^2/2M$. We note that the amplitudes of Eq.~(\ref{eq:6.2.6}) have
been split in a real contribution and a complex term containing the nucleon
poles at $\nu = \pm \nu_B \mp i\varepsilon$. The inelastic contributions are
regular functions in the complex $\nu$-plane except for cuts from $-\infty$ to
$-\nu_0$ and $+\nu_0$ to $+\infty$. The optical theorem relates the inelastic
contributions to the partial cross sections of inclusive electroproduction,
\begin{eqnarray}\label{eq:6.2.7}
{\mbox{Im}}\ g_{TT}(\nu,Q^2) &=&\frac{K}{4\pi}\, \sigma_{TT}(\nu,Q^2) \, , \\
{\mbox{Im}}\ g_{LT}(\nu,Q^2) &=& \frac{K}{4\pi}\, \sigma_{LT}(\nu,Q^2) \,,
\label{eq:6.2.8}
\end{eqnarray}
where the products $K\,\sigma_{TT/LT}$ are independent of the choice of $K$,
because they are directly proportional to the measured cross section. Comparing
the above equations with the results for real photons in
section~\ref{sec:VI.1}, we find distinct differences. In particular, the
transition from real to virtual photons is not straightforward, because the
limits $Q^2\rightarrow 0$ and $\nu\rightarrow 0$ can not be
interchanged~\cite{Ji:1993mv}. This happens because the Born amplitudes of VVCS
have poles, which also provide imaginary contributions to the amplitudes. In
more physical terms, the crucial difference between the (space-like) virtual
and the real photon is that the former can be absorbed by a charged particle
whereas the latter can only be absorbed at zero frequency, $\nu=0$. At this
point, however, the real photon amplitudes can be expanded in a Taylor series
whose leading terms are determined by the Born terms. In particular,
$g_{TT}^{\text{el}}(\nu,0)$ reproduces exactly the leading term of the
spin-flip amplitude $g(\nu)$ of Eq.~(\ref{eq:4.1.6}). The GDH sum rule is then
obtained by equating this power series to an expansion of the dispersion
integral over the imaginary parts from the inelastic processes for $\nu>\nu_0$.
The virtual photon case differs in two aspects: The imaginary parts stem from
both elastic and inelastic processes, and the real parts of the amplitudes have
two poles in the $\nu$ plane.
\subsubsection{Dispersion relations and sum rules}
\label{sec:VI.2.2}
We next turn to the sum rules for the spin dependent VVCS amplitudes.
Subtracting the pole terms from the full amplitude, we obtain the following
dispersion relation:
\begin{eqnarray}\label{eq:6.2.9}
{\mbox{Re}}~g_{TT}^\text{disp}(\nu,Q^2)& = &{\mbox{Re}}~\bigg
(g_{TT}(\nu,Q^2)-g_{TT}^{\text{pole}}(\nu,Q^2)\bigg )\\
 & = &
\frac{\nu}{2\pi^2}\,{\mathcal{P}}
\int_{\nu_0}^{\infty}\frac{K(\nu',Q^2)\,\sigma_{TT}(\nu',Q^2)} {\nu'^2-\nu^2}
\, d\nu' \, .\nonumber
\end{eqnarray}
Because this amplitude is regular below the first threshold, $\nu = \nu_0$, it
can be expanded in a Taylor series at $\nu=0$. The result is
\begin{multline}\label{eq:6.2.10}
{\mbox{Re}}~g_{TT}^\text{disp}(\nu,Q^2)=\frac{2 \alpha_\text{em}}{M^2}\,
I_{TT}(Q^2)\nu \\ +\gamma_{TT}(Q^2)\nu^3+\ldots \,
\end{multline}
where
\begin{eqnarray}\label{eq:6.2.11}
I_{TT}(Q^2) &=& \frac{M^2}{\pi e^2} \int_{\nu_0}^{\infty} \frac{K(\nu,
 Q^2) \, \sigma_{TT}(\nu , Q^2)}{\nu^2} \, d\nu \, , \\
\gamma_{TT}(Q^2) &=& \frac{1}{2\pi^2} \int_{\nu_0}^{\infty} \frac{K(\nu, Q^2)
\, \sigma_{TT}(\nu,Q^2)}{\nu^4} \, d\nu \,. \label{eq:6.2.12}
\end{eqnarray}
Comparing these expressions with Eq.~(\ref{eq:4.1.6}), we find $I_{TT} (0) =
-\kappa^2/4$ and $\gamma_{TT} (0) = \gamma_0$. The corresponding equations for
the crossing-even amplitude $g_{LT}$ are
\begin{eqnarray}\label{eq:6.2.13}
I_{LT}(Q^2) & = & \frac{M^2}{\pi e^2} \int_{\nu_0}^{\infty}\frac{K(\nu,Q^2) \,
\sigma_{LT} (\nu,Q^2)}{Q \nu} \, d\nu \,, \\
\delta_{LT}(Q^2) &=& \frac{1}{2\pi^2} \int_{\nu_0}^{\infty}\frac{K(\nu,Q^2)\,
\sigma_{LT}(\nu,Q^2)} {Q \nu^3}\, d\nu \,. \label{eq:6.2.14}
\end{eqnarray}
Both functions are finite in the real photon limit, because the factor
$\sigma_{LT}/Q$ in the integrand is finite for $Q^2\rightarrow 0$. Furthermore,
it follows from Eq.~(\ref{eq:6.2.4}) that $\sigma_{LT}$ can be replaced by $g_1
+ g_2$, the sum of the spin-dependent structure functions. In the limit of
large $Q^2$, \textcite{Wandzura:1977qf} have shown that $g_1+g_2$ can be
expressed in terms of the twist-2 spin structure function $g_1$ if the
dynamical (twist-3) quark-gluon correlations are neglected,
\begin{equation}\label{eq:6.2.15}
g_1(x, Q^2)+ g_2(x, Q^2) = \int_{x}^{1} dy \, \frac{g_{1}(y,Q^2)}{y} \, .
\end{equation}
In order to compare with the notation of DIS, we define the {\emph {inelastic}}
contributions to the first moments of the spin structure functions,
\begin{equation}\label{eq:6.2.16}
\Gamma_{1,2}^{\text{inel}}(Q^2) = \int_0^{x_0}g_{1,2}(x,\,Q^2)\,dx = \frac
{Q^2} {2M^2}I_{1,2}(Q^2)  \, ,
\end{equation}
where the integration runs over the Bjorken variable from $x=0$ (or $\nu
\rightarrow \infty$) to $x=x_0$ (or $\nu=\nu_0$). Because the above integrals
include the excited spectrum only, one has to add the elastic contribution in
order to obtain the first moment of the structure functions. At small values of
$Q^2$ the full structure functions are dominated by the elastic contributions,
which can be constructed once the nucleon form factors are known. With
increasing resolution $Q^2$ the coherent response of the many-body system
``nucleon'' decreases, whereas incoherent scattering processes on individual
constituents become more and more important. In particular, the elastic
contributions to the nucleon structure functions vanish like
$Q^{-10}$.\\

There exist two venerable sum rules for the moments of the structure functions.
The first one was originally derived from current algebra and, therefore, is
also a prediction of QCD. It deals with the isovector combination of the first
spin structure function in the limit $Q^2 \rightarrow \infty$
\cite{Bjorken:1966jh,Bjorken:1969mm},
\begin{eqnarray}
\Gamma_1^p(Q^2) &-& \Gamma_1^n(Q^2) = \int_0^{1}\bigg (
g_1^p(x,\,Q^2)-g_1^p(x,\,Q^2)
\bigg )\,dx \nonumber \\
&\rightarrow& \frac {1}{6}g_A \bigg \{
1-\frac{\alpha_s(Q^2)}{\pi}+\mathcal{O}(\alpha_s^2, \frac {M^4}{Q^4})\bigg
\}\,, \label{eq:6.2.17}
\end{eqnarray}
with $g_A$ the axial-vector coupling constant and $\alpha_s$ the running
coupling constant of the strong interaction. The other sum rule is a prediction
for the second spin structure function \cite{Burkhardt:1970ti},
\begin{equation}
\Gamma_2^N(Q^2) = \int_0^{1}\, g_2^N(x,\,Q^2)\,dx =0 \, . \label{eq:6.2.18}
\end{equation}
The Burkhardt-Cottingham (BC) sum rule relies on a ``superconvergence
relation'' for the associated VVCS amplitude such that the dispersion integral
exists not only for the odd amplitude $S_2(\nu)$ but also for the even
amplitude $\nu S_2(\nu)$. As a result the sum of the elastic and inelastic
contributions should vanish. This allows us to cast Eq.~(\ref{eq:6.2.16}) in
the form
\begin{equation}\label{eq:6.2.19}
I_2^N(Q^2) = \frac{1}{4} \, F_P^N(Q^2) \, \left( F_D^N(Q^2) + F_P^N(Q^2)
\right) \, .
\end{equation}
Provided that the assumed convergence criterion is indeed given, the BC sum
rule connects the spin structure of the excitation spectrum with ground state
properties for each value of $Q^2$. \textcite{Tsai:1975tj} have proven that
the BC sum rule is fulfilled for QED to lowest order in the fine structure
constant $\alpha_\text{em}$. Along the same lines, this sum rule is also
fulfilled in perturbative QCD to first order in $\alpha_s$
\cite{Altarelli:1994dj}. Furthermore, it is consistent with the
Wandzura-Wilczek relation, as can be proven by integrating
Eq.~(\ref{eq:6.2.15}) over all values of the Bjorken parameter $x$. However,
the BC sum rule is predicted to be valid at any $Q^2$, whereas the
Wandzura-Wilczek relation neglects dynamical (twist-3) quark-gluon correlations
and higher order terms. The different integrals discussed above are connected
by Eqs.~(\ref{eq:6.2.3}) and (\ref{eq:6.2.4}). In particular, their values at
the real photon point can be expressed by the charge and the amm of the
nucleon,
\begin{eqnarray}\begin{array}{lllll}
I_1^N(0)& = & -\frac{1}{4}\kappa_N^2 \, ,
\quad I_2^N(0) &=& \frac{1}{4}\kappa_N (e_N + \kappa_N)\, , \label{eq:6.2.20} \\
\\
I_{TT}^N(0) & = & -\frac{1}{4}\kappa_N^2\, , \quad I_{LT}^N (0) & = &
\frac{1}{4} e_N \kappa_N \ .\label{eq:6.2.21} \end{array}
\end{eqnarray}
\subsubsection{The helicity structure of the cross sections}
\label{sec:VI.2.3}
The helicity-dependent cross sections $\sigma_{TT}(\nu, Q^2)$ and
$\sigma_{LT}(\nu, Q^2)$ determine the GDH-like integrals and the
polarizabilities defined above. For momentum transfers
$Q^2\lesssim0.5$~GeV$^2$, the bulk contribution to these cross sections stems
from one-pion production, which is reasonably well known over the resonance
region. The threshold region is dominated by s-wave production ($E_{0+},\
S_{0+}$) accompanied by much smaller contributions of the p waves ($M_{1\pm},\
E_{1\pm},\ S_{1\pm}$). Low-energy theorems, the predictions of ChPT, and
several new precision experiments have provided a solid basis for the multipole
decomposition in that region. The data basis is also quite reliable in the
first resonance region. Although the leading $M_{1+}$ multipole drops with
$Q^2$ somewhat faster than the dipole form factor, it dominates that region up
to large momentum transfers. In the higher resonance regions, the multipole
decomposition is known only semi-quantitatively. In particular, there is as yet
little reliable information on $\sigma_{LT}$, except that it is generally
small. This fact is not really consoling in the context of the sum rules,
because the integral $I_{LT}$ of Eq.~(\ref{eq:6.2.13}) does not converge well.
However, great improvements in the data basis are expected from the
wealth of ongoing and planned polarization experiments.\\

\begin{figure}[]
\begin{center}
\includegraphics[width=0.75\columnwidth,angle=90]{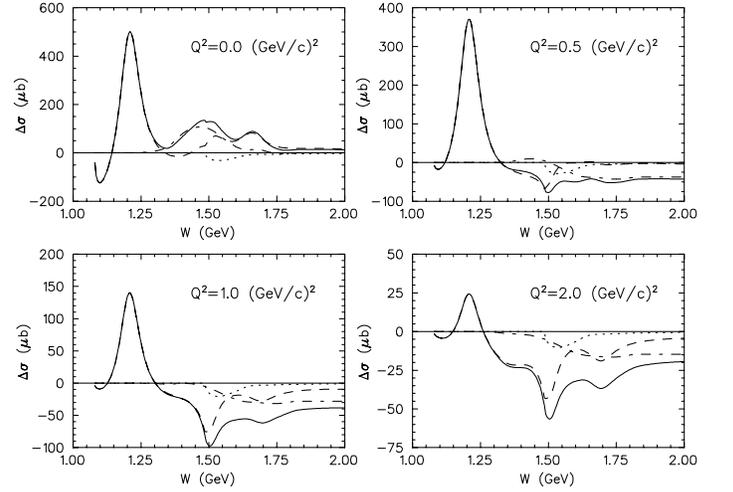}
\end{center}
\vspace{0cm} \caption{The helicity difference $\Delta\sigma
=\sigma_{3/2}-\sigma_{1/2}$ for the proton as function of the cm energy $W$ for
different values of $Q^2$ \cite{Drechsel:1998hk}. The figure shows the total
helicity difference (solid line) as well as the contributions from one-pion
(dashed line), more-pion (dashed-dotted line), and $\eta$ (dotted line)
production. Figure from \textcite{Drechsel:2004ki}.} \label{fig:SUM_sigmaTT_QQ}
\end{figure}
The helicity difference $\Delta\sigma$ of the proton is displayed in
Fig.~\ref{fig:SUM_sigmaTT_QQ} as function of the cm energy $W$ for several
values of the virtuality. The figure shows negative values near threshold,
because there the pions are produced in s waves leading to $\sigma_{1/2}$
dominance. With increasing values of $Q^2$, the s-wave production drops
rapidly. The $\Delta(1232)$ yields large positive values because of the strong
M1 transition, which aligns the quark spins (paramagnetism). For very small
values of $Q^2$, also the second and third resonance regions contribute with a
positive sign. However, this sign has changed already at $Q^2=0.5$~GeV$^2$. Let
us study this effect in more detail for the N*(1520) with multipoles $E_{2-}$
and $M_{2-}$. According to Eq.~(\ref{eq:6.1.14}) this resonance yields a term
$\Delta \sigma \sim |E_{2-}|^2 + 6\,{\mbox{Re}}\,(E_2^{\ast}M_{2-}) -
3|M_{2-}|^2$. This value is positive at the real photon point where the
electric dipole radiation (E1) dominates over the magnetic quadrupole radiation
(M2). Yet as the magnetic term increases with $Q^2$ faster than the electric
one, the helicity difference becomes negative for $Q^2 \approx 0.3$~GeV$^2$.
The latter finding is in agreement with perturbative QCD, which predicts the
dominance of helicity $\frac{1}{2}$ states at sufficiently large momentum
transfer. Figure~\ref{fig:SUM_sigmaTT_QQ} also shows an overall decrease of the
resonance structures with increasing values of $Q^2$, because the coherent
resonance effects are of long range and therefore strongly damped by form
factors. Finally, for momentum transfer beyond 4~GeV$^2$, the resonance
structures become small fluctuations on top of a broad background, the
low-energy tail of DIS.
\subsubsection{Recent data for GDH-like integrals}
\label{sec:VI.2.4}
The Bjorken sum rule of Eq.~(\ref{eq:6.2.17}) has been confirmed by a series of
experiments. A fit to all the available DIS data \cite{Anthony:2000fn} yields
the following asymptotic values: $\Gamma_1^p = 0.118 \pm 0.004 \pm 0.007$,
$\Gamma_1^n = -0.058 \pm 0.005 \pm 0.008$, and hence $\Gamma_1^p - \Gamma_1^n =
0.176 \pm 0.003 \pm 0.007$, in good agreement with the sum rule prediction of
$0.182 \pm 0.005$. The small value of $\Gamma_1^p$, on the other hand, led to
the ``spin crisis'' of the 1980's and taught us that less than half of the
nucleon's spin is carried by the quarks. With regard to the second spin
structure function, the BC sum rule predicts that $\Gamma_2(Q^2)$ vanishes
identically for all $Q^2$, and therefore the inelastic and elastic
contributions should have the same absolute value, namely
${\cal{O}}\,(Q^{-10})$ in the scaling limit.\\

\begin{figure}[]
\begin{center}
\includegraphics[width=0.95\columnwidth,angle=0]{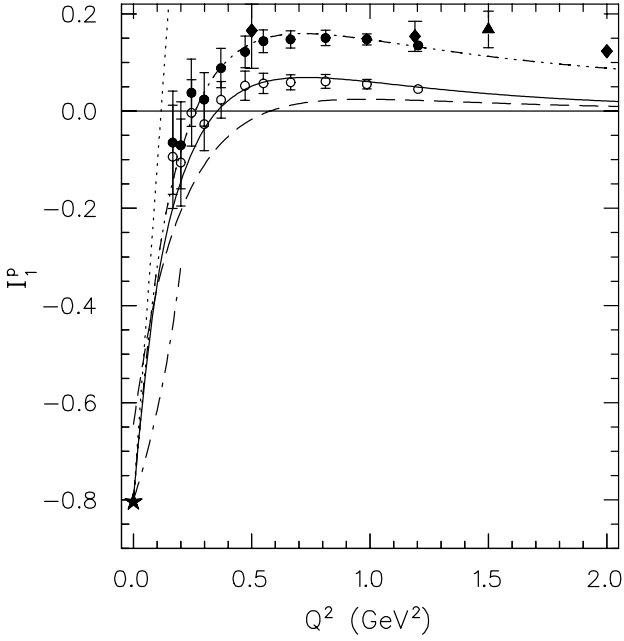}
\end{center}
\caption{The $Q^2$ dependence of the integral $I_1^p$ defined by
Eq.~(\ref{eq:6.2.16}). The open circles show the resonance contribution
($W<2$~GeV), the solid symbols also include the DIS contribution. The data are
from \textcite{Fatemi:2003yh}, CLAS Collaboration (circles),
 \textcite{Abe:1998wq}, SLAC (diamonds), and \textcite{Airapetian:2002wd},
HERMES (triangles). Full line: MAID including all channels up to $W=2$~GeV,
dashed line: one-pion channel only, dotted: ${\mathcal{O}}(p^4)$ prediction of
HBChPT \cite{Ji:1999sv}, dashed-dotted: relativistic baryon ChPT
\cite{Bernard:2002pw,Bernard:2002bs}, dash-dot-dotted: interpolating formula of
\textcite{Anselmino:1988hn}, asterisk: sum rule value at $Q^2=0$. Figure from 
\textcite{Drechsel:2004ki}.}
\label{fig:SUM_I1p}
\end{figure}
\begin{figure}[]
\begin{center}
\includegraphics[width=0.95\columnwidth,angle=0]{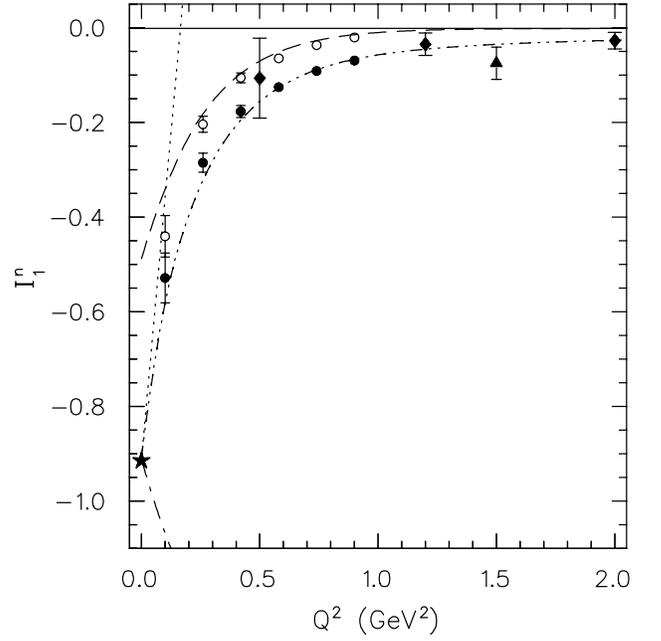}
\end{center}
\caption{The $Q^2$ dependence of the integral $I_1^n$ defined by
Eq.~(\ref{eq:6.2.16}). The data are from \textcite{Amarian:2002ar}, JLab
E94-010 Collaboration (circles). For further notation see
Fig.~\ref{fig:SUM_I1p}. Figure from \textcite{Drechsel:2004ki}.}
\label{fig:SUM_I1n}
\end{figure}
Figure~\ref{fig:SUM_I1p} shows the $Q^2$ dependence of $I_1^p$. The rapid
increase from large negative values near the real photon point to positive
values in the DIS region is particularly striking. The JLab data of
\textcite{Fatemi:2003yh} clearly confirm the sign change of $I_1^p$ at $Q^2
\approx 0.3$~GeV$^2$. These data are in good agreement with the MAID estimate
which covers the same energy region. However, with increasing momentum transfer
the DIS contributions at the higher energies become more and more important.
Altogether we see a rather dramatic transition from resonance-dominated
coherent processes at low $Q^2$ to incoherent partonic contributions at large
$Q^2$. This physics is driven by (I) the strong damping of the long-range
coherent effects by form factors, and (II) the change from $\frac{3}{2}$ to
$\frac{1}{2}$ helicity dominance in the second and third resonance regions. The
neutron integral as displayed in Fig.~\ref{fig:SUM_I1n} shows a similar rapid
increase with $Q^2$, except that $I_1^n$ approaches zero right away. The MAID
prediction for the one-pion channel is in reasonable agreement with the
resonance data \cite{Amarian:2002ar} except for the region of very small
momentum transfer. This disagreement may have its origin in uncorrected binding
effects of the ``neutron target'' in the MAID multipoles and/or the data
analysis. In Fig.~\ref{fig:SUM_I1p} we also display the predictions from heavy
baryon ChPT to ${\mathcal O}(p^4)$ of \textcite{Ji:1999sv} and relativistic
baryon ChPT to ${\mathcal O}(p^4)$ of \textcite{Bernard:2002pw,Bernard:2002bs}.
The figure shows that the chiral expansion can be only applied in a very
limited range of $Q^2 \lesssim 0.05$~GeV$^2$. The value at the real photon
point is, of course, obtained by inserting the amm as a low-energy constant.
However, the slope and the curvature of $I_1(Q^2)$ come about by a complicated
interplay of s-wave pion and resonance production. Whereas the former process
is well described by the pion loops of ChPT, the transition form factors and
widths of the resonances require a dynamical treatment of both the
$\Delta(1232)$ and the higher resonance region. However, ChPT is able to
describe the difference $I_1^p - I_1^n$ over a much larger $Q^2$ range
\cite{Burkert:2000qm}, because the contributions of the $\Delta(1232)$ and
other isospin $\frac{3}{2}$ resonances drop out in the isovector combination.
This opens the possibility to bridge the
gap between the low and high $Q^2$ regimes, at least for this particular observable.\\

\begin{figure}[]
\begin{center}
\includegraphics[width=1.0\columnwidth,angle=0]{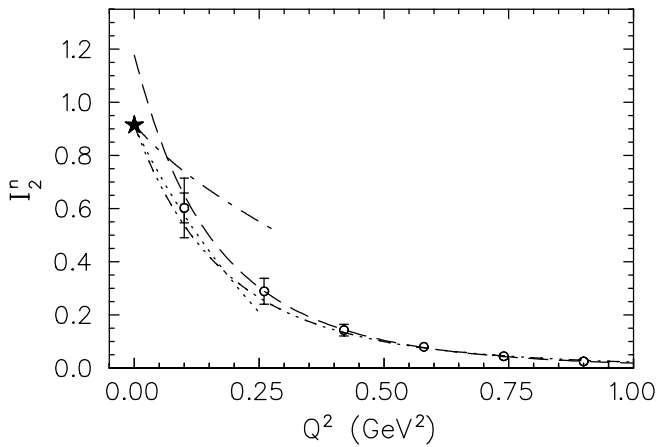}
\end{center}
\caption{The $Q^2$ dependence of the integral $I_2^n$ defined by
Eq.~(\ref{eq:6.2.16}). The neutron data were obtained by the JLab E94-010
Collaboration \cite{Amarian:2003jy}. Dash-dot-dotted line: Burkhardt-Cottingham
sum rule of Eq.~(\ref{eq:6.2.16}). For further notation see
Fig.~\ref{fig:SUM_I1p}. Figure from \textcite{Drechsel:2004ki}.} 
\label{fig:SUM_I2n}
\end{figure}
\begin{figure}[]
\begin{center}
\includegraphics[width=1.0\columnwidth,angle=0]{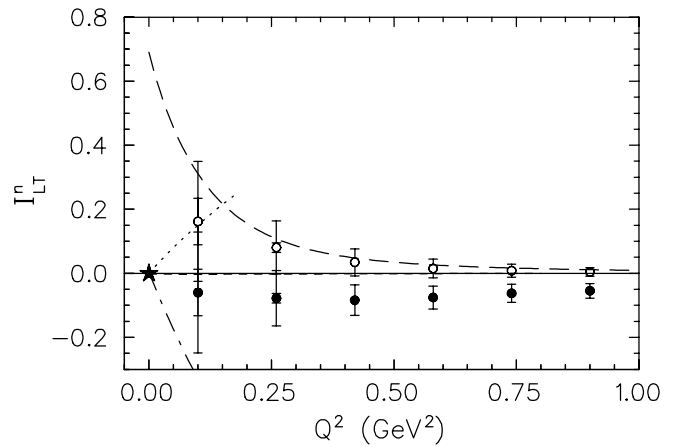}
\end{center}
\caption{The $Q^2$ dependence of the neutron integral $I_{LT}^n$ defined by
Eq.~(\ref{eq:6.2.13}). Open circles: the resonance contribution ($W<2$~GeV)
measured by  \textcite{Amarian:2002ar,Amarian:2003jy} at JLab, full circles:
resonance contribution plus estimate for DIS region. See Fig.~\ref{fig:SUM_I1p}
for further notation. Figure from \textcite{Drechsel:2004ki}.} 
\label{fig:SUM_ILT}
\end{figure}

Figure~\ref{fig:SUM_I2n} compares the neutron data of the JLab E94-010
Collaboration \cite{Amarian:2002ar} and the MAID prediction for $I_2^n(Q^2)$
with the BC sum rule. The MAID result overshoots the BC sum rule at small $Q^2$
but agrees with the data for $Q^2\gtrsim0.1$~GeV$^2$. At the much larger
momentum transfer of $Q^2=5$~GeV$^2$, the SLAC E155 Collaboration has recently
evaluated the BC integral in the measured region of $0.02\le x\le0.8$. The
results indicate a small deviation from the sum rule, but this could well be
compensated by contributions from the unmeasured region. The integrals
$I_{TT}^n$ and $I_{LT}^n$ have been derived from the $^3$He data of the JLab
E94-010 Collaboration \cite{Amarian:2002ar,Amarian:2003jy} and corrected for
nuclear effects according to  \textcite{CiofidegliAtti:1996cg}. The data for
$I_{TT}^n$ show qualitatively the same behavior as the discussed integral
$I_1^n$. The integral $I_{LT}^n$ is displayed in Fig.~\ref{fig:SUM_ILT}. This
observable deserves particular attention, because it samples the information
from the longitudinal-transverse cross section. As indicated by
Eq.~(\ref{eq:6.2.13}), the convergence of $I_{LT}^n$ requires that
$\sigma_{LT}$ drop faster than $1/\nu$ at large $\nu$. Because the
longitudinal-transverse interference involves a helicity flip, this is likely
to happen at sufficiently large $\nu$. However, there is little experimental
information on $\sigma_{LT}$ over the whole energy region, and therefore the
phenomenological description is on shaky ground. The zero of $I_{LT}^n$ at
$Q^2=0$ is particularly interesting, because this requires a complete
cancelation of resonance and DIS contributions. The agreement between the new
JLab data \cite{Amarian:2003jy} and MAID in the resonance region $(W<2$~GeV) is
quite satisfactory, except for the real photon point where both the
experimental and the theoretical error bars increase. Furthermore, the
contribution of DIS is known to be large and negative over the full $Q^2$
region, which brings the integral much closer to zero at $Q^2=0$. Concerning
the ChPT calculations \cite{Ji:1999sv,Bernard:2002pw}, the zero value at the
photon point is of course taken for granted, but the steep slope is a
prediction.
\subsubsection{Generalized polarizabilities}
\label{sec:VI.2.5}
\begin{figure}[]
\begin{center}
\includegraphics[width=0.9\columnwidth,angle=0]{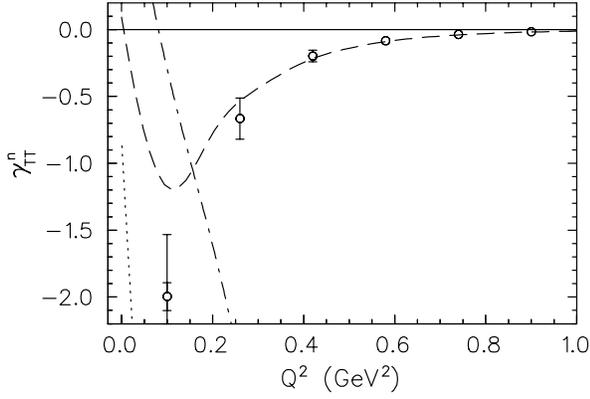}
\end{center}
\caption{The $Q^2$ dependence of the generalized neutron polarizability
$\gamma_{TT}^n$ defined by Eq.~(\ref{eq:6.2.12}). The open circles are the data
of the JLab E 94-010 Collaboration \cite{Amarian:2004yf}. See
Fig.~\ref{fig:SUM_I1p} for further notation. Figure from
\textcite{Drechsel:2004ki}.}
\label{fig:SUM_gamma}
\end{figure}

\begin{figure}[]
\begin{center}
\includegraphics[width=0.9\columnwidth,angle=0]{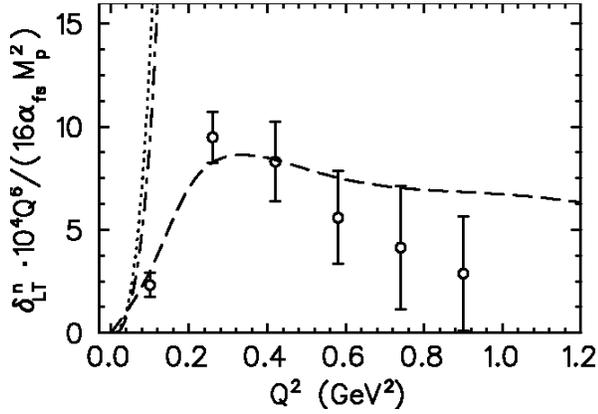}
\end{center}
\caption{The $Q^2$ dependence of the generalized neutron polarizability
$\delta_{LT}^n$ defined by Eq.~(\ref{eq:6.2.14}). Note that $\delta_{LT}^n$ has
been multiplied by a factor of $10^4Q^6/(16\alpha_\text{em} M_p^2)$ in order to
compensate for its rapid decrease with increasing $Q^2$. The open circles are
the data of the JLab E 94-010 Collaboration \cite{Amarian:2004yf}. See
Fig.~\ref{fig:SUM_I1p} for further notation. Figure from
\textcite{Drechsel:2004ki}.} 
\label{fig:SUM_delta}
\end{figure}

In section~\ref{sec:VI.1} we have discussed the delicate cancelation between
negative and positive contributions to the GDH-like integrals, and in
particular the rapid change of the integrals as functions of momentum transfer.
For the generalized polarizabilities, the integrands are weighted by an
additional factor $\nu^{-2}$ or $x^2$, which enhances the importance of the
threshold region relative to resonance excitations, and suppresses the
contributions of the DIS continuum above $W=2$~GeV. Figures~\ref{fig:SUM_gamma}
and \ref{fig:SUM_delta} display the transverse-transverse ($\gamma_{TT}$)
and longitudinal-transverse ($\delta_{LT}$) polarizabilities as function of 
$Q^2$. As in the case of $I_{TT}^n$, the data for $\gamma_{TT}^n$ 
\cite{Amarian:2004yf} show considerably more strength at small $Q^2$ than 
predicted for the one-pion contribution. However, the agreement for 
$Q^2\gtrsim 0.4$~GeV$^2$ is again quite satisfactory. In view of the
additional weight factor towards the low-energy region, this behavior is 
another indication that the ``neutron problem'' near the real photon point 
should be related to low-energy and long-range phenomena. A comparison with 
the predictions of ChPT shows that $\gamma_{TT}^n$ is a particularly sensitive
observable because of the cancelation between  s-wave pion production and
$\Delta$ resonance excitation, as is apparent from Eq.~(\ref{eq:6.1.14}). 
In fact, the additional weight factor $\nu^{-2}$ increases this cancelation 
considerably relative to the integral $I_{TT}$. It is therefore no big
surprise that ChPT cannot describe $\gamma_{TT}$ without including the 
$\Delta$ resonance. And indeed, the ${\mathcal{O}}(p^3)$ and 
${\mathcal{O}}(p^4)$ approximations of HBChPT change from positive to negative 
values, and also the newly developed Lorentz invariant version of ChPT misses 
the real photon point. This behavior changes for the longitudinal-transverse 
polarizability shown in Fig.~\ref{fig:SUM_delta}, because $\delta_{LT}$ is 
dominated by the s-wave term $S_{0+}^{\ast}E_{0+}$ in the multipole
expansion. The contribution of the $\Delta$ resonance is proportional to 
$S_{1+}^{\ast}E_{1+}$, which is much suppressed because both amplitudes are 
small. As a consequence $\delta_{LT}$ decreases rapidly as function of $Q^2$ 
without showing any pronounced resonance structures. In other words, the 
longitudinal-transverse polarization takes place in the outer regions of the 
nucleon and is mostly due to the pion cloud. This notion is well supported by 
the fact that the ChPT prediction for $\delta_{LT}$ is much better than for 
$\gamma_{TT}$.\\

\section{conclusion}
\label{sec:VII}

In this review we have concentrated on the bulk properties of hadrons seen at
low momentum transfer, such as shape, polarizability, and low-energy excitation
spectrum. We have presented many new precision data implying the compositeness
of the hadrons based on the interaction among the constituents, quarks and
gluons, as prescribed by the QCD Lagrangian. Experiments at high momentum
transfer have established this theory as the fundamental theory of the strong
interactions, with previously unexpected features like asymptotic freedom and
chiral symmetry. The great success of QCD at high $Q^2$ has become possible by
the decrease of the strong coupling constant with increasing $Q^2$, which
allows for a perturbative treatment of QCD in this region. This is definitely
not possible in the $Q^2$ region below 1\,GeV$^2$, which defines the realm of
non-perturbative QCD. Therefore, QCD has not yet passed its final tests in the
low-energy domain, in which new phenomena show up, as for instance the
confinement of the many-body system ``hadron'' and the spontaneous breaking of
the chiral symmetry. In fact, QCD encounters very fundamental conceptional
problems at low $Q^2$: One has to deal with the relativistic many-body aspect,
the theoretically not yet understood confinement, and the fundamental strong
interaction at the same time. The simplifications of the parton model typical
for the high $Q^2$ physics do not work in the non-perturbative region, and
neither is it possible to describe the bound many-body system of the light $u$
and $d$ quarks by basically non-relativistic physics as is done for systems
of heavy quarks.\\

This review has two punch lines. On the experimental side one has to insist on
measurements with the utmost precision. As it became evident only a new
generation of cw accelerators together with modern detectors and data
acquisition allowed for progress after almost two decades of stagnation.
However, since the most interesting observables like spin observables are
still difficult to get at, the experiments have to be well chosen. They 
have to be ``significant'' for contributing to the understanding of hadrons in
the framework of QCD. For this purpose a profound theoretical guidance is
needed. As promising examples of such significant experiments we mention:
\begin{itemize}
\item
A more complete and even more precise study of meson threshold production, in
particular through a study of spin observables (see section~\ref{sec:V.1}) and
the production of strange quarks, for example by kaon electroproduction.
\item
Real and virtual Compton scattering in order to disentangle all the spin
polarizabilties of the nucleon by double-polarization experiments (see
section~\ref{sec:IV}).
\item
A fresh approach to investigating the excitation spectrum of the nucleon, with
more attention given to the resonance-``background'' separation (see
section~\ref{sec:V.2}). Once more, such dedicated experiments require an
intense study of the spin observables. Such investigations are also the basis
for further progress in understanding the sum rules discussed in
section~\ref{sec:VI}.
\item
The nucleon form factors remain a challenge at both low and high $Q^2$ (see
section~\ref{sec:III}). The improvement of the data base at low $Q^2$ would
however take a very strong effort with the existing facilities. It is in
place here to regret the closure of the MIT/Bates storage ring which offered
the best means to perform such experiments at low $Q^2$.
\end{itemize}

On the theoretical side the challenges are not smaller. In this review we have
met several effective models such as constituent quark models, dynamical baryon
models, and effective meson field models. All are effective in the sense that
they describe the data more or less well and provide an insight into the
physics. They also provide the necessary guidance for the experiments. However,
these models are only ``inspired'' by QCD but not fundamentally based on this
theory. As of today only two schemes exist that are both directly based on QCD
and able to describe the low-energy region: chiral perturbation theory (ChPT)
and lattice gauge theory (LGT). Ironically the first one works best for very
small quark masses, whereas the second one requires large quark masses, at
least with the computer power of today. The extrapolation between these two
approximations is therefore still on shaky ground. However, the ongoing efforts
promise decisive turns in the near future:
\begin{itemize}
\item
The possibilities for calculating loops beyond the leading order in ChPT
is improving continously, and the necessary low-energy constants get better and
better constrained by experiment and, maybe, in the near future also by LGT.
\item
Lattice gauge theory is progressing through more effective algorithms and the
growing performance of the computers. It appears that the used quark masses
will soon approach sufficiently low values to make contact with ChPT by a
chiral extrapolation of the numerical results from LGT. This would indeed mark
an essential breakthrough in low-energy hadron physics.
\end{itemize}

As became evident in this review, the agreement between experiment and theory
is not yet satisfactory in quite a few cases, which leaves major challenges
in hadron physics for both sides.\\

We conclude that the study of hadrons at low energy and momentum transfer is a
unique possibility to unravel the structure of a relativistic many-body system
in the realm of QCD, one of our most fundamentally based quantum field
theories, as is highlighted by the quote from \textcite{Wilczek:1999id}: ``QCD
is our most perfect physical theory.''

\vspace*{5mm}

{\bf {Acknowledgement}}:  The authors are grateful to J.~Ahrens, R.~Beck,
M.~Distler, J.~Friedrich, H.~W.~Hammer, B.~Pasquini, F.~Maas, H.~Merkel,
U.~M\"uller, S.~Scherer, and L.~Tiator for their help and many stimulating
discussions. This work was supported by the Deutsche Forschungsgemeinschaft
(SFB 443). We thank the following publishers for their permission to reproduce
figures: Annual Reviews for the Annual Reviews of Nuclear and Particle
Science, Springer and the Italian Physical Society for the European Physical 
Journal A and C, and Elsevier for Physics Reports and Physics Letters. 

\bibliography{literature}

\end{document}